\documentclass[a4paper,11pt]{article}
\pdfoutput=1

\usepackage{jheppub}

\usepackage{subfig}
\usepackage{xspace}
\usepackage[countmax]{subfloat}

\usepackage{multirow}
\usepackage{makecell}
\usepackage{slashed}

\usepackage{booktabs}

\usepackage{array}

\makeatletter
\newcommand{\thickhline}{%
    \noalign {\ifnum 0=`}\fi \hrule height 1pt
    \futurelet \reserved@a \@xhline
}

\newcolumntype{"}{@{\hskip\tabcolsep\vrule width 1pt\hskip\tabcolsep}}

\makeatother

\newcommand{\ecf}[2]{e_{#1}^{(#2)}}

\newcommand{\Dobs}[2]{D_{#1}^{(#2)}}

\DeclareRobustCommand{\Sec}[1]{Sec.~\ref{#1}}

\DeclareRobustCommand{\App}[1]{App.~\ref{#1}}
\DeclareRobustCommand{\Tab}[1]{Table~\ref{#1}}

\DeclareRobustCommand{\Fig}[1]{Fig.~\ref{#1}}
\DeclareRobustCommand{\Figs}[2]{Figs.~\ref{#1} and \ref{#2}}
\DeclareRobustCommand{\Eq}[1]{Eq.~(\ref{#1})}

\DeclareRobustCommand{\Ref}[1]{Ref.~\cite{#1}}
\DeclareRobustCommand{\Refs}[1]{Refs.~\cite{#1}}

\newcommand{\Nsub}[2]{\tau_{#1}^{(#2)}}
\newcommand{\Nsubnobeta}[1]{\tau_{#1}}

\newcommand{\pythia}[1]{\textsc{Pythia\xspace #1}}

\newcommand{\fastjet}[1]{\textsc{FastJet\xspace #1}}
\newcommand{\herwig}[1]{\textsc{Herwig\xspace #1}}

\newcommand{\delphes}[1]{\textsc{Delphes\xspace #1}}

\preprint{MIT--CTP 4651\\
\vspace{-0.87cm}
\begin{flushright}
CP3-15-06\\
MCnet-15-05
\end{flushright}
}

\title{Tracking down hyper-boosted top quarks}

\author[a]{Andrew J. Larkoski,}
\author[b]{Fabio Maltoni,}
\author[b]{and Michele Selvaggi}

\affiliation[a]{Centre for Theoretical Physics, Massachusetts Institute of Technology, Cambridge, MA 02139, USA}
\affiliation[b]{Centre for Cosmology, Particle Physics and Phenomenology (CP3), Universit\'e catholique de Louvain, Chemin du Cyclotron 2, B-1348 Louvain-la-Neuve, Belgium
}

\emailAdd{larkoski@mit.edu}
\emailAdd{fabio.maltoni@uclouvain.be}
\emailAdd{michele.selvaggi@uclouvain.be}

\abstract{
The identification of hadronically decaying heavy states, such as vector bosons, the Higgs, or the top quark, produced with large transverse boosts has been and will continue to be a central focus of the jet physics program at the Large Hadron Collider (LHC).   At a future hadron collider working at an order-of-magnitude larger energy than the LHC,  these heavy states would be easily produced with transverse boosts of several TeV. At these energies, their decay products will be separated by angular scales comparable to individual calorimeter cells, making the current jet substructure identification techniques for hadronic decay modes not directly employable. In addition, at the high energy and luminosity projected at a future hadron collider, there will be  numerous sources for contamination including initial- and final-state radiation, underlying event, or pile-up which must be mitigated.  We propose a simple strategy to tag such ``hyper-boosted'' objects that defines jets with radii that scale inversely proportional to their transverse boost and combines the standard calorimetric information with charged track-based observables. By means of a fast detector simulation, we apply it to top quark identification and demonstrate that our method efficiently discriminates hadronically decaying top quarks from light QCD jets up to transverse boosts of 20 TeV. Our results open the way to tagging heavy objects with energies in the multi-TeV range at present and future hadron colliders.}

\begin{document} 
\maketitle

\section{Introduction}
\label{sec:intro}

The discovery of the Higgs boson \cite{Aad:2012tfa,Chatrchyan:2012ufa} represents a landmark in the exploration of the electroweak scale started with Run I of the LHC. The forthcoming Run II will extend the energy frontier for the direct searches further into the multi-TeV range, providing the first data that will test the up-to-now complete Standard Model (SM) to higher energies and to higher degree of precision. Among  the best probes available to study the electroweak scale and the underlying mechanism responsible for its breaking are the vector bosons, the Higgs, and the top quark.  Because of their large masses with respect to the electroweak scale, this implies a ``strong" interaction with the Higgs field and therefore possibly enhanced couplings to new resonances and a higher sensitivity to deviations from SM predictions. In particular, the heavy states in the SM are responsible for large radiative corrections to the Higgs mass leading to the hierarchy problem.  New physics models that solve the hierarchy problem and feature states at multi-TeV scales therefore will presumably significantly couple to heavy bosons and the top quark.

This sensitivity to new physics makes the efficient identification of vector bosons, the Higgs, and the top quark an important goal for the LHC and future colliders. Two simple factors need to be taken into account in  the quest for the optimal tagging strategies. The first is that the largest decay rates of all the electroweak heavy states of the SM are into hadronic final states, producing multiple jets. For instance, the top quark  features a two-body (semi-weak) decay into a $b$ quark and a $W$ boson, the latter decaying to a jet pair around 67\% of the time. For the Higgs boson the branching ratio into jets approaches 80\%. The inability to select and identify such objects when decaying hadronically would therefore severely limit the final statistics available for physics studies. The second factor is that all methods for identification and the relevant backgrounds are highly sensitive to the $p_T$ of the heavy state.  For example, top quarks produced near threshold decay to widely-separated and relatively low $p_T$ jets, with dominant backgrounds from multi-jet production in QCD.  This low-$p_T$ regime presents significant challenges, both because QCD backgrounds can be enormous and because the combinatorics of determining the set of jets which came from a single top quark decay can be very inefficient. The same argument applies to the heavy bosons of the SM.

At moderate boosts, with $p_T$ up to a few times the heavy state mass, the decay products begin to merge, and become collimated in the detector.  In this regime, combinatoric ambiguities and QCD backgrounds are greatly reduced.  All jets produced from the decay of a heavy state can be clustered into the same jet, with a large jet radius, and so backgrounds are limited to high-$p_T$ jets in QCD which have masses around that of the  heavy state.  Further, one can study the substructure of the fat jets to identify prongy structure that would be a signature of a decay, and be highly unlikely in a QCD jet.   With this motivation, the past several years have seen substantial development and implementation of observables for identifying boosted hadronically decaying heavy objects  produced at the LHC \cite{Abdesselam:2010pt,Altheimer:2012mn,Altheimer:2013yza}.  Several of the most powerful techniques have been validated on data \cite{CMS:2011xsa,Miller:2011qg,ATLAS-CONF-2012-066,Chatrchyan:2012mec,Aad:2012meb,ATLAS-CONF-2012-065,ATLAS:2012am,Aad:2013gja,Aad:2013fba,TheATLAScollaboration:2013tia,TheATLAScollaboration:2013sia,TheATLAScollaboration:2013ria,TheATLAScollaboration:2013pia,CMS:2013uea,CMS:2013kfa,CMS:2013wea,CMS-PAS-QCD-10-041,Aad:2014gea,LOCH:2014lla,CMS:2014fya,CMS:2014joa,CMS:2014ata,atlas_pu14} and are now standard tools for jet analysis at ATLAS and CMS.

Both because of the relatively low boosts and the fine granularity of the detectors, techniques for identifying heavy states at the LHC have impressive performance by exploiting all of the radiation from an event.  However, looking forward to Run II of the LHC and beyond, many of the techniques used thus far will be significantly limited for several reasons.  First, at higher luminosities there are more secondary proton collisions (pile-up) that contaminate the primary hard collision event.  Tracking can be used to identify the charged particles that originated from the hard collision, but observables that depend on calorimetry will be significantly degraded by pile-up contamination.  Grooming the jet \cite{Butterworth:2008iy,Ellis:2009su,Ellis:2009me,Krohn:2009th,Soyez:2012hv,Krohn:2013lba,Dasgupta:2013ihk,Larkoski:2014wba,Berta:2014eza,Cacciari:2014gra,Bertolini:2014bba} to remove that radiation that most likely came from pile-up will be required.  Also, as the centre-of-mass energy of the LHC increases, more heavy bosons and top quarks at higher transverse momenta will be created.  With a transverse momentum of a few TeV, the  decay products begin to merge into a single calorimeter cell with the current resolution of ATLAS and CMS.  At higher transverse momenta, unless the angular granularity increases significantly, using the calorimeter for identification of weak-scale particles will be essentially impossible.  In this ``hyper-boosted'' regime, where $p_T$s approach and exceed 10 TeV, it may be possible to access multi-boson/multi-top signatures such as $VVV$, $t\bar t H/V$, four top quarks, or even more exotic final states. Such final states produce huge numbers of low $p_T$ jets and it would to be extremely difficult to reconstruct the individual heavy states, separate them from background, or resolve combinatorics.

\begin{table}[t]
\begin{center}
\begin{tabular}{ll"l|l|l}
\midrule\midrule
& & \multicolumn{3}{c}{{ Cross section at $pp$, $\sqrt{s}=100$ TeV}}\\
\cline{3-5}
& Process &\makecell{$p_T > 1$ TeV \\ (pb)}& \makecell{$p_T > 5$ TeV \\ (fb)}& \makecell{$p_T > 10$ TeV \\ (ab)}\\
\midrule

\parbox[t]{4mm}{\multirow{9}{*}{\rotatebox[origin=c]{90}{{\bf Standard Model}}}} \parbox[t]{2mm}{\multirow{6}{*}{\rotatebox[origin=c]{90}{Signals}}} &$pp\to t\bar{t}$ & 12  &2.8  & 24 \\
& $pp\to t\bar{t}j$ &52  & 14  & 94  \\
& $pp\to t  j$ &0.67 & 0.46 & 0.76  \\
& $pp\to t\bar{t}V$ &0.40  & 0.30  & 3.7  \\
& $pp\to t\bar{t}H$ &0.19  & 7.4e-02 &0.65 \\
& $pp\to t\bar{t}t\bar{t}$ &0.17 &8.5e-02  &0.51 \\

\cline{2-5}
\hspace{4mm}\parbox[t]{-40mm}{\multirow{2}{*}{\rotatebox[origin=c]{90}{Bkgds}}} & $pp\to jj$ &3500  &1000 &11000  \\
& $pp\to jjV$ & 110  & 130  & 2200  \\
\midrule

\parbox[t]{6mm}{\multirow{6}{*}{\rotatebox[origin=c]{90}{{\bf BSM}}}} & $pp\to Z' \to t \bar{t}$  ($m_{Z'}$ = 3 TeV) & 4.6  & -  & -   \\
& $pp\to Z' \to t \bar{t}$  ($m_{Z'}$ = 15 TeV) &  7.1e-03 &  4.7 & -   \\
& $pp\to Z' \to t \bar{t}$  ($m_{Z'}$ = 30 TeV) &  7.1 e-05 &  6.5e-02 & 48   \\
& $pp\to \tilde t \tilde t \to t\bar{t} + \slashed{E}_T$ ($m_{\tilde t}$ = 1 TeV) & 0.49  & 7.8e-03  &  -  \\
& $pp\to \tilde t \tilde t \to t\bar{t} + \slashed{E}_T$ ($m_{\tilde t}$ = 5 TeV)& 7.5e-04  & 0.063 & - \\
& $pp\to \tilde t \tilde t \to t\bar{t} + \slashed{E}_T$ ($m_{\tilde t}$ = 10 TeV)& 4.4e-06  & 0.27e-03 & 0.024  \\
& $pp\to \tilde g \tilde g \to t\bar{t}t\bar{t} + \slashed{E}_T$ ($m_{\tilde g}$ = 2 TeV)& 2.5  & 0.94  & - \\
& $pp\to \tilde g \tilde g \to t\bar{t}t\bar{t} + \slashed{E}_T$ ($m_{\tilde g}$ = 5 TeV)& 2.7e-02  & 1.5 & 11  \\
& $pp\to \tilde g \tilde g \to t\bar{t}t\bar{t} + \slashed{E}_T$ ($m_{\tilde g}$ = 10 TeV)& 1.9e-04  & 0.12 & 4.5  \\
\midrule\midrule
\end{tabular}
\end{center}
\caption{
Inclusive leading-order cross sections for Standard Model and beyond-Standard Model ($\text{BR} = 1$) processes with at least one top quark with $p_T >$  1, 5, 10 TeV at a 100 TeV future collider. For the Standard Model backgrounds, the momentum requirement is imposed on one of the jets.  Omitted entries have cross sections which are too small to be relevant.
}
\label{tab:scaling}
\end{table}

With the recent excitement for a long-term goal of a 100 TeV future circular collider (FCC) \cite{Gershtein:2013iqa,cern100,china100}, both of these issues become more acute.  For example, at such high collision energies, weak bosons and the top quark could be produced with transverse momenta of order 10 TeV, with decay products separated by angles much smaller than the resolution of any foreseeable (sufficiently compact) calorimeter.  To be more concrete, in \Tab{tab:scaling}, we collect representative inclusive cross sections for several SM and beyond-SM (BSM) processes at a 100 TeV collider calculated to leading order.  With very strong $p_T$ cuts on jets in the final state, backgrounds of jets from QCD can be dramatically reduced, while probing heavy mass resonances.  For example, the decay of a SM-like $Z'$ of 15 TeV to top quarks (assuming a branching fraction equal to 1) would produce about 5 events per fb$^{-1}$ at 100 TeV with $p_T >$  5 TeV, while the dominant background, from $pp\to jj$ events, produces about 1000 events per fb$^{-1}$.  Further predictions at next-to-leading order in QCD for cross sections with final state  $W,Z,H$ bosons and top quarks can found in \Ref{Torrielli:2014rqa}. 

The 100 TeV collider is projected to run at luminosities that are orders of magnitude greater than the LHC and so pileup will be a huge issue and grooming techniques will need to be understood at transverse momenta well beyond their current range of validity.  In addition, even  resolving these issues, for a QCD jet with large transverse momentum, a perturbative emission can affect the jet mass by hundreds of GeV,  comparable to the electroweak scale.  The mass of a jet of radius $R\simeq 1$ and transverse momentum $p_T$ is approximately given by  
\begin{equation}
m^2\simeq p_T\,p_{T}^{\rm soft} R^2 \ ,
\end{equation}
where $p_T^{\rm soft}$ is the transverse momentum of a soft emission clustered in the jet.  For example, at transverse momenta of $p_T\gtrsim 6$ TeV, an emission of only $p_T^{\rm soft}= 5$ GeV would contribute to the jet mass an amount greater than the mass of the top quark.

Therefore, to push identification of electroweak-scale states to higher and higher energies requires controlling all of these effects.
In this paper, we propose a new method as a robust and relatively $p_T$-independent procedure for identifying heavy SM states at very high boosts produced at a future collider and apply it to the case of the top quark.  Our approach to boosted object discrimination at very high $p_T$ consists of combining three elements:

\begin{enumerate}

\item[I.] {\bf Global-jet calorimetric information}\\ We are interested in the extreme limit where the energy deposit of a jet is confined to a single, or only a few, calorimetric cells. We assume that the information on the total energy of the jet  (possibly determined also using the track information as in a particle-flow algorithm~\cite{CMS:2009nxa}) is available, but detailed information about the energy distribution inside the jet is not.

\item[II.] {\bf Inner-jet charged track information}\\ We exploit the high angular resolution of the tracking system to define  observables sensitive to the internal structure of the jet only using charged particle tracks.

\item[III.] {\bf Dynamic contamination removal}\\ Contamination from initial state radiation (ISR), underlying event (UE) and pile-up is proportional to area of the jet. In addition for a coloured particle such as the top quark, for a fixed jet radius more perturbative QCD radiation is emitted  as the $p_T$ increases and is collected by the jet algorithm. Such an effect degrades the  accuracy of the jet mass reconstruction and therefore the efficiency of tagging the boosted object. On the other hand, the typical angular size of a jet generated by a particle of mass $m$ scales as $m/p_T$, i.e., inversely proportional to the transverse momentum. To mitigate all these effects  we propose use of a jet radius that scales inversely with the jet $p_T$.  This is similar to the variable $R$ jet algorithm \cite{Krohn:2009zg}, with the important difference that in our method the clustering metric is not modified.

\end{enumerate}

To develop and test our identification strategy, we employ  \delphes{} \cite{deFavereau:2013fsa,Selvaggi:2014mya}, a modular and  fast detector simulation framework.  With \delphes, we are able to make qualitative, and (thanks to especially designed modifications/improvements) to some extent also quantitative statements about  finite resolution effects on the final discrimination power.

This paper is organised as follows.  In \Sec{sec:obs_meth}, we introduce the methods and observables used for the identification of heavy state jets at very high $p_T$ explicitly applied to the case of a top quark.  In addition to identifying top quarks through the jet invariant mass reconstruction, we measure $N$-subjettiness and energy correlation function observables to identify the substructure characteristic of a top quark decay. In \Sec{sec:onept}, we apply our top quark finding procedure to a single jet $p_T$ range, presenting a detailed analysis and comparison between LHC detectors and projected performance of detectors in a future collider.  We also show that, even in the hyper-boosted regime, top quark identification efficiencies can be 40\%, while rejecting over 90\% of the dominant background. We draw our conclusions and discuss the outlook of tagging hyper-boosted objects in \Sec{sec:conclusions}.  Appendices provide details of the \delphes{}fast detector simulation used throughout this paper and plots of top tagging results for a wide range of jet $p_T$.

\section{Observables and methodology}\label{sec:obs_meth}

In this section, we introduce the methods and observables that we will use throughout this paper to identify boosted hadronically decaying states. While the approach is quite general and, as it will be clear in the following, could be applied to tag other heavy states such as colour-singlet bosons,  in this work we will focus on top quark identification.  

\subsection{Jet finding and definition}\label{sec:jetdef}

For a given jet radius $R$, the angular size of the top quark decay products scales as $m_\text{top}/p_T$ and therefore  significantly shrinks as the $p_T$ increases. On the other hand even in the absence of pile-up, the amount of radiation in the event, both from the initial and final state, generically increases with the jet $p_T$.  Our aim will be to show that mitigating both initial- and final-state radiation effects can be efficiently achieved by dynamically scaling the jet radius by the jet $p_T$.

\begin{figure}
\begin{center}
\includegraphics[width=8.5cm]{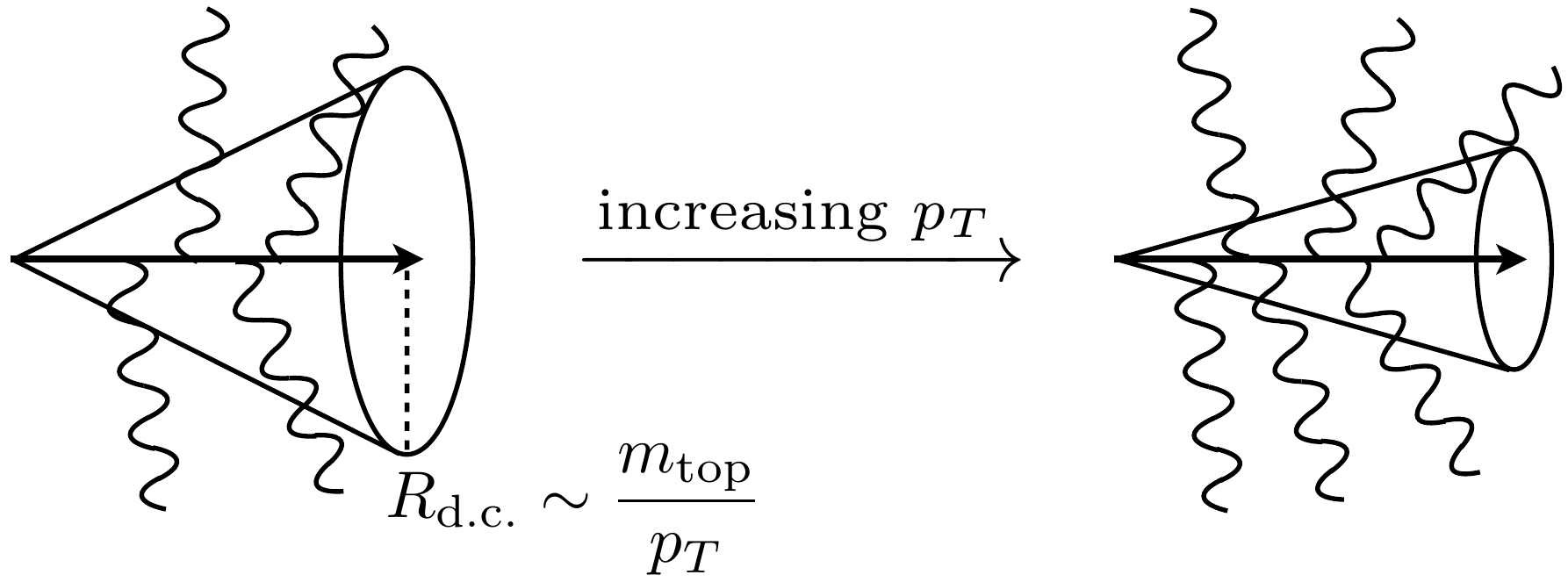}
\end{center}
\caption{
Illustration of the dead-cone effect.  Final-state radiation from a top quark is suppressed within an angular region scaling like $R_\text{d.c.}\sim m_\text{top}/p_T$, called the dead-cone region.  As the $p_T$ of the top quark increases, the amount of radiation emitted from the top quark in a fixed angular region increases both due to the increase in energy and the decrease in size of the dead-cone region.
}
\label{fig:deadcone}
\end{figure}

Because the top quark is unstable, final-state radiation (FSR) arises from two possible sources: either before it decays, or from the the daughter bottom quark after the decay.\footnote{We remind the reader that the notion of initial- and final-state radiation and that of associating emissions to individual coloured particles is an approximation valid only in the collinear limit in the narrow-width approximation.  For the top quark, the notion of radiation before or after its decay is meaningful when $1/E_{\rm rad} \simeq \tau_{\rm rad} < \tau_{\rm top} = 1/\Gamma_{\rm top}$, where $E_{\rm rad}$ is the characteristic energy scale of the radiation. By contrast, for a colour-singlet like the $W$ boson, radiation from its decay products are localised about the direction of the $W$, in a region scaling like the characteristic angular size of the decay, $m_W/p_T$.}  Ideally, for an optimal invariant mass reconstruction, the former type of radiation should not be clustered into the top quark jet while the latter should.   At low $p_T$s, radiation from  top quark is suppressed, as it must vanish when the top is at rest.  This is a general property of QCD (and QED) radiation from any massive particle that often phrased in terms of a dead cone:  radiation about the direction of motion of the top quark is suppressed in a cone of angular size $m_\text{top}/p_T$; see \Fig{fig:deadcone}.  As the momentum of the top quark increases, the dead-cone shrinks and more phase space opens up for radiation from the top quark.  

The contribution from ISR (or UE) to the $p_T$ of the jet scales like $R^2$ (proportional to the area of the jet) while the effect on the squared mass scales like $R^4$ because ISR is approximately uncorrelated with final state jets.  At very high transverse momenta, ISR will only distort the $p_T$ spectrum of the jets slightly, but can affect the jet mass substantially.  To see this behavior, consider a top quark jet of radius $R$ with a single emission near the boundary of the jet.  As long as the energy of this emission is small with respect to the energy of the top, its effect on the jet mass $m$ is approximately given by
\begin{equation}
m^2 \simeq m_\text{top}^2+p_{T}\,p_{T}^{\rm ISR} R^2 \ .
\end{equation}
Therefore, an ISR emission in the jet can contribute a mass comparable to the mass of the top quark when
\begin{equation}
p_{T}^{\rm ISR} \simeq \frac{m_{\rm top}^2} {p_{T} R^2} \ .
\end{equation}
At moderate boosts, typical of the LHC, where $p_{T}\sim\text{ few}\times m_\text{top}$, ISR of 50 GeV or so can affect the jet mass by an ${\cal O}(1)$ amount (with $R\sim 1$).  While this is quite hard radiation at the LHC, its effect must be mitigated in  top quark mass measurements; numerous methods have been introduced to groom jets so as to remove contaminating radiation in the jet \cite{Butterworth:2008iy,Ellis:2009su,Ellis:2009me,Krohn:2009th,Soyez:2012hv,Krohn:2013lba,Dasgupta:2013ihk,Larkoski:2014wba,Berta:2014eza,Cacciari:2014gra,Bertolini:2014bba}.  However, at the ranges of $p_T$s accessible by a future collider, the jet $p_T$ can be several to tens of TeV.  In this regime, even emissions of a few GeV can change the jet mass by an ${\cal O}(1)$ amount.  Thus, it is absolutely necessary to consider these effects at a future collider.

One is therefore lead to consider whether grooming methods that have been successfully applied at the LHC to mitigate contamination from both FSR and ISR  could be applied at higher energies and luminosities.  So far the standard grooming techniques have not been studied in detail in such an extreme environment\footnote{Some studies of extreme pile-up or jet grooming at energies and luminosities beyond current LHC applications are \cite{Avetisyan:2013onh,Anderson:2013kxz,Larkoski:2014bia,CMS:2014ata,atlas_pu14}.} and so we will consider an alternative (simpler and possibly more robust) approach.  

To this aim, we note that the ISR effects scale like the jet radius to a positive power and the dead-cone effect suppresses FSR in a region of angular size $m_\text{top}/p_T$. We can therefore reduce the contamination from radiation by  appropriately scaling the jet radius by (the inverse of) its $p_T$.  The specific procedure that we employ for boosted top quarks is the following.  We first find jets with the anti-$k_T$ algorithm \cite{Cacciari:2008gp} and the Winner-Take-All (WTA) recombination scheme \cite{Bertolini:2013iqa,Larkoski:2014uqa,Larkoski:2014bia,Salambroadening} with a fixed jet radius, which we take to be $R=1.0$. We then recluster the jet with the anti-$k_T$ algorithm where we set the subjet radius to:
\begin{equation}\label{eq:scaling_radius}
R = C\frac{ m_{\rm top} }{p_{T}} \ ,
\end{equation}
where $C$ is a constant and $p_{T}$ is defined by the original, fixed-radius, jet.  We then keep only the hardest subjet found with this procedure and promote it to be the boosted top quark jet.  With this prescription, the resulting jet is parametrically affected by contamination from ISR as
\begin{align}
m^2 & \simeq m_{\rm top}^2+p_{T} \,  p_{T}^ {\rm ISR} 
\left(\frac{C \, m_{\rm top} } { p_{T} }\right)^2 \nonumber\\
&\simeq m_{\rm top}^2 \left( 1+C^2\frac{p_{T}^{\rm ISR}}{p_{T}} \right) \ ,
\end{align}
whose effects are suppressed by the (small) ratio $p_{T}^ {\rm ISR}/p_{T}$.  Similarly, because we choose the jet radius to scale with jet $p_T$, the dead-cone effect suppresses contamination from FSR over a fraction of the area of the jet that is independent of jet $p_T$.  To ensure that the scaled radius jet captures the decay products of the top quark, in the remainder of the paper, we will adopt $C=4$, corresponding to including above 95\% of the energy fraction of the top quark decay products.\footnote{\Ref{Han:2014iha} proposed a shrinking cone algorithm with a radius determined by the demanding that a fixed fraction of final state radiation was captured into the jet.} We note that this coefficient should  be optimised in an actual analysis. However, as our aim is to illustrate the principle, we prefer to choose a conservative value so to be sure to cluster as much radiation from the top quark decay products as possible.

This approach is similar to variable $R$ jets \cite{Krohn:2009zg} in which the clustering metric for the $k_T$ class of algorithms \cite{Catani:1993hr,Ellis:1993tq,Dokshitzer:1997in,Wobisch:1998wt,Wobisch:2000dk,Cacciari:2008gp} is modified to be
\begin{align}
d_{ij}&=\min[p_{Ti}^{2n},p_{Tj}^{2n}] R_{ij}^2\,, &d_{iB} = p_{Ti}^{2n}\frac{\rho^2}{p_{Ti}^2} \ ,
\end{align}
where $\rho$ is a dimensionful constant.  The beam distance $d_{iB}$ sets the effective jet radius to scale inversely with the $p_T$ of the jet.  In contrast to variable $R$ jets, our procedure does not modify the jet algorithm.  Because we first find a fat jet and then recluster to find the hardest subjet, our procedure effectively defines an infrared and collinear (IRC) safe seeded cone jet algorithm with a radius that scales inversely with $p_T$.

\subsection{Track-based observables}\label{sec:track}

Scaling the jet radius inversely with $p_T$ decreases the amount of contamination radiation in the jet, though it does so at the cost of reducing the number of calorimeter cells in the detector that contribute to the jet.  As a consequence, the calorimetric angular resolution is also reduced, though this effect was to be  expected anyway because the angular size of the decay products decreases as the $p_T$ increases. Here, we explore recovering high angular resolution by measuring jet observables from charged particle tracks.  We propose to use standard calorimetric information (together with tracks in the case of particle-flow algorithms) to determine the total energy of the jet and use tracking information only for its substructure.  For observables like the jet mass, when measured on tracks, this systematically biases the mass to lower values because charged particles do not contain the full energy of the jet.  However, the bulk of this bias can be removed and the jet mass restored to its ``nominal'' value by simply rescaling the track-based mass by the ratio of the total jet $p_T$ to the $p_T$ of its charged tracks.\footnote{We thank Gavin Salam for suggesting this procedure.  A related technique was employed in the charged-track version of the HEPTopTagger \cite{Plehn:2009rk,Plehn:2010st} of \Ref{Schaetzel:2013vka}.}  That is, in this paper we define the reconstructed jet mass as 
\begin{equation}\label{eq:rescaled_tracks}
m = \frac{p_{T}}{p_{T}^{\text{tracks}}} m_{{\text{tracks}}} \,,
\end{equation}
where $m_{{\text{tracks}}}$ is the mass as measured on charged tracks and $p_{T}^{\text{tracks}}$ is the transverse momentum of the tracks.
We will also compare observables measured on current and projected future calorimetry to the track-based measurements.

\begin{figure}
\hspace*{-0.5cm}
\subfloat[]{\label{fig:mass_pt25}
\includegraphics[width=0.51\textwidth]{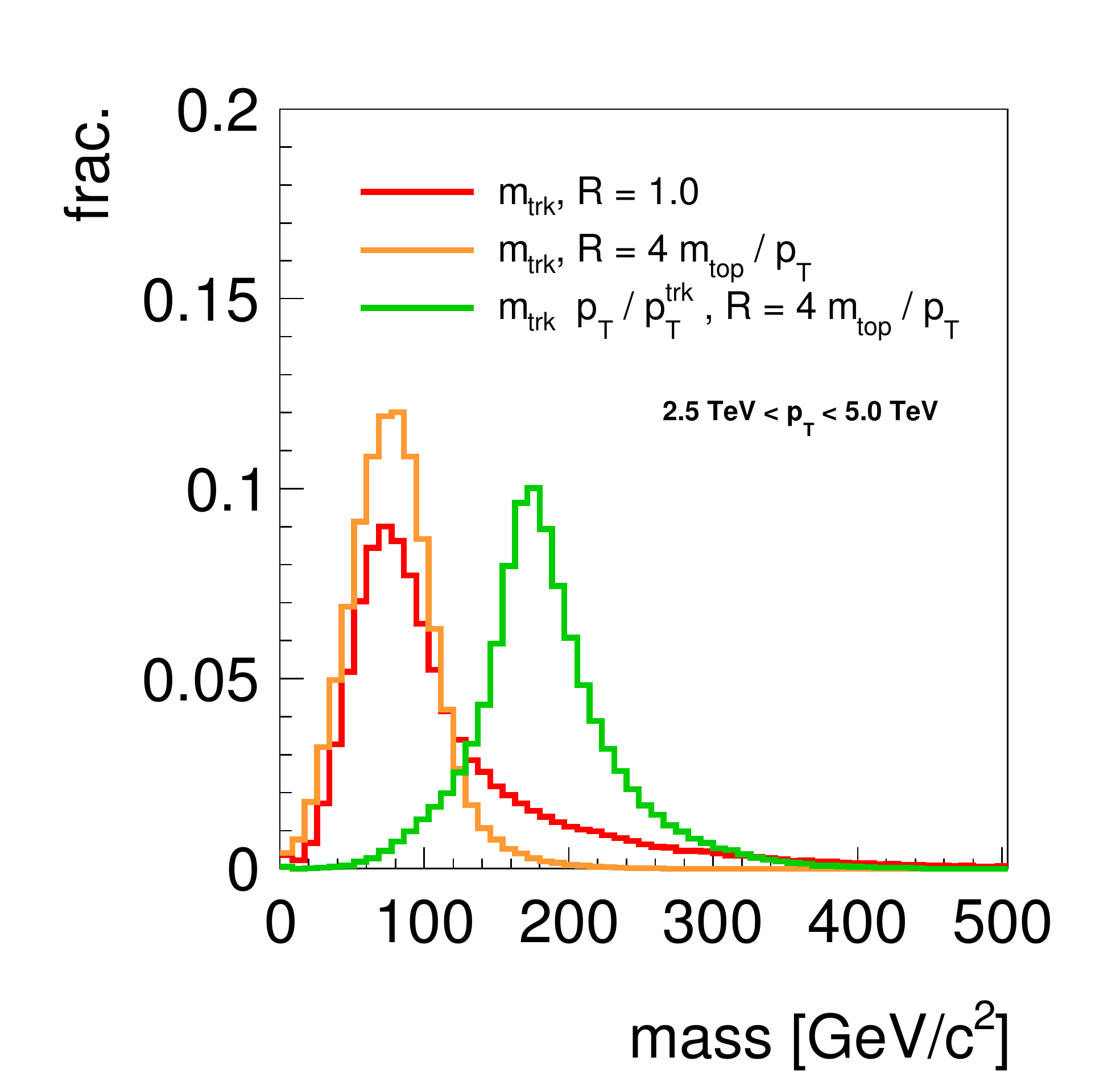}
}
\subfloat[]{\label{fig:mass_pt100}
\includegraphics[width=0.51\textwidth]{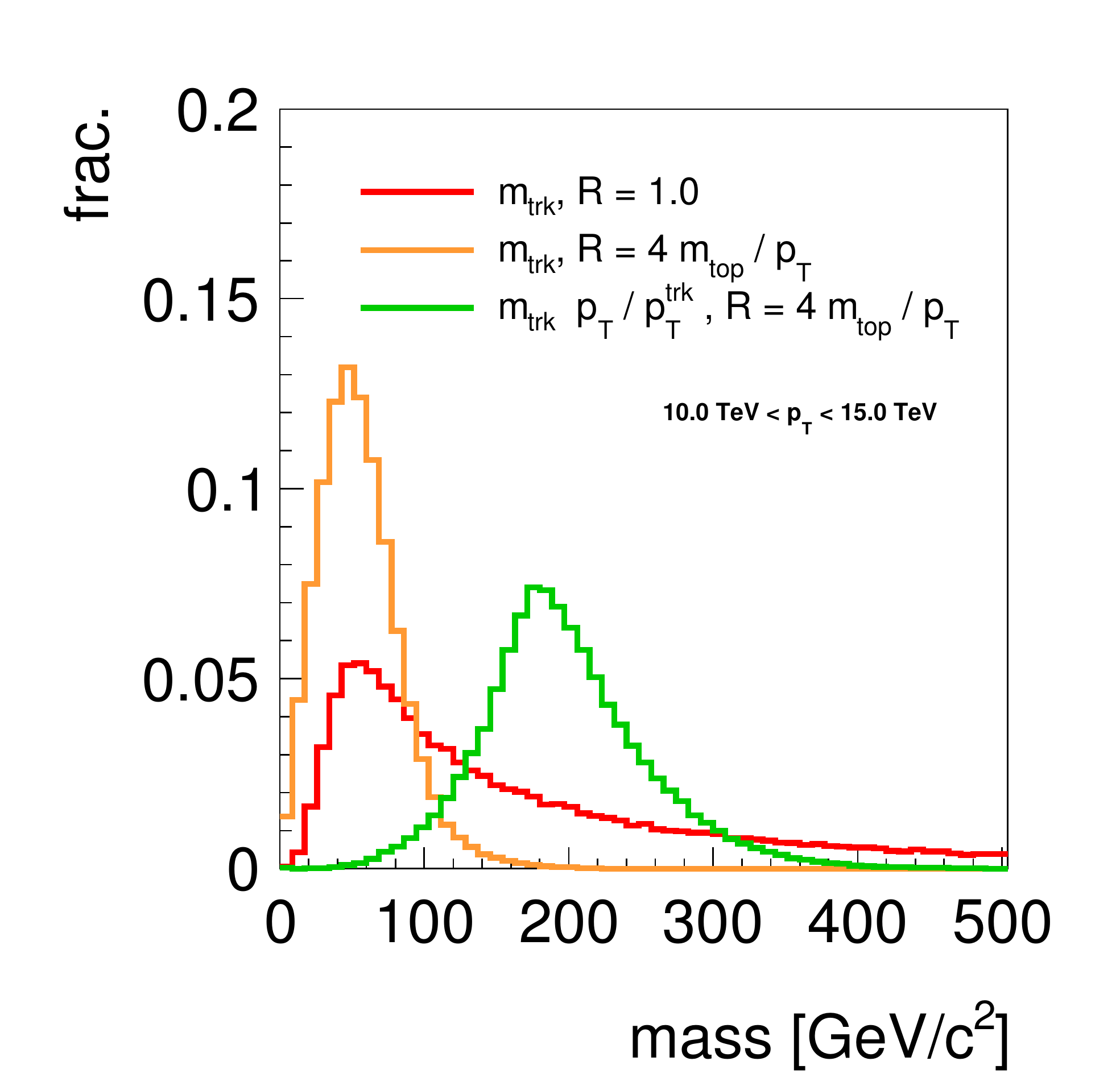}
}

\caption{Distributions of the jet mass defined in three different ways in two $p_T$ bins: $2-5$ TeV (left) and  $10-15$ TeV (right).  The red curves are the mass of jets with radius $R=1.0$ measured on tracks; orange curves are the mass of jets with scaled radius $R= 4m_\text{top}/p_T$ measured on charged tracks; green curves are the track mass of jets rescaled by the ratio of total jet $p_T$ to the $p_T$ of tracks with the scaled jet radius $R= 4m_\text{top}/p_T$.
}
\label{fig:scaledmass}
\end{figure}

In \Fig{fig:scaledmass}, we plot the top quark jet mass distribution measured on charged tracks to illustrate the effectiveness of our procedure, for two jet energy ranges. For these plots, we have simulated $pp\to t\bar t$ at leading order at 100 TeV center-of-mass energy with the {\sc MadGraph5\_aMC@NLO 2.2.2}~\cite{Alwall:2014hca}, \pythia{6.4} \cite{Sjostrand:2006za,Sjostrand:2007gs}  and \delphes{3.1.2} simulation chain.  Jets are clustered with \fastjet{3.0.6}~\cite{Cacciari:2011ma} using the anti-$k_T$ algorithm and WTA recombination scheme.  We identify the jet in the event with the largest $p_T$ that lies in the appropriate bin.  We plot results corresponding to three different  jet mass definitions: the track jet mass measured on jets with fixed jet radius $R=1.0$, the track jet mass measured on jets with radius that scales with $p_T$ as described earlier, and the rescaled track jet mass defined in \Eq{eq:rescaled_tracks}.  Both bare track jet masses are, as expected, systematically smaller than the true top quark mass.  As the jet $p_T$ increases from a few TeV to over 10 TeV, the track mass as measured on fixed-radius jets significantly drifts to higher values, due to the increased contamination from ISR/FSR.  On the other hand, the rescaled track mass measured on the scaled-radius jet peaks around the top quark mass, independent of the jet $p_T$ bin.

While the measurement of the jet mass is certainly a key observable to successfully identify  a boosted top quark, it has to be kept in mind that large masses can also be generated by perturbative soft and collinear emissions in a QCD jet initiated by light quarks or gluons at high transverse momentum.  In the collinear approximation, the average squared QCD jet mass is \cite{Ellis:2007ib,Salam:2009jx}
\begin{equation}
\langle m^2 \rangle \simeq a_i\frac{\alpha_s}{\pi}p_T^2 R^2 \,,
\end{equation}
where $a_i$ is a constant that depends on the jet algorithm and is proportional to the colour of the initiating parton.  While this provides a rule-of-thumb for the location of the peak, the QCD jet mass distribution is very wide.  Nevertheless, at sufficiently high $p_T$ with a fixed jet radius, QCD jets can have masses comparable to and even exceeding that of the top quark.  This approximately occurs when 
\begin{equation}
m_\text{top}^2 \lesssim a_i\frac{\alpha_s}{\pi}p_T^2 R^2 \,,
\end{equation}
or when $p_T \gtrsim 600$ GeV, assuming $R\simeq a_i/\pi \simeq 1$.  Using our scaled jet radius procedure, the mass of QCD jets is instead modified to
\begin{equation}
\langle m^2   \rangle \simeq a_i\frac{\alpha_s}{\pi}C^2m_\text{top}^2 \,,
\end{equation}
independent of the jet $p_T$, but comparable to the mass of the top quark.  Therefore a mass cut is not sufficient to efficiently reduce QCD backgrounds, and observables that are independent of the jet mass must be used. 

Particularly  sensitive observables for boosted top quark identification are those that measure the prongy-ness of the jet.  QCD jets dominantly consist of a single hard core of radiation, while hadronically decaying top quark jets typically have a 3-prong substructure. Several such observables have been proposed and studied~\cite{Abdesselam:2010pt,Altheimer:2012mn,Altheimer:2013yza}.  In this work  we employ the $N$-subjettiness observables $\tau_N^{(\beta)}$ and the $n$-point energy correlation functions $\ecf{n}{\beta}$.  The $N$-subjettiness observables $\tau_N^{(\beta)}$ are defined as
\begin{equation}
\tau_N^{(\beta)} = \sum_{i\in J} p_{Ti}\min \left\{
R^\beta_{i1},\dotsc,R^\beta_{iN}
\right\}\,,
\end{equation}
where the sum runs over particles in the jet $J$, $p_{Ti}$ is the transverse momentum of particle $i$ with respect to the beam, and $R_{iK}$ is the boost-invariant angle between particle $i$ and subjet axis $K$, and $\beta>0$ is an angular exponent whose value can be used to control sensitivity to wide-angle radiation.  In our study, we use the exclusive $k_T$ algorithm \cite{Catani:1993hr} with the WTA recombination scheme on the jet to define the $N$ subjet axes.  For the identification of 3-prong top quark jets, it has been shown \cite{Thaler:2010tr,Thaler:2011gf} that the ratio 
\begin{equation}
\tau_{3,2}^{(\beta)} = \frac{\tau_3^{(\beta)}}{\tau_2^{(\beta)}} \,,
\end{equation}
is an efficient observable for top quark identification and is widely used in both ATLAS and CMS experiments.  In our analysis, we will measure  $\tau_{3,2}^{(\beta)}$ using only track information and on jets with a scaled jet radius, as described in \Sec{sec:jetdef}.

The (dimensionless) $n$-point energy correlation functions  are defined as (for $n=2,3,4$) \cite{Larkoski:2013eya}:
\begin{align}
\ecf{2}{\beta} &=\frac{1}{p_{T}^2}\sum_{i<j \in J}   p_{Ti}\, p_{Tj} R_{ij}^\beta  \,, \nonumber \\
\ecf{3}{\beta} &=\frac{1}{p_{T}^3}\sum_{i<j<k \in J}   p_{Ti}\, p_{Tj}\,p_{Tk} \left(R_{ij} R_{ik} R_{jk}\right)^\beta  \,, \nonumber \\
\ecf{4}{\beta} &=\frac{1}{p_{T}^4}\sum_{i<j<k<l \in J}   p_{Ti} \,p_{Tj}\,p_{Tk}\, p_{Tl} \left(R_{ij} R_{ik} R_{il} R_{jk} R_{jl} R_{kl}\right)^\beta \,,
\end{align}
where $R_{ij}$ is the boost invariant angle between particles $i$ and $j$.  Employing power counting of the soft and collinear regions of phase space, \Ref{Larkoski:2014zma} showed that the particular combination
\begin{align}\label{eq:D3_def}
D_3^{(\alpha,\beta,\gamma)}\equiv  \frac{   \ecf{4}{\gamma}   \left ({\ecf{2}{\alpha}}\right)^{\frac{3\gamma}{\alpha}}  }{     \left( \ecf{3}{\beta}\right )^{\frac{3\gamma}{\beta}}     } +x \frac{   \ecf{4}{\gamma}   \left (\ecf{2}{\alpha}\right)^{\frac{2\gamma}{\beta}-1}  }{      \left (\ecf{3}{\beta}\right )^{\frac{2\gamma}{\beta}}    }   +y \frac{   \ecf{4}{\gamma}   \left (\ecf{2}{\alpha}\right)^{\frac{2\beta}{\alpha}-\frac{\gamma}{\alpha}}  }{      \left (\ecf{3}{\beta}\right )^{2}    }\,,
\end{align}
for angular exponents $\alpha,\beta,\gamma$ is the optimal observable for identification of 3-prong jets formed from the energy correlation functions.  Here, 
\begin{equation}
x=\kappa_1 \left (\frac{(p^{\text{cut}}_T)^2}{m_{\text{top}}^2}\right )^{\left (\frac{\alpha \gamma}{\beta}-\frac{\alpha}{2} \right)},\quad y=\kappa_2   \left(\frac{(p^{\text{cut}}_T)^2}{m_{\text{top}}^2}\right )^{\left (\frac{5\gamma}{2}-2\beta \right)}\,,
\end{equation}
where $p_{T}^\text{cut}$ is a proxy for the $p_T$ bin of the jet sample of interest.  We will take $\kappa_1=\kappa_2=1$ for simplicity in the following, keeping in mind that varying their values may provide improved discrimination.  We will find that $\tau_{3,2}^{(\beta)}$ and $D_3^{(\alpha,\beta,\gamma)}$ are complementary observables for boosted top identification.

\section{Detailed studies at fixed $p_T$}\label{sec:onept}

In this section, we present a detailed study of the methods introduced before in a restricted jet $p_T$ range relevant for top quark identification at a future collider.  The results quantify the discrimination power of our top quark tagging methods, show that projected calorimetry of a future collider is insufficient, and tracked-based information can provide the needed complementary information. \App{app:allpt} collects plots and analyses for a different set of jet $p_T$ bins demonstrating that our arguments have a wide range of validity.

Here,  we consider jets in the $p_T$  range of $7.5-10$ TeV.  The reason for this choice is the following.  
First, the $p_T$ is sufficiently hard that standard methods show clear limitations. The angular size of the top quark decay products in this bin is approximately
\begin{equation}
R_\text{top} \sim  \frac{2 m_\text{top}}{p_T} \lesssim \frac{2\cdot 175 \text{ GeV}}{7.5\text{ TeV}} \approx 0.05 \,,
\end{equation}
which is comparable to the resolution of the electromagnetic calorimeters at ATLAS and CMS.  Therefore, the reconstruction of top quark jets produced in this $p_T$ bin will be very sensitive to the resolution of the detector and can provide a valuable benchmark for future collider detector resolution goals.

\begin{figure}
\begin{center}
\subfloat[]{\label{fig:mass_cms_calo_fix}
\includegraphics[width=7.5cm]{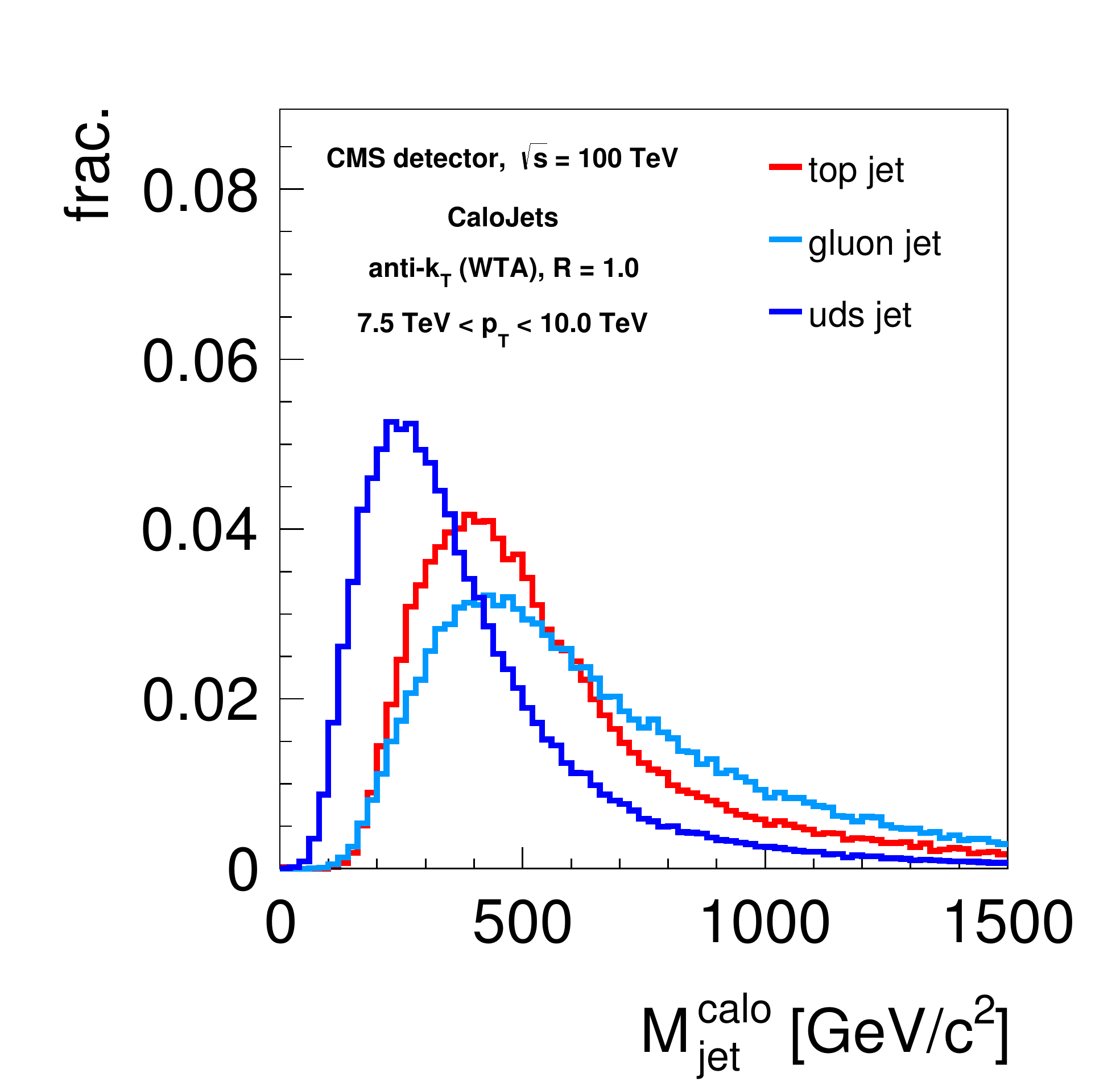}
}
\subfloat[]{\label{fig:mass_fcc_calo_fix}
\includegraphics[width=7.5cm]{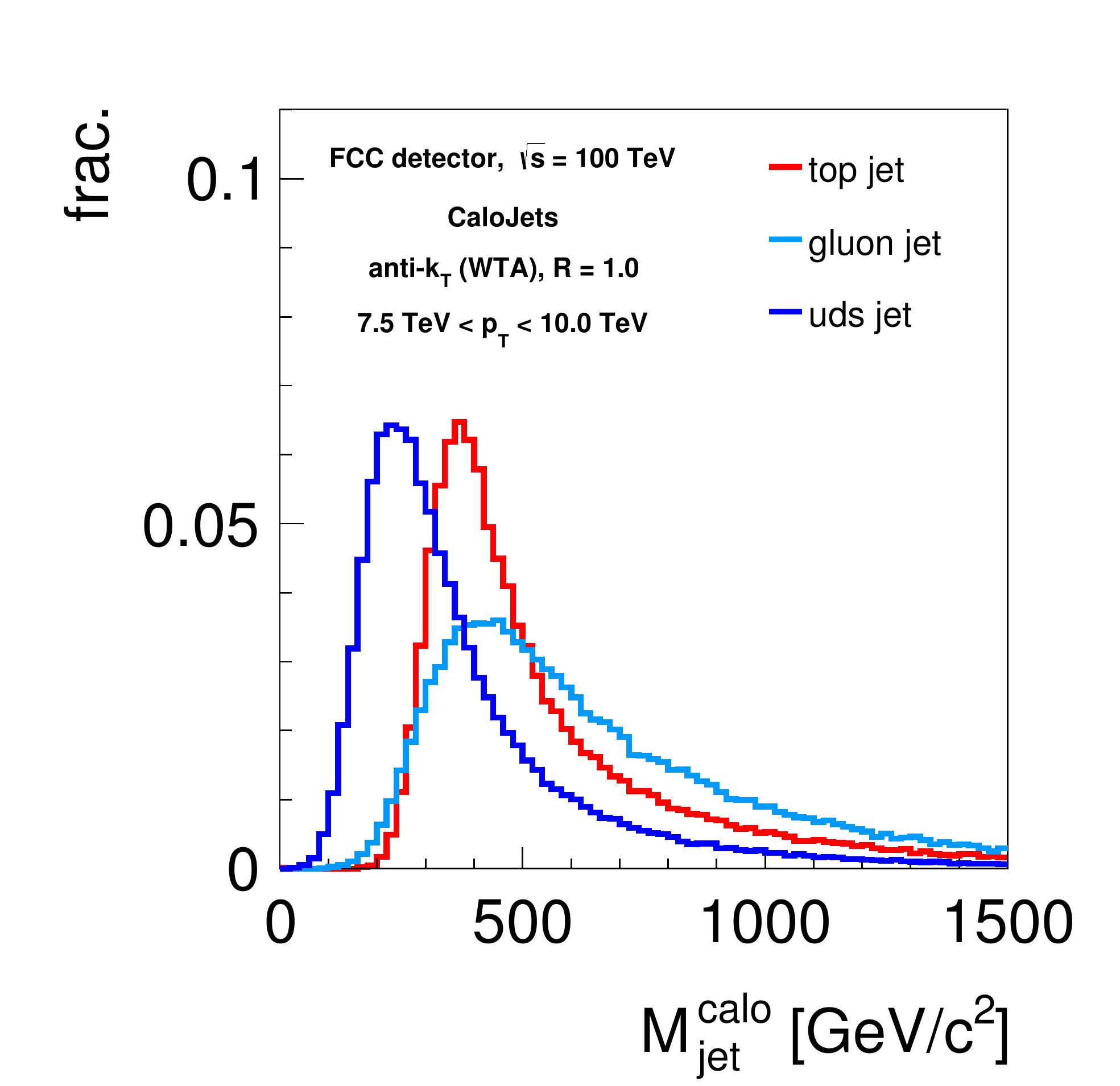}
}
\end{center}
\caption{
Distributions of the jet mass as measured on anti-$k_T$ jets with radius $R=1.0$ and $p_T \in [7.5,10]$ TeV on boosted top jets and QCD jets from light quarks and gluons.  (a) Mass distribution as measured from the CMS detector's calorimeter.  (b) Mass distribution as measured from a future collider detector's calorimeter. 
}
\label{fig:calo_fix_mass}
\end{figure}

To do this, we will compare the efficiency to identify boosted top quark jets by using the standard calorimetric information only and by using our track-based method in a simulated CMS-like detector and in a projected future 100 TeV collider using \delphes.  The parameters for the fast detector simulation are presented in \App{app:det_sim} and provide, we believe, rather conservative estimates for the resolution of detectors of a future collider.  All event simulation is done with the {\sc MadGraph5\_aMC@NLO 2.2.2}, \pythia{6.4} and \delphes{3.1.2} simulation chain.  We study signal samples of boosted $Z'\to t\bar t$ events, where the mass of the $Z'$ is varied depending on the $p_T$ of the top quarks, and background samples of the partonic SM processes $q\bar q \to q' \bar q'$ and $gg\to gg$ at a 100 TeV $pp$ collider.  Jets are reconstructed using the anti-$k_T$ algorithm with \fastjet{3.0.6}, with either a fixed jet radius or using the jet radius that scales with $p_T$, defined in \Eq{eq:scaling_radius}.  We ignore details and subtleties of defining quark flavor and gluon flavor jets, and pragmatically define them as what results from showering quark or gluon partons.

\subsection{Mass Distributions}

We begin by presenting the distribution for the jet mass in these samples as measured from calorimeter cells with a fixed jet radius $R=1.0$.  Distributions of the signal and background jet masses for the CMS and future collider detectors are presented in \Fig{fig:calo_fix_mass}.  Because of the large jet radius, there is a significant contamination from radiation in the jets, resulting in all mass distributions, either at CMS or a future collider, peaking at masses substantially greater than the top quark mass.  Additionally, the light quark and gluon background distributions straddle the signal top quark distribution, showing that any background QCD jet sample  would be essentially indistinguishable from a top quark sample from these mass distributions alone.

\begin{figure}
\begin{center}
\subfloat[]{\label{fig:mass_cms_calo_scale}
\includegraphics[width=7.5cm]{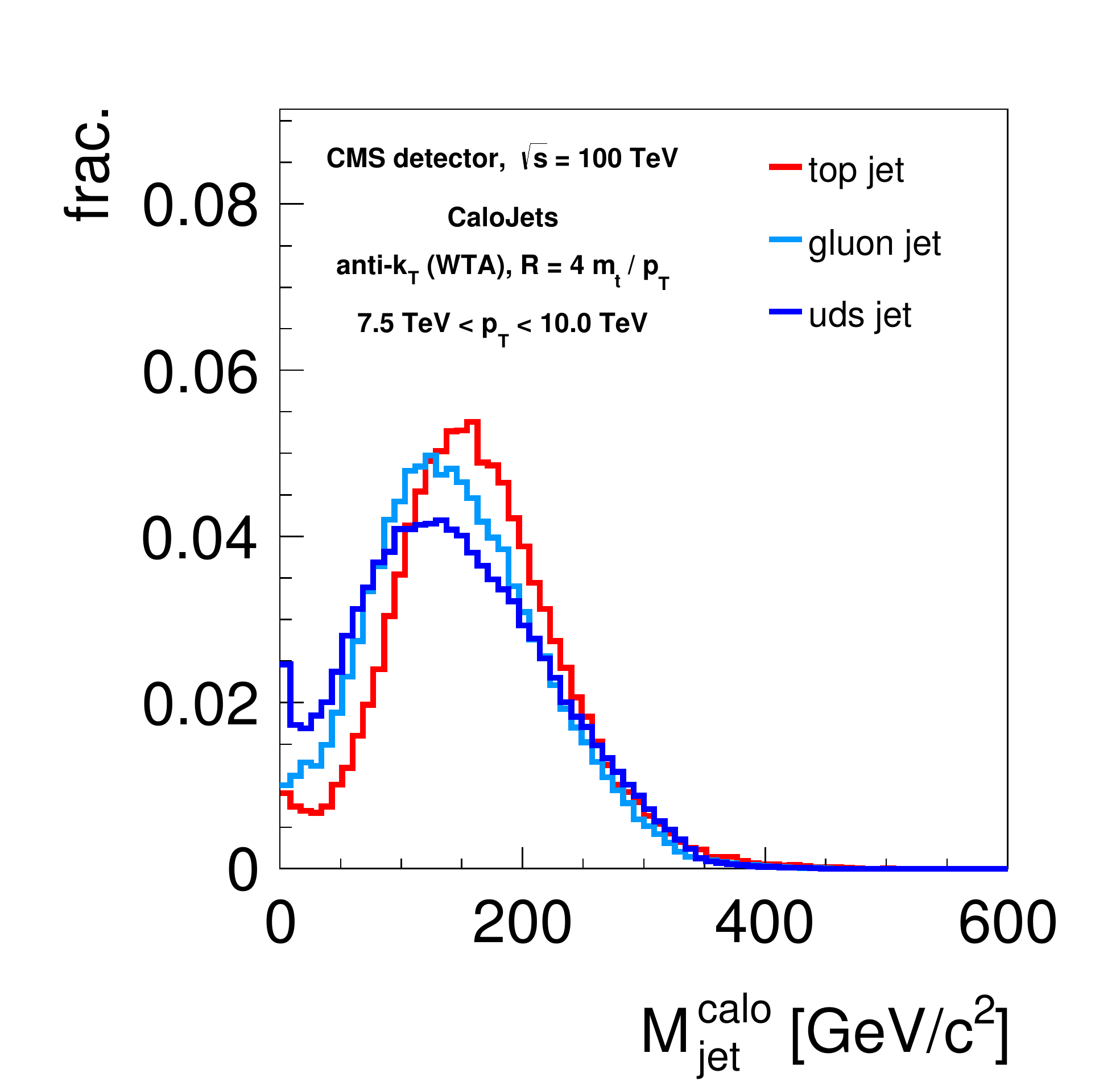}
}
\subfloat[]{\label{fig:mass_fcc_calo_scale}
\includegraphics[width=7.5cm]{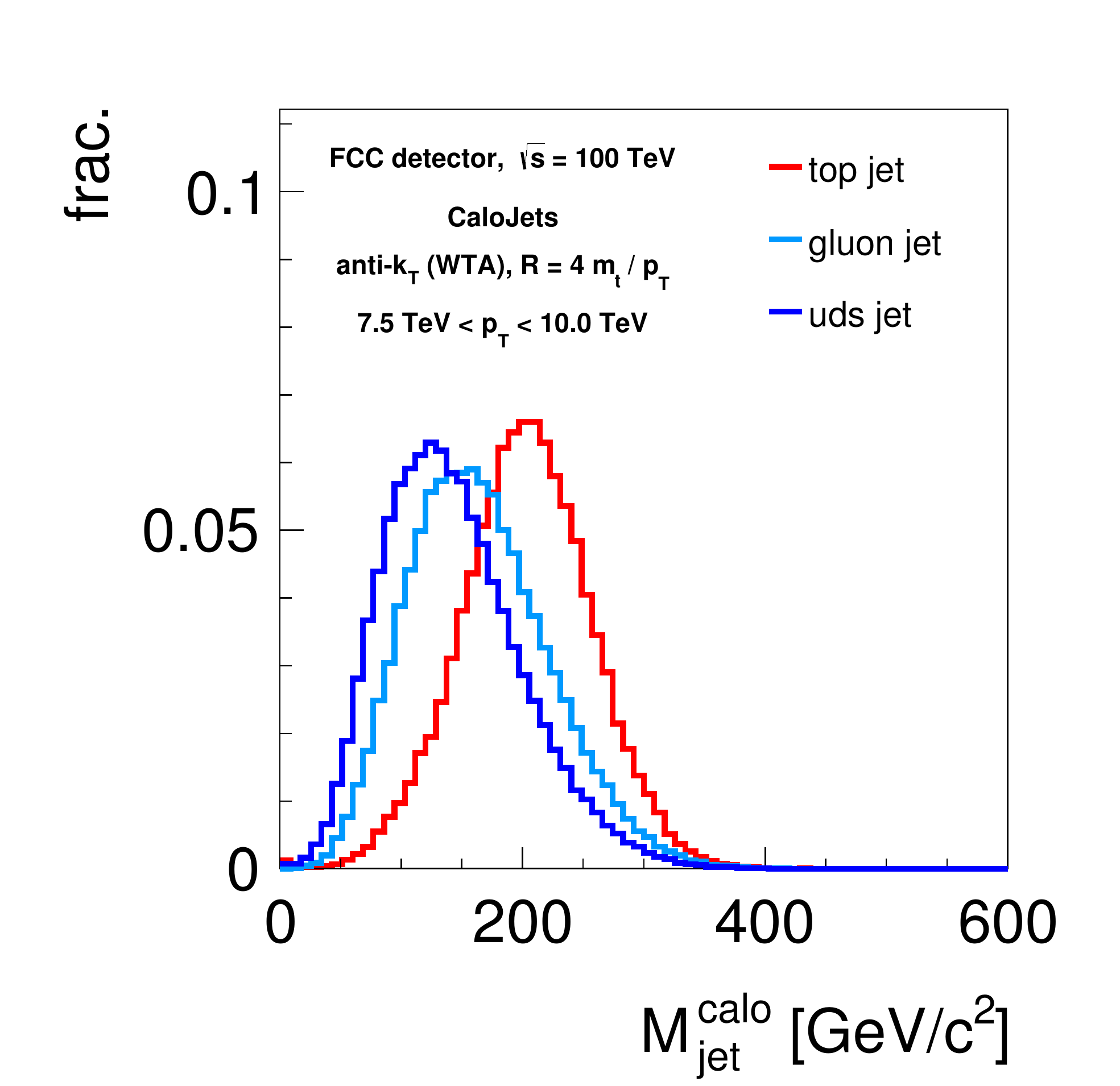}
}
\end{center}
\caption{
Distributions of the jet mass as measured on anti-$k_T$ jets with radius $R=4 m_\text{top}/p_T$ and $p_T \in [7.5,10]$ TeV on boosted top jets and QCD jets from light quarks and gluons.  (a) Mass distribution as measured from the CMS detector's calorimeter.  (b) Mass distribution as measured from a future collider detector's calorimeter. 
}
\label{fig:calo_scale_mass}
\end{figure}

In \Fig{fig:calo_scale_mass}, we present the distributions of the jet mass as measured on calorimeter jets with a scaled jet radius $R ={4 m_\text{top}}/{p_T}$.  With the scaled jet radius, the amount of contamination in the jet is greatly reduced; however, the low resolution degrades  the mass distributions measured in the CMS detector.  Signal and background distributions overlap and all samples have a peak at zero mass indicating that the jet consists of a single calorimeter cell.  Detector resolution effects are also significant in the mass distributions measured at a future collider.  While the angular resolution of the calorimeter for the future collider scenario is $\times$2 finer than for CMS (see App. \ref{appsub:calorimetry}) and so these distributions do not have a peak at zero mass, the overlap of signal and background is substantial, and a cut on the mass would only result in a marginal top quark tagging efficiency.

\begin{figure}
\begin{center}
\subfloat[]{\label{fig:mass_cms_trk_scale}
\includegraphics[width=7.5cm]{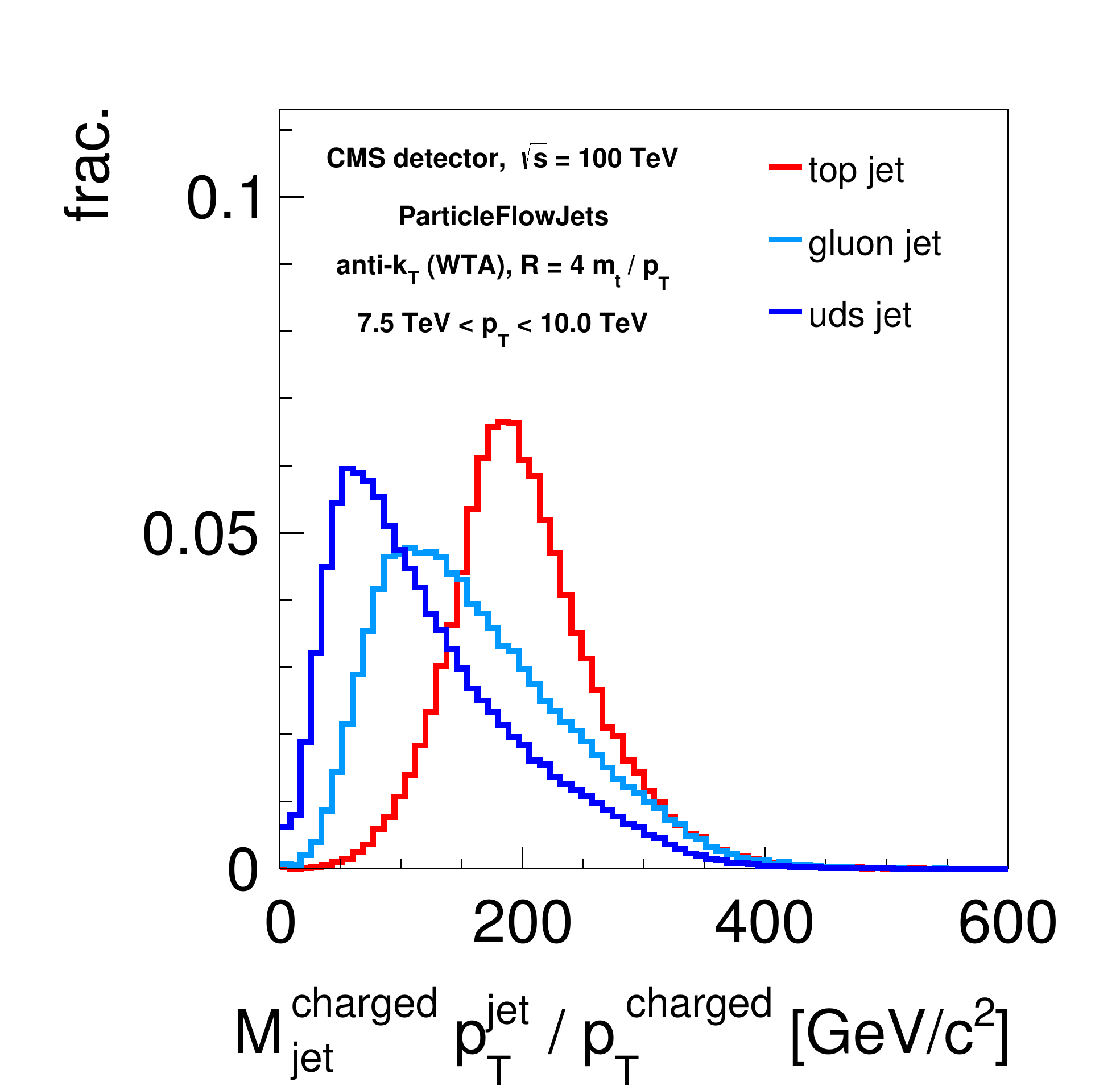}
}
\subfloat[]{\label{fig:mass_fcc_trk_scale}
\includegraphics[width=7.5cm]{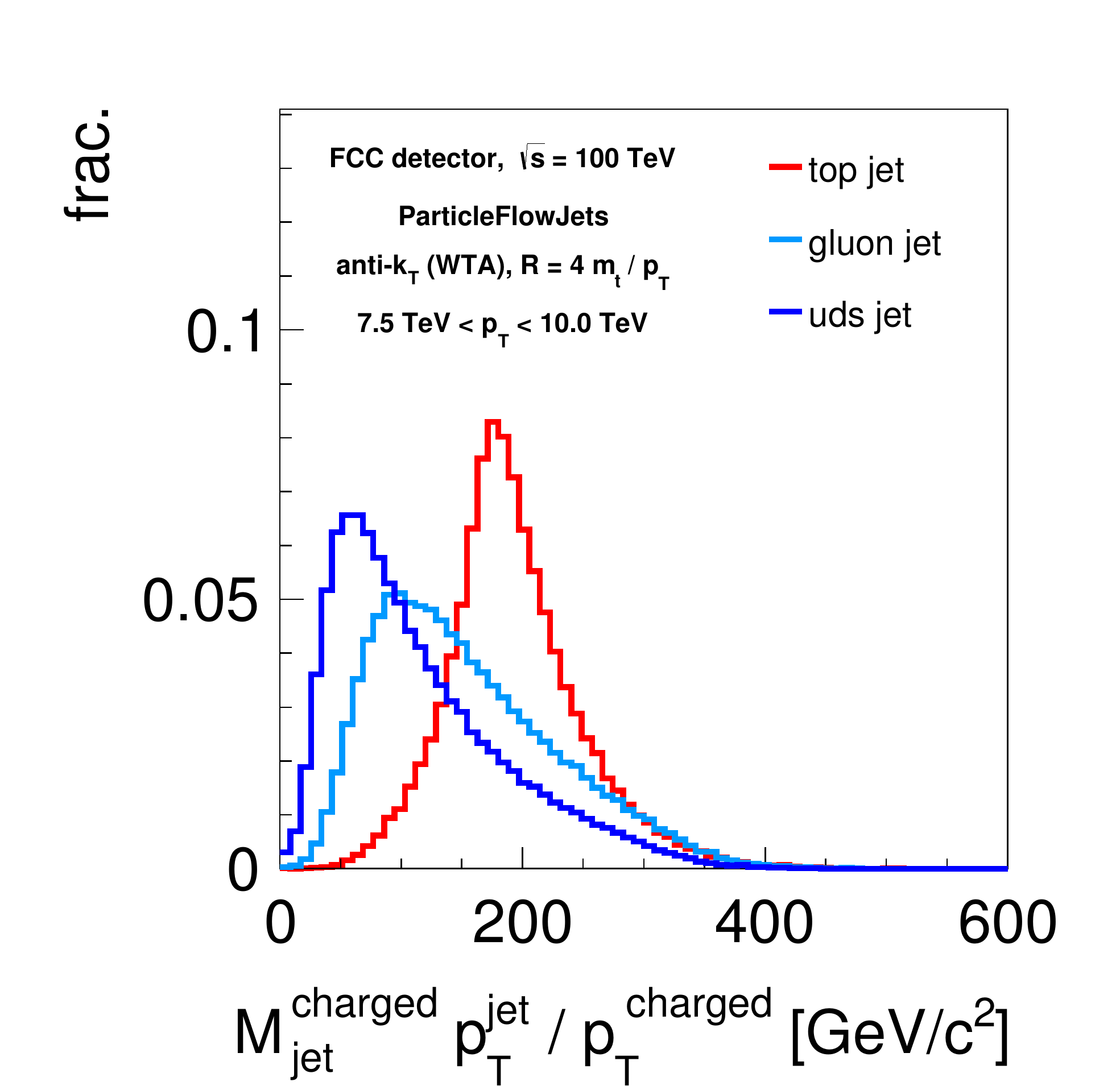}
}
\end{center}
\caption{
Distributions of the rescaled track-based jet mass as measured on anti-$k_T$ jets with radius $R=4 m_\text{top}/p_T$ and $p_T \in [7.5,10]$ TeV on boosted top jets and QCD jets from light quarks and gluons.  (a) Distribution as measured from the CMS detector's tracking system.  (b) Distribution as measured from a future collider detector's tracking system. 
}
\label{fig:track_scale_mass}
\end{figure}

Finally, in \Fig{fig:track_scale_mass}, we present the distributions of the jet mass as measured from tracks with the jet radius $R ={4 m_\text{top}}/{p_T}$ and rescaling by the ratio of the total jet $p_T$ to the track $p_T$.  Unlike the jet mass as measured from the calorimeter, the rescaled track mass accurately reproduces the top mass for both the CMS and future collider detectors and barely suffers from resolution effects.  Additionally, the mass distributions of QCD backgrounds are pushed to small values, below the mass of the top quark for both detectors.  A cut in a window around the top quark mass would therefore robustly and efficiently discriminate boosted top quark jets from the QCD background.  Larger rejection of QCD background can be accomplished by measuring additional substructure observables on the jet.

\subsection{Substructure observables}

As mentioned earlier, a jet mass comparable to the mass of the top quark can be reconstructed in several ways: for signal, by the decay into hard subjets and no FSR radiation; while for background by the emission of significant amounts of soft radiation in the jet.  For improved rejection of the background, in addition to the mass determination,  observables on the jet that are sensitive to the 3-prong substructure of the top decay must to be measured.  We therefore consider  $N$-subjettiness ratio $\Nsub{3,2}{\beta}$ and the energy correlation function ratio $\Dobs{3}{\alpha,\beta,\gamma}$.  While a full study of the efficiency as a function of the angular exponents and other parameters in these observables may result in improved discrimination power, we choose to use the parameters for boosted top tagging as recommended by the original investigations.  That is, we consider the observables
\begin{align}
&\Nsubnobeta{3,2}\equiv \Nsub{3,2}{\beta=1}\,, &D_3\equiv \Dobs{3}{\alpha=2,\beta=0.8,\gamma=0.6} \,,
\end{align}
as measured on jets, which lie in the rescaled track-based mass window $m_J\in [120,250]$ GeV in the following. Even without optimisation of parameters, we find impressive signal to background efficiency rates for jets produced at a 100 TeV collider.

\begin{figure}[t]
\begin{center}
\vspace{-0.5cm}
\hspace{-0.5cm}
\subfloat[]{\label{fig:mass_cms_trk_r32_2}
\includegraphics[width=7.5cm]{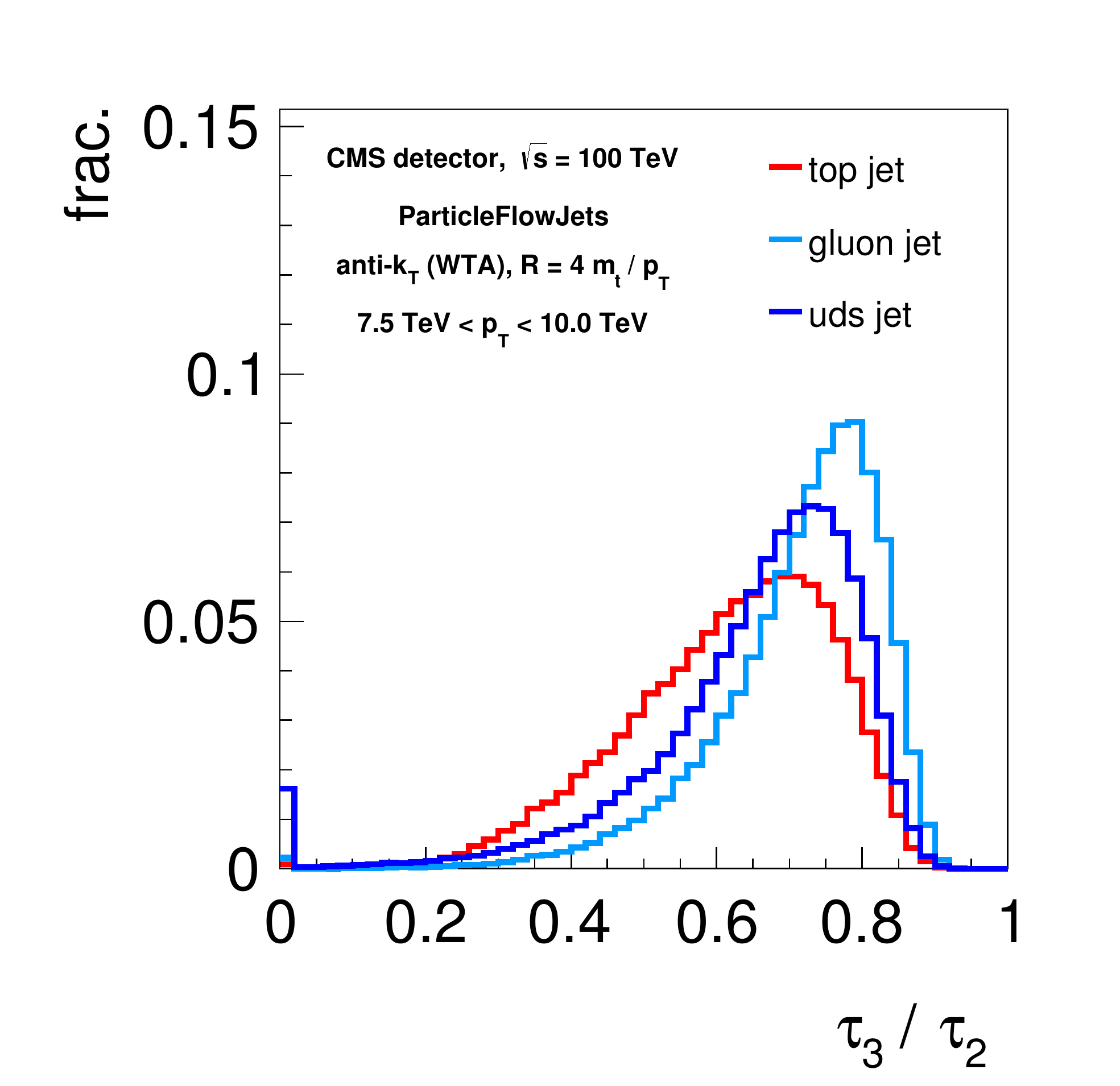}
}
\subfloat[]{\label{fig:mass_fcc_trk_r32_2}
\includegraphics[width=7.5cm]{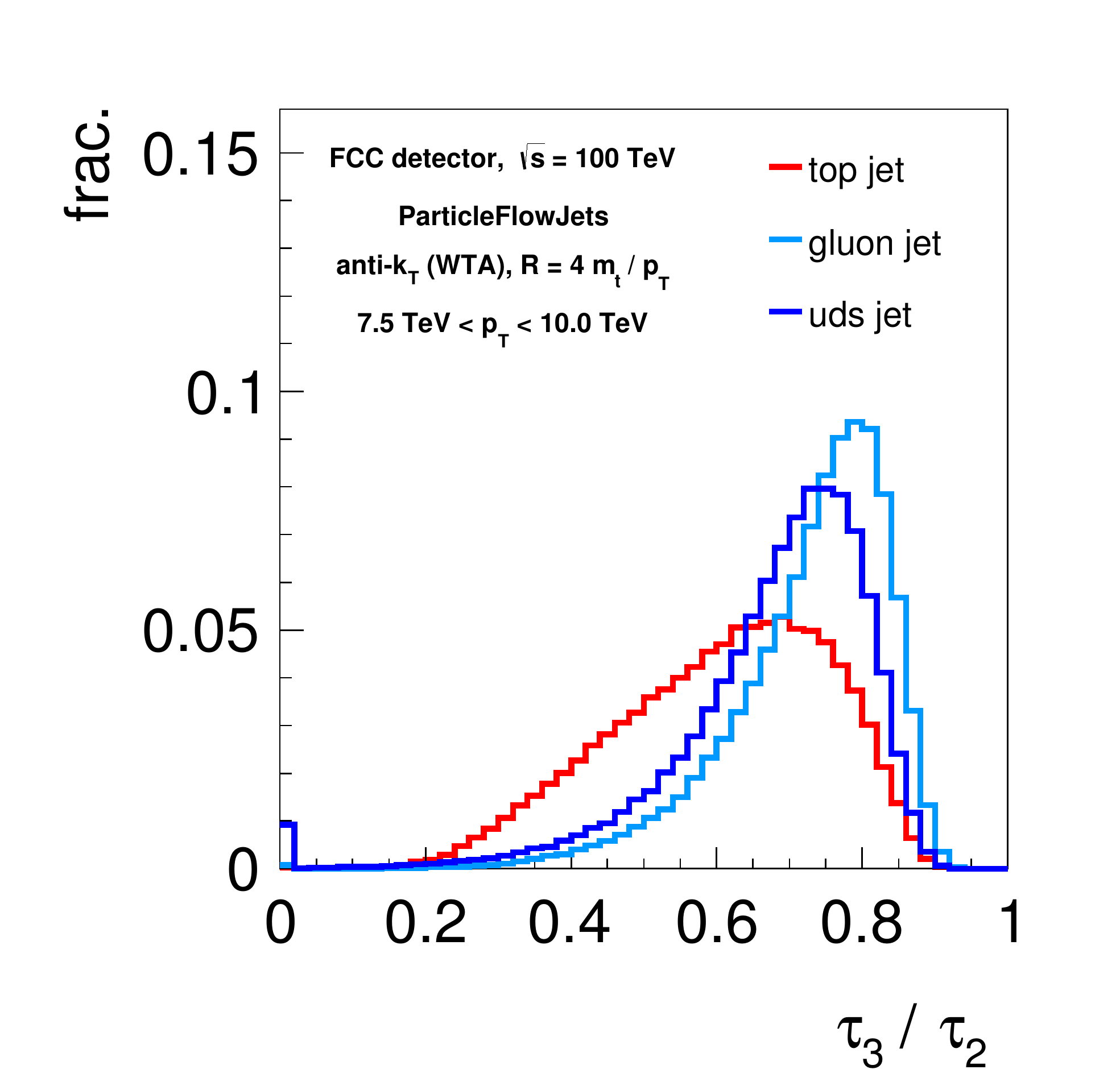}
}
\\
\vspace{-0.4cm}
\hspace{-0.5cm}
\subfloat[]{\label{fig:mass_cms_trk_d3_2}
\includegraphics[width=7.5cm]{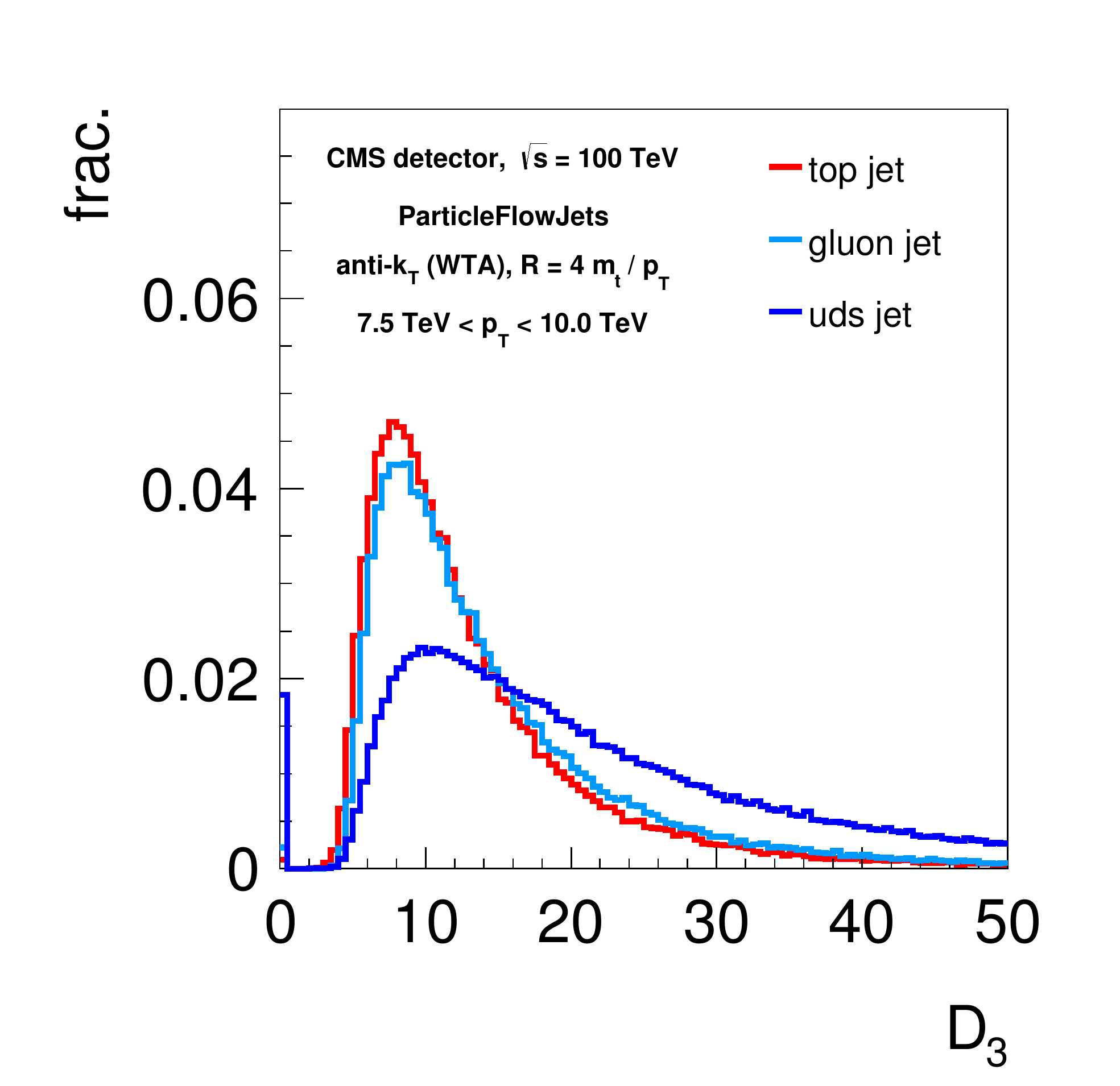}
}
\subfloat[]{\label{fig:mass_fcc_trk_d3_2}
\includegraphics[width=7.5cm]{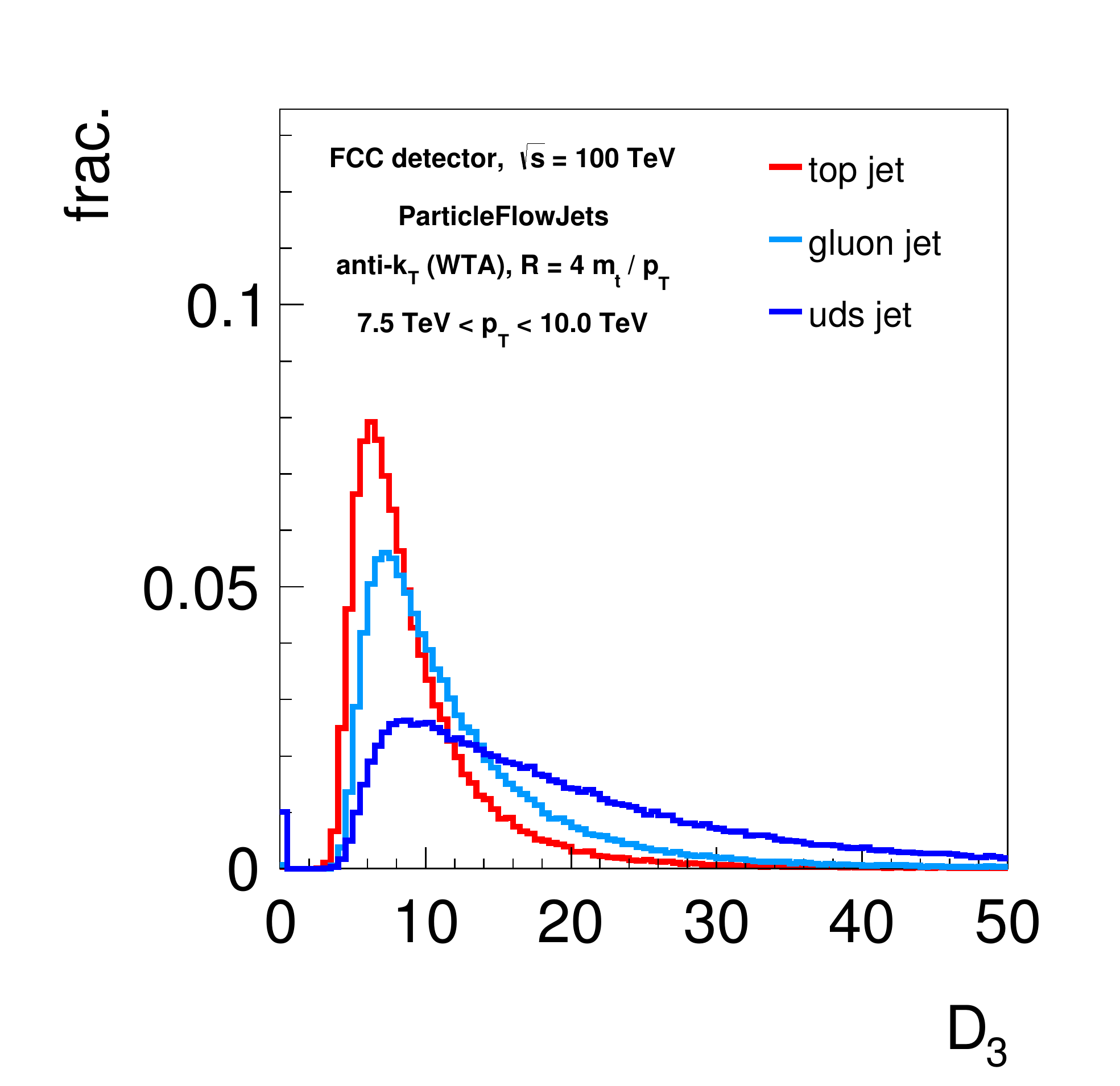}
}
\end{center}
\caption{
Distributions of the jet $\tau_3 / \tau_2$ (top) and $D_3$ (bottom) as measured on anti-$k_T$ jets with radius $R=4 m_\text{top}/p_T$ and $p_T \in [7.5,10]$ TeV on boosted top jets and QCD jets from light quarks and gluons.  Additionally, we require that the rescaled track-based mass lie in the window $m\in [120,250]$ GeV.  (left) Distributions as measured from the CMS detector's tracking system.  (right) Distributions as measured from a future collider detector's tracking system.
}
\label{fig:track_scale_r32_1}
\end{figure}

In \Fig{fig:track_scale_r32_1}, we present the distributions of $\tau_{3,2}$ and $D_3$ measured on signal and background jets.  We only consider these observables as measured on tracks and with jet radius that scales inversely with $p_T$.  Though we only measure on tracks, because both $\tau_{3,2}$ and $D_3$ are dimensionless, we do not rescale them by the ratio of the total jet $p_T$ to the charged track $p_T$.  Both $\tau_{3,2}$ and $D_3$ provide excellent separation of signal and background jet samples, and in a complementary way.  For $\tau_{3,2}$, the top quark distribution overlaps with the light quark distribution more than it does with gluons, while for $D_3$ the opposite occurs.  While we do not have a complete theoretical understanding of this behaviour, this suggests that simultaneous measurement of $\tau_{3,2}$ and $D_3$ on jets would provide further discrimination power, depending on the light quark and gluon composition of the background sample.

\begin{figure}[t]
\vspace{-0.5cm}
\hspace{-1.5cm}
\subfloat[]{\label{fig:roc_gluon_r32}
\includegraphics[width=9cm]{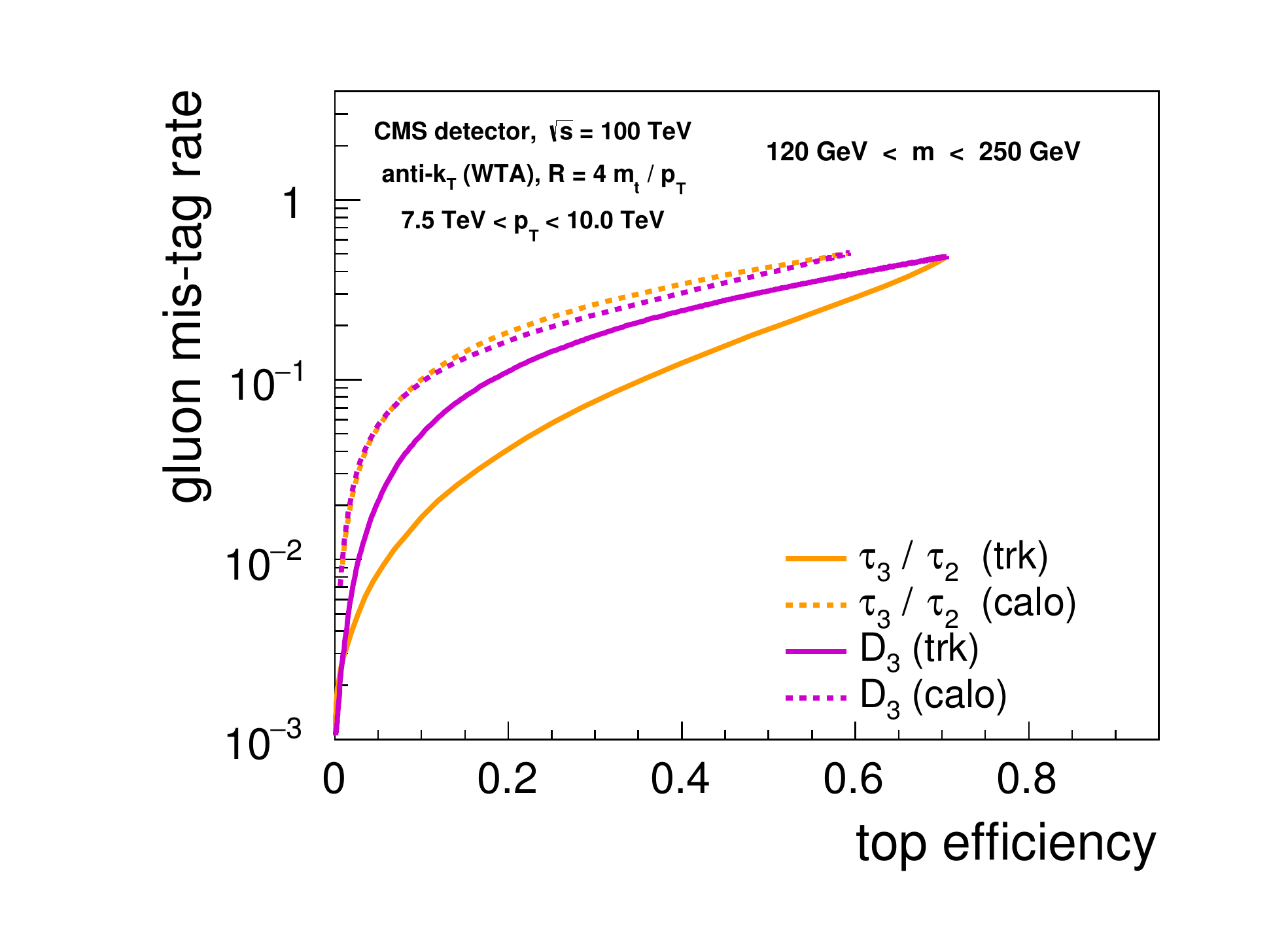}
}
\hspace{-1.2cm}
\subfloat[]{\label{fig:roc_quark_r32}
\includegraphics[width=9cm]{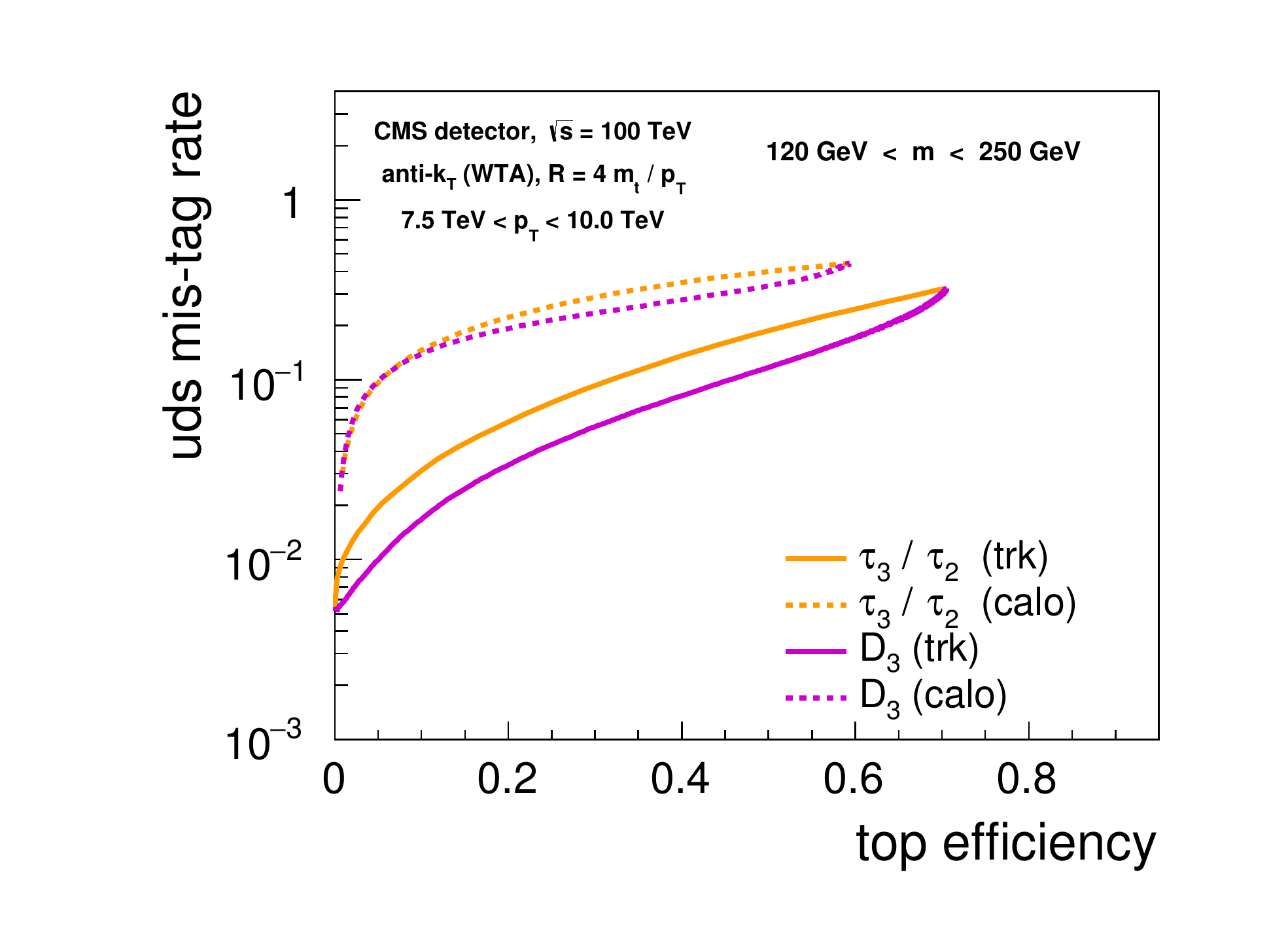}
}
\\
\vspace{-0.5cm}
\hspace{-1.5cm}
\subfloat[]{\label{fig:roc_gluon}
\includegraphics[width=9cm]{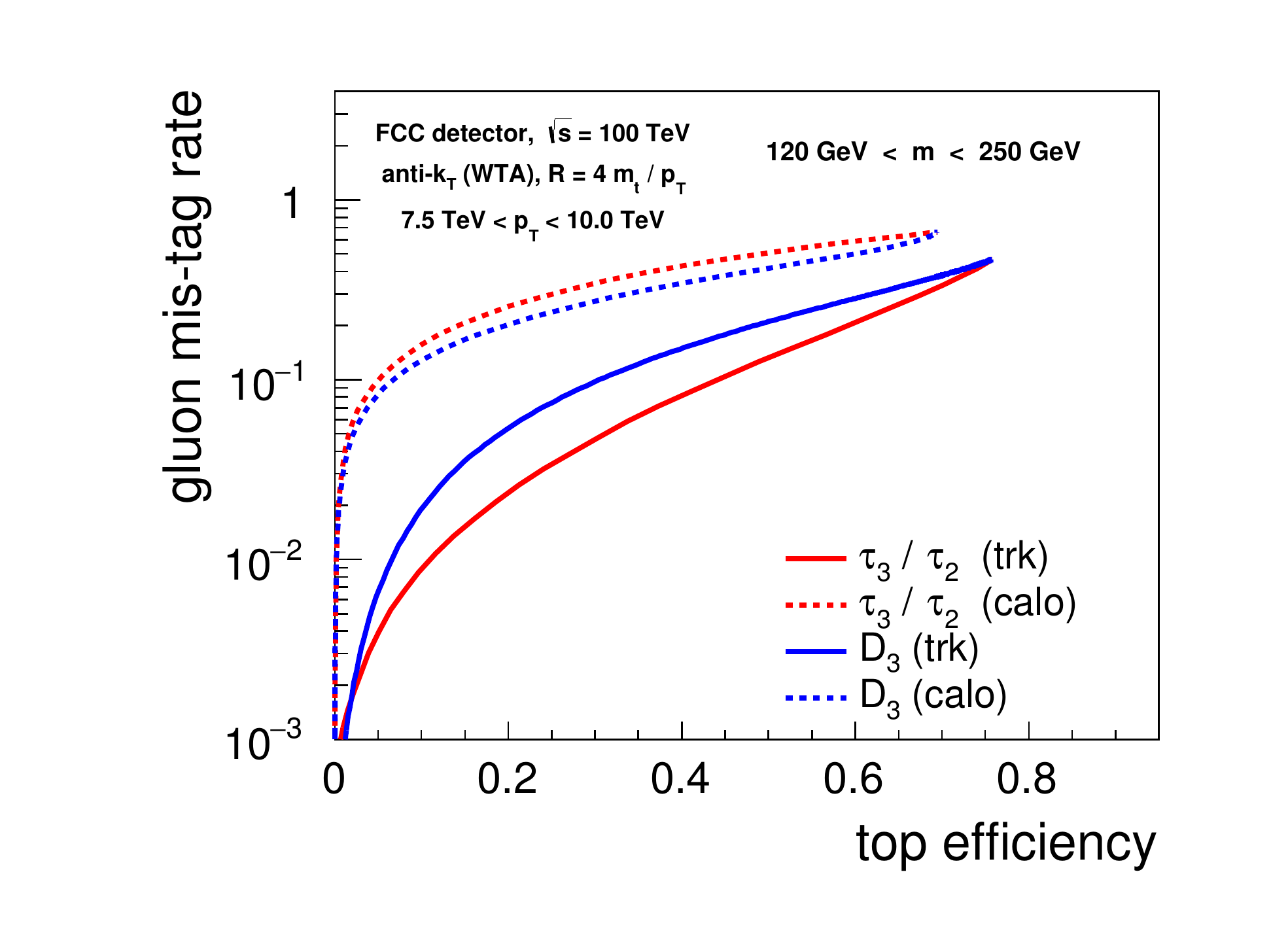}
}
\hspace{-1.2cm}
\subfloat[]{\label{fig:roc_quark}
\includegraphics[width=9cm]{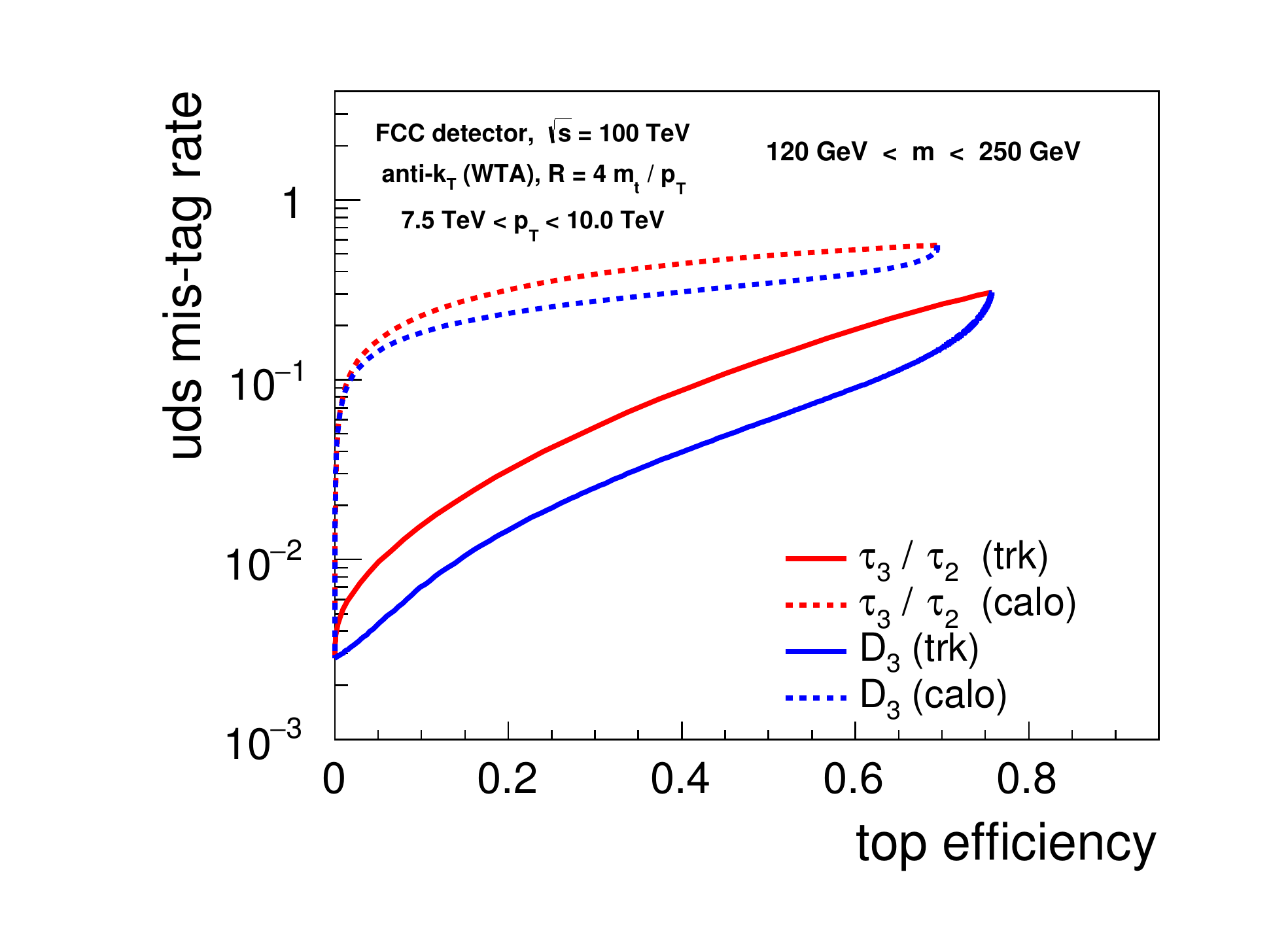}
}
\caption{
Signal vs.~background efficiency (ROC) curves for top quark identification from QCD background utilising $\tau_{3,2}$ and $D_3$: (left) top quarks vs.~gluon jets, (right) top quarks vs.~light quark jets.  The ROC curves as measured in the calorimeter are dashed lines and as measured on tracks are solid lines for (top) the CMS detector and (bottom) the FCC detector.  The cut on the jet mass of $m\in[120,250]$ GeV is included in the efficiencies.
}
\label{fig:roc_curves}
\end{figure}

To emphasise the discrimination power provided by $\tau_{3,2}$ and $D_3$, in \Fig{fig:roc_curves} we show signal versus background efficiency curves (ROC curves) produced by making a sliding cut in either observable.  As we require jets to lie in the rescaled track mass window $m\in [120,250]$, the efficiencies in \Fig{fig:roc_curves} include the effect of the mass cut.  As anticipated, measuring these observables on tracks provides significant improvement in discrimination power as compared to including calorimeter information alone.  Depending on the quark or gluon composition of the background sample, $\tau_{3,2}$ or $D_3$ measured on tracks provides better discrimination power.  For example, at a future collider in this $p_T$ bin at 50\% top quark jet efficiency, 83\% of gluon jets can be rejected by cuts on the mass and $\tau_{3,2}$,  while 94\% of light quark jets can be rejected by a mass cut and $D_3$.  Thus, these observables exhibit a nice complementarity in how they can reject QCD background.

We stress that for the distributions in \Fig{fig:track_scale_r32_1} and their corresponding ROC curves in \Fig{fig:roc_curves}, no optimisation has been performed.  These observables have been applied ``out of the box'', using the values for their parameters as recommended in the original studies.  Therefore, it may be possible to substantially improve boosted top quark discrimination power by optimising the free parameters that enter the definition of the observables.  Nevertheless, even with this na\"ive application of a cut on the track mass and cuts on $\tau_{3,2}$ or $D_3$ impressive discrimination can be obtained.  Additionally, while $\tau_{3,2}$ and $D_3$ are both sensitive to the 3-prong substructure, their relative performance rejecting gluon and light quark jets is not well understood.  The motivation for constructing these observables relies on the behavior of QCD in the parametric soft and collinear limits.  However, gluon and light quark jets are not parametrically different objects, and so, without detailed calculation, it may not be possible to understand the performance.  Further complication for detailed theoretical understanding occurs because we only measure these observables on charged tracks.  Quark and gluon partons fragment to charged hadrons differently, and this may also affect the discrimination power.  A detailed study of these effects and their impact on discrimination power employing track-based methods of \Refs{Waalewijn:2012sv,Chang:2013rca,Chang:2013iba} would be certainly welcome, even though clearly beyond the scope of this paper.

\section{Conclusions}\label{sec:conclusions}

\begin{figure}[t]
\hspace{-1cm}
\subfloat[]{\label{fig:money_calo}
\includegraphics[width=8.5cm]{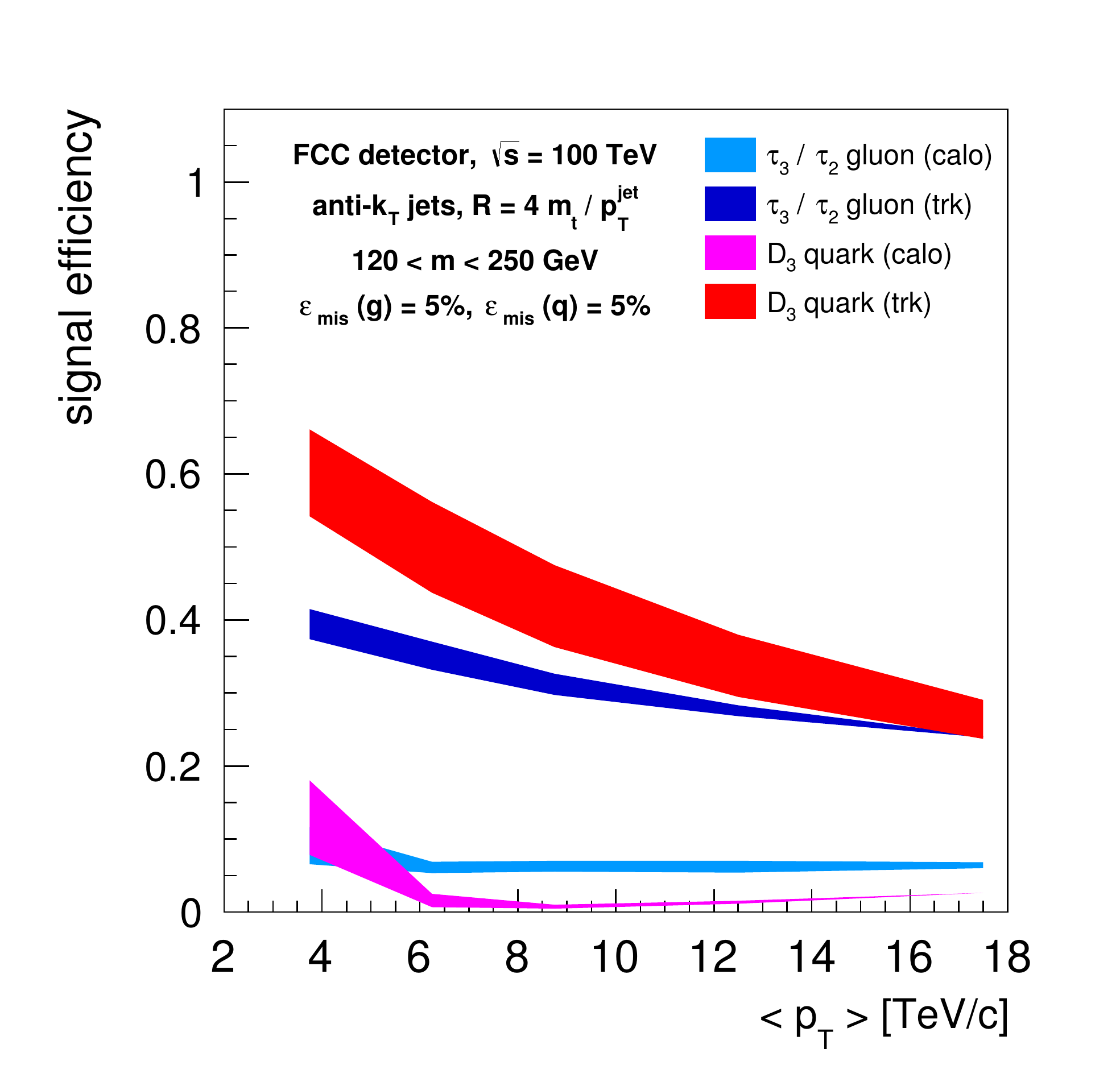}
}
\hspace{-0.7cm}
\subfloat[]{\label{fig:money_track}
\includegraphics[width=8.5cm]{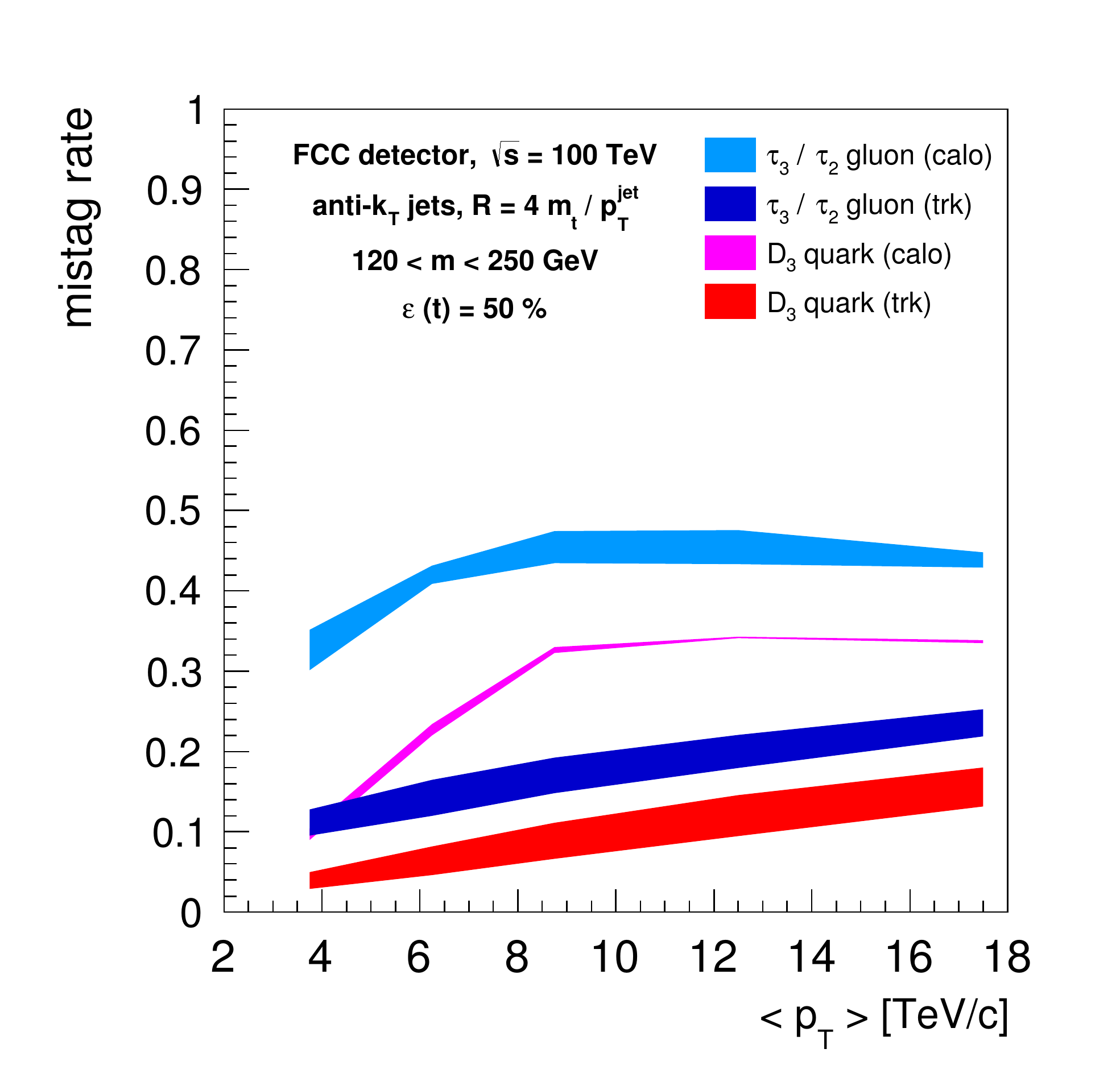}
}
\caption{
Plots illustrating the efficiency for tagging top quarks and rejecting QCD background as a function of jet $p_T$ using tracking (blue, red) versus calorimetry (light blue, magenta) at a future collider.  For identifying three-prong substructure the observables $\tau_{3,2}$ and $D_3$ have been used.
(a) Top tagging efficiency for a fixed light quark and gluon mistag rate of 5\%.
(b) Light quark and gluon mistag rate at fixed top quark efficiency of 50\%.
The bands represent the envelope of efficiencies spanned by the Monte Carlo simulations (\herwig{6} and \pythia{6.4}) that we use.
}
\label{fig:money}
\end{figure}

We summarize the results of this paper in \Fig{fig:money}, illustrating the potential discrimination power for identifying boosted top quarks at the detector of a future high energy proton collider modeled with \delphes.  On the left, we show the hadronically-decaying top quark tagging efficiency as a function of jet $p_T$ at fixed mistag rate for jets produced from light quarks and gluons of 5\% comparing our method using tracking versus using calorimetry exclusively.  The bands represent the envelope of efficiencies from using either \herwig{6}  \cite{Marchesini:1991ch,Corcella:2000bw,Corcella:2002jc} or \pythia{6.4} Monte Carlo simulations.  On the right, we plot the efficiency for mis-tagging jets initiated by light quarks or gluons as top quarks at fixed signal efficiency of 50\%.  Our procedure enables significant rejection rates at $p_T$s approaching 20 TeV, while using calorimetry alone struggles to reject more background than signal.  Exploiting tracking enables impressive signal efficiency, comparable to that of taggers used at the LHC, whose performance is relatively independent of jet $p_T$ and extends well beyond 10 TeV.

In this paper, we have presented a procedure for the identification of top quarks in the multi-TeV energy range as those that could be produced at a future 100 TeV proton collider.  High-resolution tracking information was required for identification of the prongs produced in the top quark decay and contamination due to initial- and final-state radiation, underlying event, pile-up, or other sources can be reduced significantly by dynamically scaling the jet radius inversely proportional to the transverse momentum of the jets.  By applying a simple jet selection on the track-based jet mass, $N$-subjettiness ratio, and energy correlation function based jet substructure observables, very high efficiency for discriminating boosted top quarks from light QCD jets can be accomplished, and with rates that are relatively independent of the boost.  These results are encouraging for the prospects of precision studies of the electroweak sector at a future collider.

Though we have employed a fast detector simulation, we believe that the analysis presented here can provide a useful benchmark for performance and response of detectors at a future collider.  The relatively conservative parameters in the fast detector simulation should provide a realistic goal for the resolution of a future detector.  Nevertheless, there were several effects not included in this analysis. For example, we have not considered backgrounds to boosted top quarks from electroweak boson emission from light quarks. The main reason is that vector bosons in energetic jets are typically soft~\cite{Christiansen:2014kba,Krauss:2014yaa} and therefore when decaying hadronically their effect is very similar to that of QCD radiation yet subleading due to the weak coupling. However, they may be relevant in high purity (low signal efficiency) samples.  In addition, while rescaling the jet radius inversely proportional to the transverse momentum and using charged tracks reduces contamination from pile-up, we did not include any simulation of pile-up in our analysis.  The effects of pile-up on the efficiency for tagging top quarks would be important to understand in a dedicated analysis, but we believe that the effects would be minimal  because we use a recoil-free jet algorithm as well as scaling the jet radius.

Besides these experimental issues, our procedure for tagging highly boosted top quarks raises some interesting theoretical questions.  Because of our prescription for scaling the jet radius, the angular size of the jets we consider becomes very small at sufficiently high transverse momenta.  By restricting the radiation in the jet to such small angular regions, non-global logarithms \cite{Dasgupta:2001sh}, clustering logarithms \cite{Banfi:2005gj}, or logarithms of the jet radius itself \cite{Seymour:1997kj,Gerwick:2012fw,Dasgupta:2014yra} could become important for a theoretical understanding.  For high angular resolution, we also require measurements on charged tracks, and not on all radiation in the jet.  Therefore, a model of the fragmentation of partons to charged hadrons is required to predict the distributions of the jet substructure observables on these jets.  An understanding of track-based observables \cite{Waalewijn:2012sv,Chang:2013rca,Chang:2013iba} may also explain the difference that we observed between the performance of $\tau_{3,2}$ and $D_3$ on light quark or gluon jets and could potentially predict the optimal parameters of those observables for discrimination.

We also emphasize that the methods presented here apply more broadly than to top quarks alone.  Studies for the identification of hyper-boosted 2-prong jets from hadronic decays of $W$, $Z$, or $H$ bosons can benefit from these techniques.  Contamination radiation can be controlled using the scaled jet radius, though, because these objects are colour-singlets and narrow resonances, unlike for top quarks, there is no contaminating FSR produced.  Because the masses of the electroweak bosons are smaller than the top, using the high resolution of the tracker becomes even more important, as the structure of these jets lies at smaller angular scales.  For identification of 2-prong substructure, instead of $\tau_{3,2}$ and $D_3$, one would use observables such as $\tau_{2,1}$ $N$-subjettiness ratio and $D_2$ formed from the energy correlation functions \cite{Larkoski:2014gra}.

Looking forward to a future hadron collider during the era of the LHC provides context and motivation for collider physics studies in extreme environments.  Additionally, analyses at the LHC itself can benefit from efforts for a future collider.  For example, while the energies and luminosities at the LHC are an order of magnitude smaller than a proposed future collider, our proposal for jet radius rescaling and track-based measurements could be useful for analyses at the LHC in a high pile-up environment or over a large energy range.  Looking forward to Run 2 at the LHC, a new, higher energy regime will be explored and will require the use of new techniques to push forward.

\begin{acknowledgments}
We thank Michelangelo Mangano, Gavin Salam, and Jesse Thaler for helpful discussions, and the FCC working group for a very stimulating environment.  A.~L.~is supported by the U.S. Department of Energy (DOE) under grant Contract Number  DE-SC00012567. This work is supported in part by the MISTI MIT-Belgium Program, the ERC Grant No. 291377 "LHCTheory�,  by the Research Executive Agency (REA) of the European Union under the Grant Agreement number  PITN-GA-2012-315877 (MCNet), and the National Fund for Scientific Research (F.R.S.-FNRS Belgium), by the IISN ``MadGraph'' convention 4.4511.10, by the IISN ``Fundamental interactions'' convention 4.4517.08, and in part by the Belgian Federal Science Policy Office through the Interuniversity Attraction Pole P7/37. 

\end{acknowledgments}

\appendix

\section{Parametrised detector simulation}\label{app:det_sim}

The ability to disentangle the structure of jets is largely dependent on detector resolution effects, such as the angular and energy-momentum resolution of the jet constituents. In addition, the finite geometrical acceptance of the detector, and the presence of a magnetic field can have an influence on the measurement of jet properties. Such important detector effects have been taken into account by means of the fast simulation framework \delphes{} \cite{deFavereau:2013fsa}. For this study, two detector setups have been used: the CMS detector \cite{Bayatian:2006zz}, as a reference for present performance, and a hypothetical detector for a future circular collider. The FCC detector measures particles created from $pp$ collisions with $\sqrt{s} =$ 100 TeV. Since no prototype of such a detector currently exists yet, the performance of existing detectors needs to be extrapolated to future scenarios. We will briefly describe the modeling of the detector response for these two configurations in this appendix. 

\subsection{Tracking}\label{appsub:tracking}

After collision, parton showering, hadronisation, and decays, the first step of \delphes{}is the propagation of long-lived particles inside the tracking volume within a uniform axial magnetic field parallel to the beam direction. The magnetic field strength $B$, the size of the tracking radius $L$ and the single hit spatial resolution $\sigma_{r\phi}$ are the main parameters that constrain the resolution on the track transverse momentum:
 \begin{equation}
\frac{\sigma(p_T)}{p_{T}} \approx \frac{\sigma_{r\phi}}{B\cdot L^2}\,.
\end{equation}
As a benchmark, we assume that the radius of the FCC inner detector is twice that of CMS and that the magnetic field is increased by a factor of $3/2$.  We also assume that the spatial resolution $\sigma_{r\phi}$ is two times smaller than at CMS, which is possible by designing a more granular pixel detector~\cite{CMS:2012sda}. These improvements in an FCC detector would allow measurements of 1 TeV charged hadrons with a precision of $\sigma(p_T)/p_T  \simeq 1-2\%$.

In \delphes, charged tracks are reconstructed assuming an infinite angular resolution, which is a good approximation for moderately boosted objects, but not optimal for the energy regime that is considered in this study. A module that performs a smearing on the track direction was therefore developed specifically for this study. We assume an overall factor 2 with respect to present CMS performance, which is mainly driven by a more granular detector, and can eventually improved further by increasing the number a tracker layers, which is possible thanks to a larger tracker volume. This leads to a conservative estimate for the angular resolution of the FCC tracker detector  $\sigma (\theta, \phi) \simeq 10^{-3}$ rad. 

\begin{figure}
\begin{center}
\includegraphics[width=8.5cm]{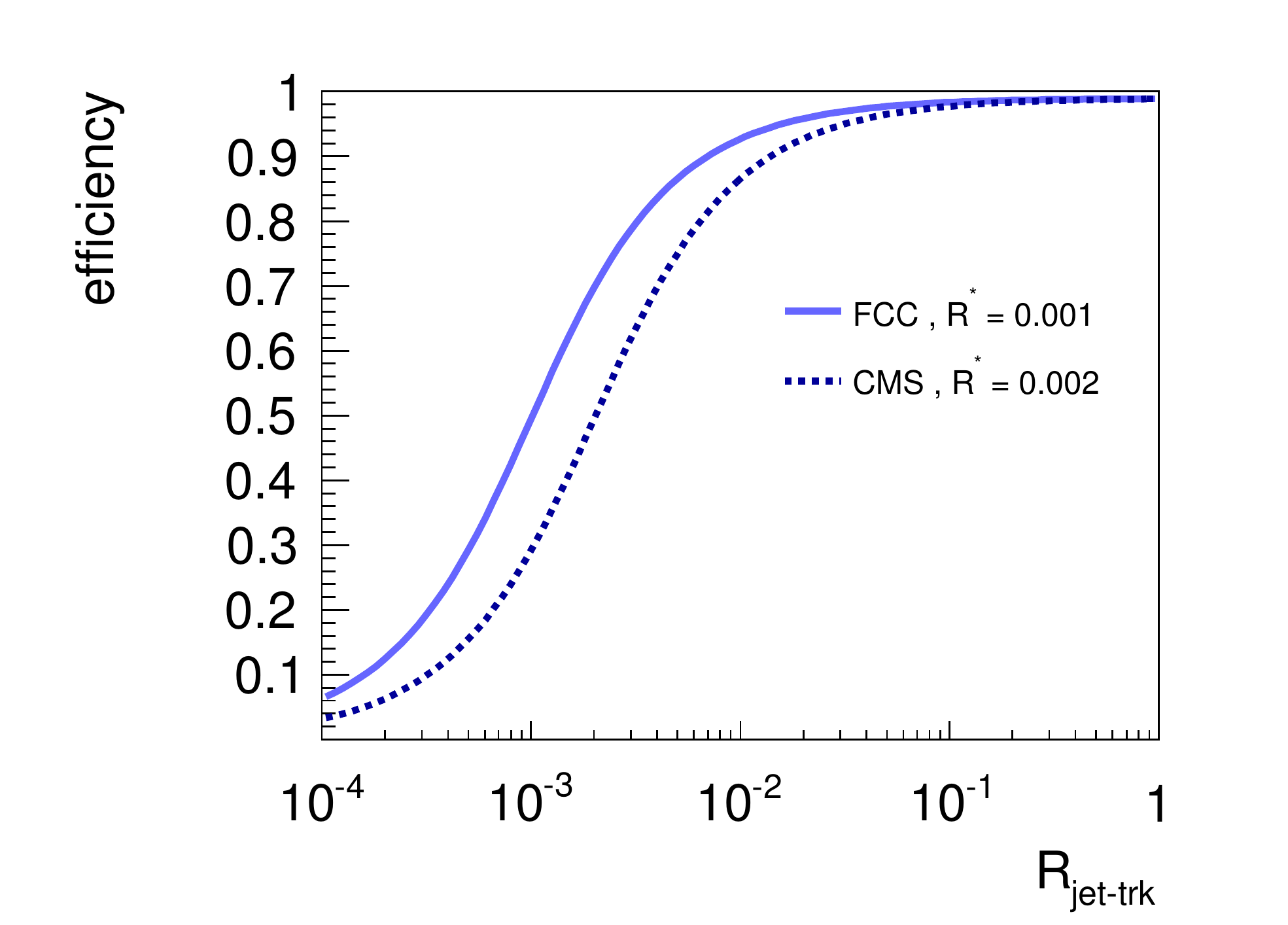}
\end{center}
\caption{Tracking efficiency as a function of the transverse distance between the track direction and the jet center.
}
\label{fig:trkeff}
\end{figure}

Accounting for the finite angular resolution on the track direction is not enough in a high occupancy regime. Charged particles confined inside a highly boosted jet can be extremely collimated, resulting in unresolvable tracker hits, especially in the innermost tracking layers. Although an accurate description of this feature would require a full event reconstruction by means of a GEANT-based simulation~\cite{Agostinelli:2002hh}, we construct a simple model of this effect by parametrising the track reconstruction efficiency as a function of the transverse distance between a track and the jet center.  The jet center has the highest density of charged particles, and so this should describe the dominant effect on the resolution. For tracks a distance $R$ from the jet center, we define the track resolution efficiency
\begin{equation}
\epsilon(R) = \frac{2\epsilon_0}{\pi} \arctan\left(\frac{R}{R^*}\right) \,,
\label{eq:trkeff}
\end{equation}
which is plotted in \Fig{fig:trkeff}.

Inspired from CMS public tracker performance~\cite{CMS:2014aa}, we choose $\epsilon_0 = 90\%$ (95\%) and $R^*$ = 0.002 (0.001) for the CMS (FCC) detector. We stress that in principle the tracking angular resolution depends on the track momentum and pseudorapidity, but for simplicity we assume a constant (conservative) value. For a robust determination of the jet center in the case of light quarks and gluon jets we rely on the WTA recombination scheme. On the other hand, given that top quark jets often consists of three hard prongs, parametrising the efficiency as a function of the distance between the track and the jet center would result in an underestimate of track losses for tracks that belong to the subleading prongs. Therefore, in order to realistically simulate track losses inside top jets we first run the $N$-jettiness clustering algorithm ~\cite{Stewart:2010tn,Thaler:2011gf} at particle level to help us identify the three dominant subjet axes. We then require tracks to pass the efficiency criterion from \Eq{eq:trkeff} with respect to each subjet core.   A summary of all the tracking parameters used for the tracking parametrisation in \delphes{}is given in \Tab{tab:trk_param}.
 
\begin {table}
\begin{center}  
\begin{tabular}{l||c|c}
& CMS & FCC \\
  \hline
  \hline 
$B_z$ $(T)$ &  3.8 & 6.0 \\ 
  \hline
Length $(m)$ & 6 & 12 \\
 \hline
Radius $(m)$ & 1.3 &  2.6 \\
 \hline
 \hline
$\epsilon_0$ & 0.90 & 0.95\\
 \hline
 $R^*$ & 0.002 & 0.001  \\
 \hline
 \hline
$\sigma(p_T)/p_T $ &  $0.2\cdot p_T$ (TeV/c) &  $0.02\cdot p_T$ (TeV/c)\\
   \hline  
$\sigma(\eta,\phi)$ & 0.002 & 0.001 
\end{tabular}
\caption{Tracking-related parameters for the CMS and FCC setup in Delphes.}
\label{tab:trk_param}
\end{center}
\end{table}

\begin {table}
\begin{center}  
\begin{tabular}{l||c|c}
& CMS & FCC \\
  \hline
  \hline  
$\sigma(E)/E $ (ECAL)& 7\%/$\sqrt{E} \oplus 0.7\%$ &  3\%/$\sqrt{E} \oplus 0.3\%$\\
   \hline  
$\sigma(E)/E $ (HCAL)& 150\%/$\sqrt{E} \oplus 5\%$  &  50\%/$\sqrt{E} \oplus 1\%$\\
  \hline
  \hline
$\eta \times \phi$ cell size (ECAL)& $(0.02\times0.02)$ &  $(0.01\times0.01)$\\
  \hline  
 $\eta \times \phi$ cell size  (HCAL)& $(0.1\times0.1)$ & $(0.05\times0.05)$ 
\end{tabular}
\caption{Calorimeter parameters for the CMS and FCC setup in Delphes.}
\label{tab:cal_param}
\end{center}
\end{table}

\subsection{Calorimetry}\label{appsub:calorimetry}

After propagating within the magnetic field, long-lived particles reach the electromagnetic (ECAL) and hadronic (HCAL) calorimeters. Since these are modeled in \delphes{}by two-dimensional grids of variable spacing, the calorimeter deposits natively include finite angular resolution effects. In order to accurately model the angular resolution on reconstructed jets, separate grids for ECAL and HCAL has been adopted for this study. The ECAL and HCAL maps for the CMS detector are taken from~\Ref{Bayatian:2006zz}.

For the FCC detector we assume that the same elementary calorimeter cells will be used, but placed at twice the CMS distance from the interaction point, leading to an improved angular resolution by a factor 2. In order to obtain the best possible energy resolution, the best nominal performance between the CMS \cite{Bayatian:2006zz} and ATLAS \cite{Aad:2009wy} calorimeters was chosen. The calorimeter performance for the CMS and FCC setups are summarised in \Tab{tab:cal_param}.

\begin{table}[t]
\begin{center}
\begin{tabular}{cc|ccccc}
\multicolumn{7}{c}{{\bf 20\% Top Efficiency}}\\
 &$p_T$ cut&$[2.5,5] $ TeV & $[5,7.5] $ TeV & $[7.5,10] $ TeV& $ [10,15] $ TeV&$ [15,20] $ TeV\\
\toprule
{\multirow{2}{*}{gluons} } &CMS& 2\% & 3\%& 4\%& 5\%& 6\%\\
&FCC& 1\% & 2\%& 2\%& 3\%& 4\%\\
\cline{2-7}
{\multirow{2}{*}{quarks} } &CMS& 1\% & 2\%& 3\%& 5\%& 7\%\\
&FCC& 0.5\% & 1\%& 1.5\%& 2\%& 4\%\\
\end{tabular}
\begin{tabular}{cc|ccccc}
\multicolumn{7}{c}{{\bf 40\% Top Efficiency}}\\
 &$p_T$ cut&$[2.5,5] $ TeV & $[5,7.5] $ TeV & $[7.5,10] $ TeV& $ [10,15] $ TeV&$ [15,20] $ TeV\\
\toprule
{\multirow{2}{*}{gluons} } &CMS& 7\% & 9\%& 10\%& 14\%& 17\%\\
&FCC& 5\% & 6\%& 7\%& 10\%& 12\%\\
\cline{2-7}
{\multirow{2}{*}{quarks} } &CMS& 3\% & 5\%& 7\%& 11\%& 17\%\\
&FCC& 1.5\% & 2.5\%& 4\%& 5\%& 8\%\\
\end{tabular}
\begin{tabular}{cc|ccccc}
\multicolumn{7}{c}{{\bf 60\% Top Efficiency}}\\
 &$p_T$ cut&$[2.5,5] $ TeV & $[5,7.5] $ TeV & $[7.5,10] $ TeV& $ [10,15] $ TeV&$ [15,20] $ TeV\\
\toprule
{\multirow{2}{*}{gluons} } &CMS& 18\%  & 20\%& 24\%& 30\%& 38\%\\
&FCC& 13\% & 15\%& 20\%& 24\%& 25\%\\
\cline{2-7}
{\multirow{2}{*}{quarks} } &CMS& 7\% & 10\%& 15\%& 22\%& 30\%\\
&FCC& 4\% & 6\%& 8\%& 11\%& 15\%\\
\end{tabular}
\end{center}
\caption{
Table of background rejection rates at fixed signal efficiencies for jet $p_T$s ranging from $2.5$ TeV to $20$ TeV at the CMS or FCC detector.  For gluon (quark) jet backgrounds, efficiencies are determined from cuts on $\tau_{3,2}$ ($D_3$) measured on tracks. The cut on the rescaled track-based jet mass of $m_J\in[120,250]$ GeV is included in the efficiencies.
}
\label{tab:rej_pt}
\end{table}

\section{$p_T$ scan and results}\label{app:allpt}

\begin{figure}[t]
\begin{center}
\subfloat[]{
\includegraphics[width=6cm]{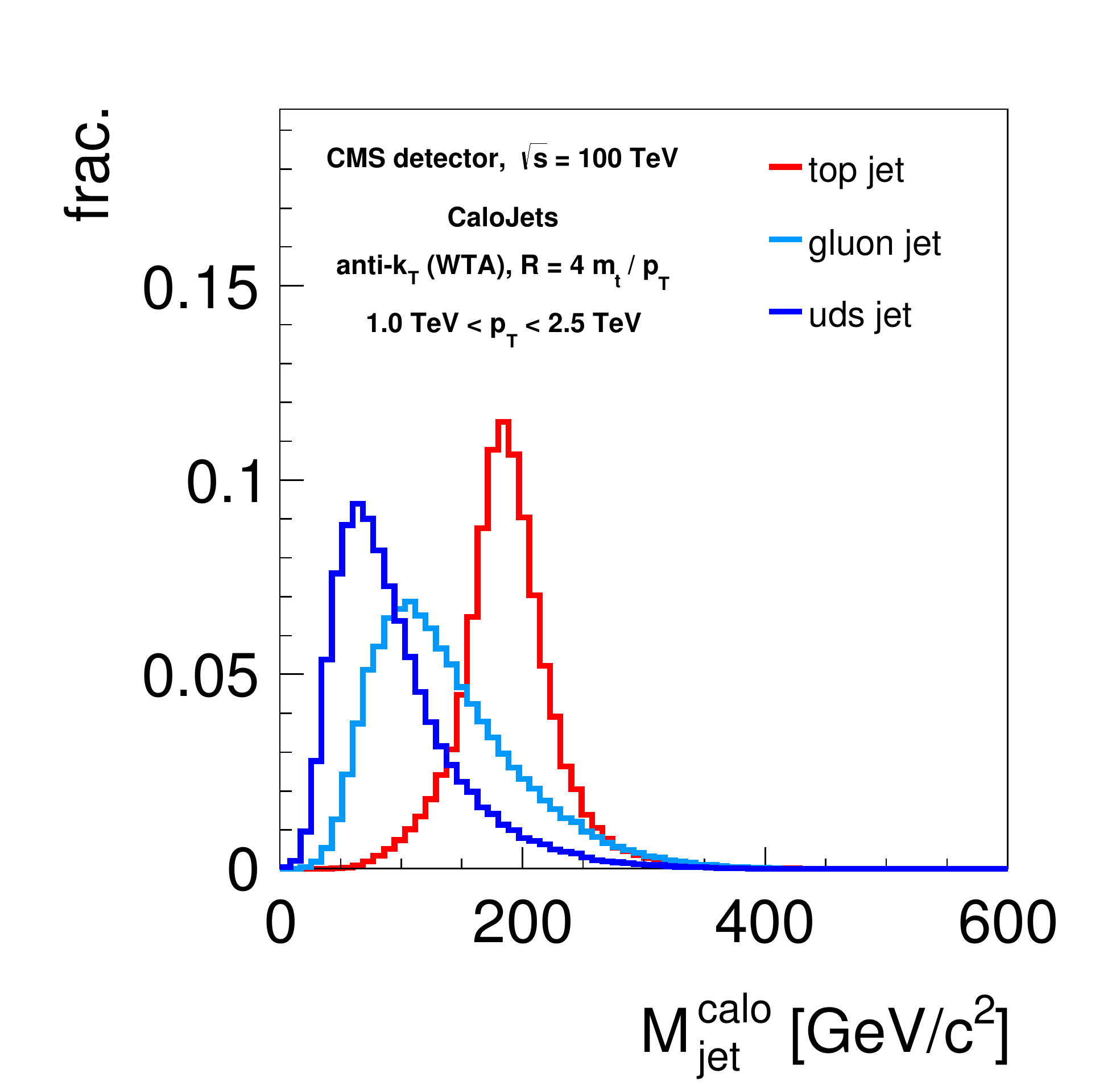}
}
\subfloat[]{
\includegraphics[width=6cm]{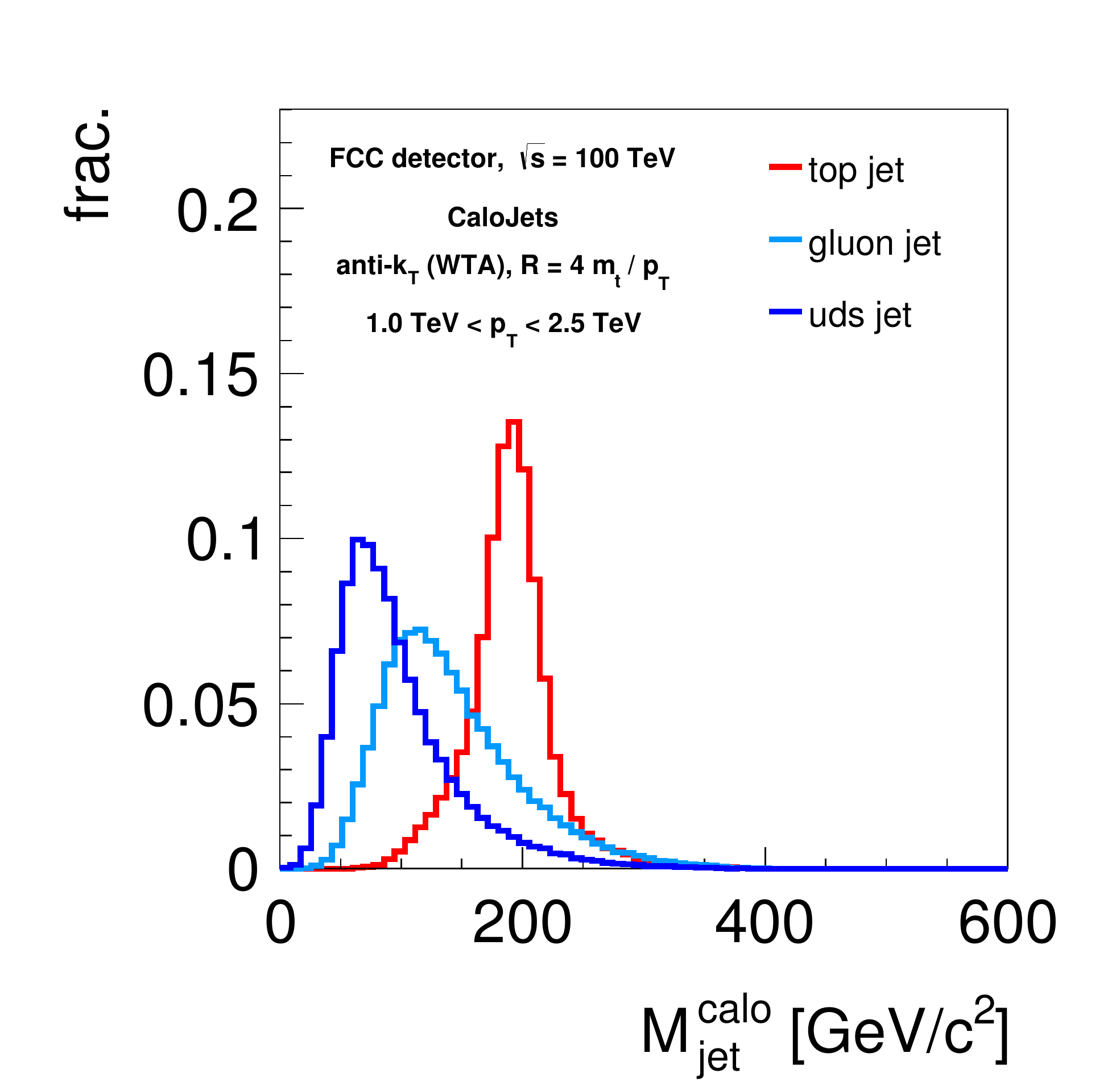}
}
\\
\subfloat[]{
\includegraphics[width=6cm]{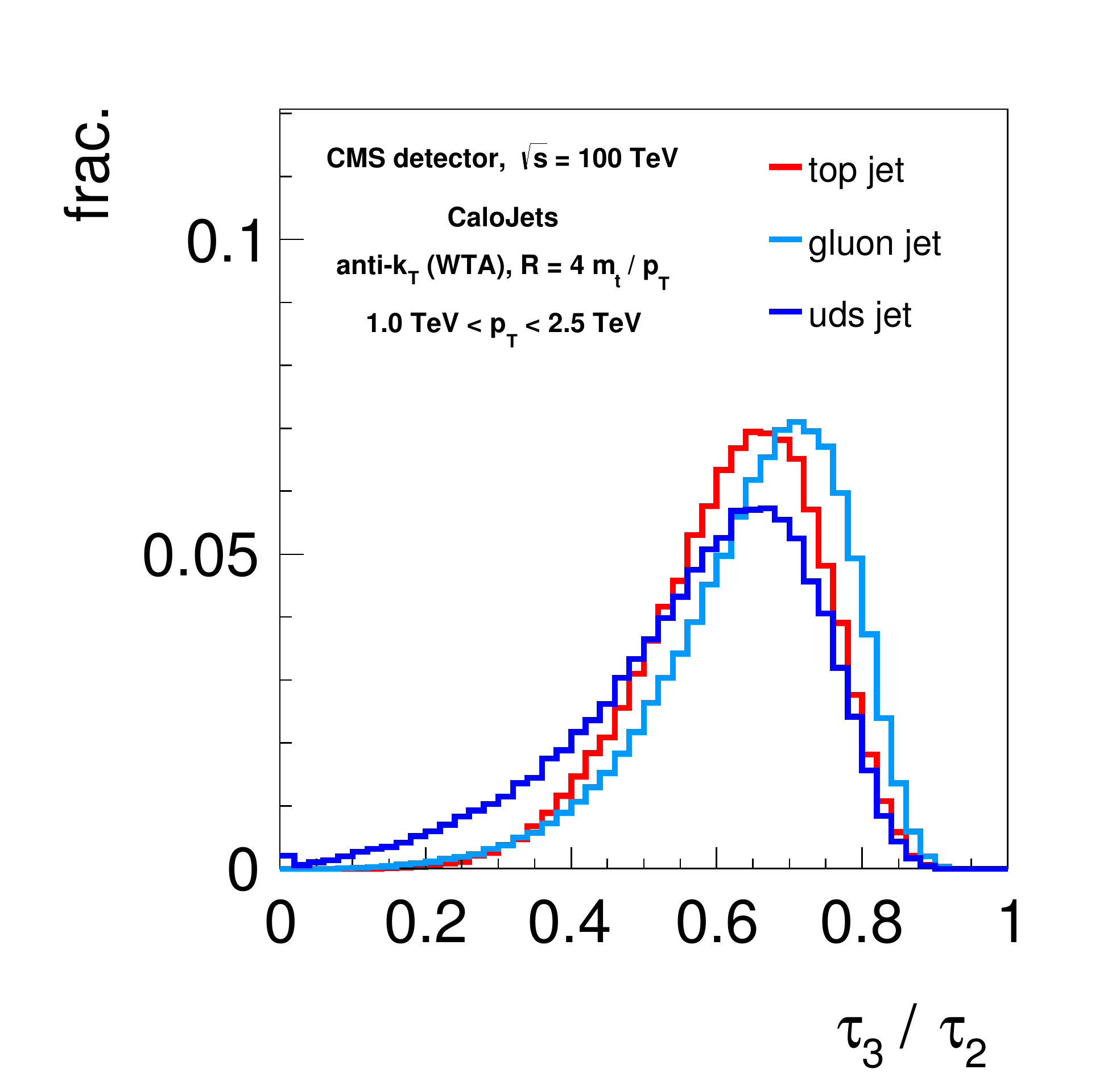}
}
\subfloat[]{
\includegraphics[width=6cm]{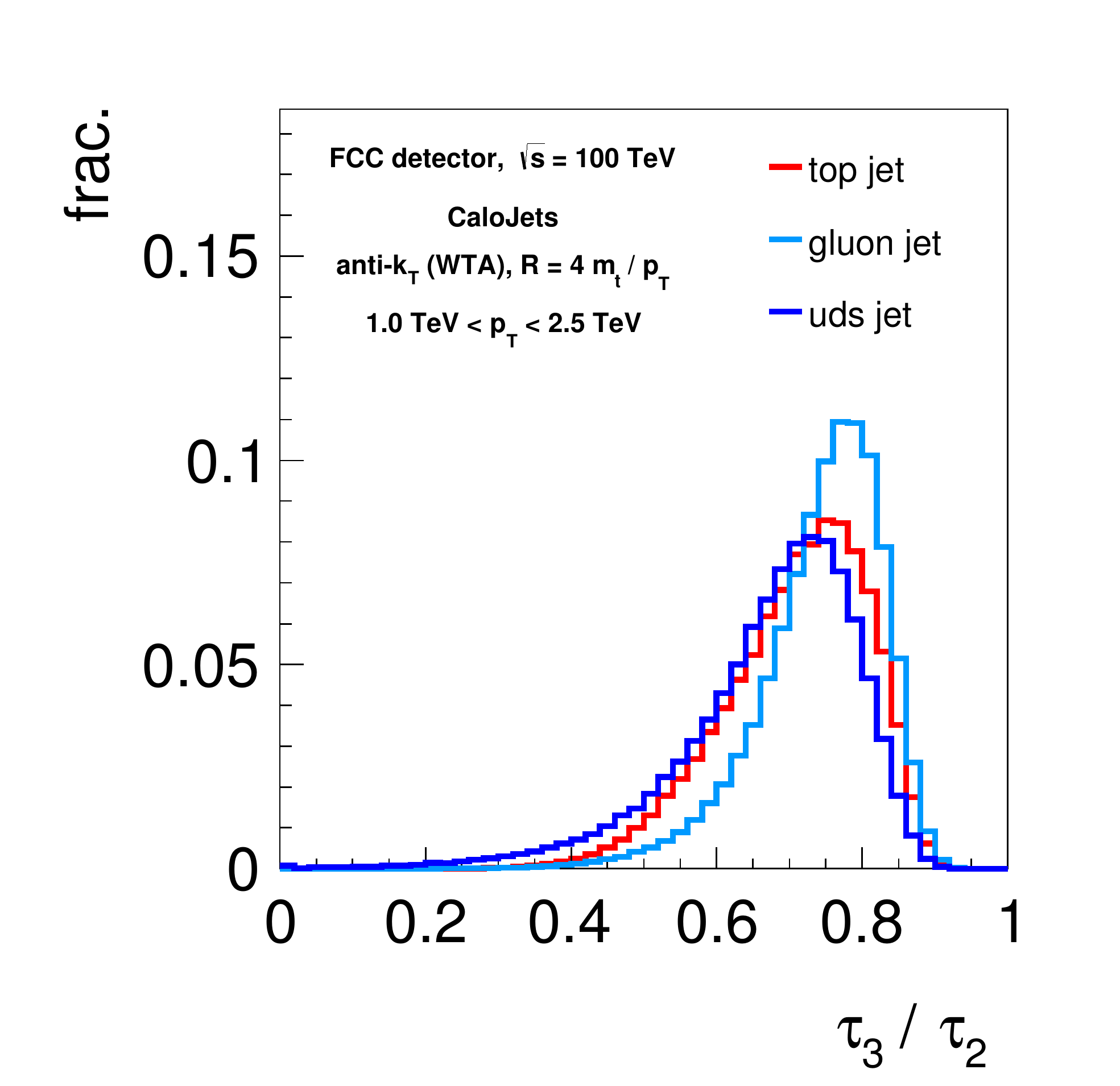}
}
\\
\subfloat[]{
\includegraphics[width=6cm]{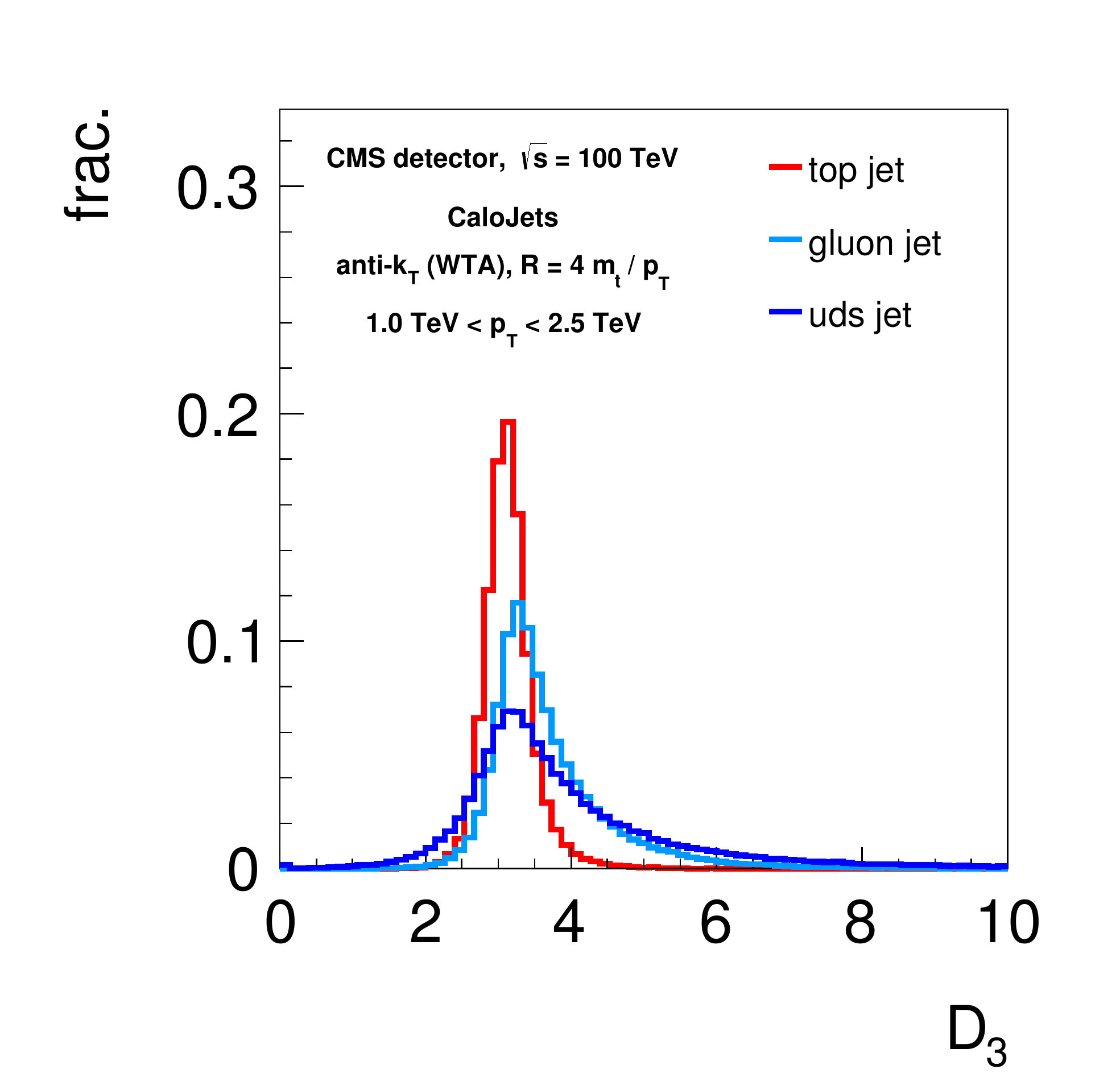}
}
\subfloat[]{
\includegraphics[width=6cm]{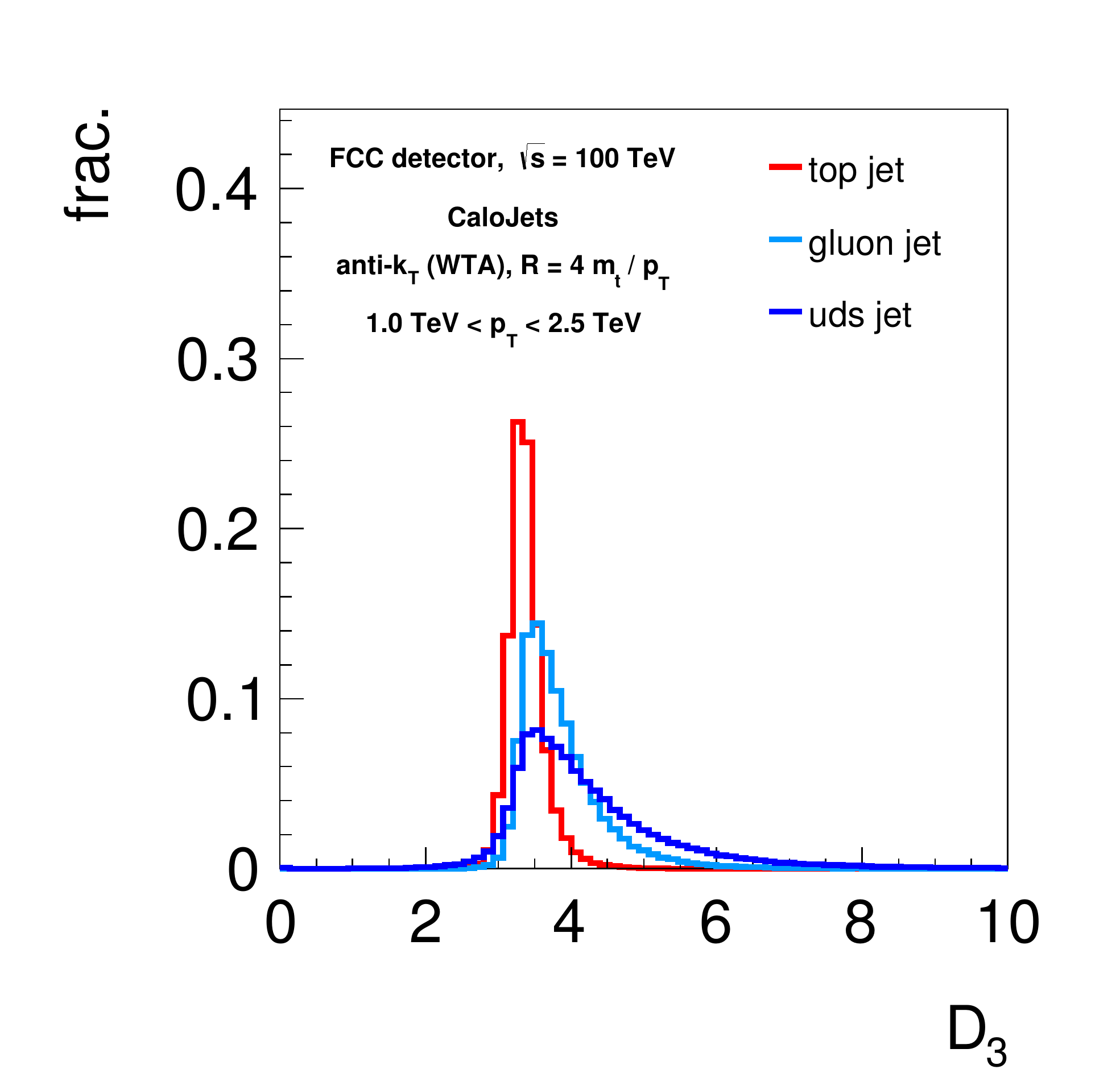}
}
\end{center}
\caption{
Distributions of the jet mass (top), $\tau_3 / \tau_2$ (middle), and $D_3$ (bottom) as measured on anti-$k_T$ jets with radius $R=4 m_\text{top}/p_T$ and $p_T \in [1.0,2.5]$ TeV on boosted top jets and QCD jets from light quarks and gluons.  (left) Distributions as measured from the CMS detector's calorimeter system.  (right) Distributions as measured from a future collider detector's calorimeter system.
}
\label{fig:cal_scale_r32_1}
\end{figure}

\begin{figure}[t]
\begin{center}
\subfloat[]{
\includegraphics[width=6cm]{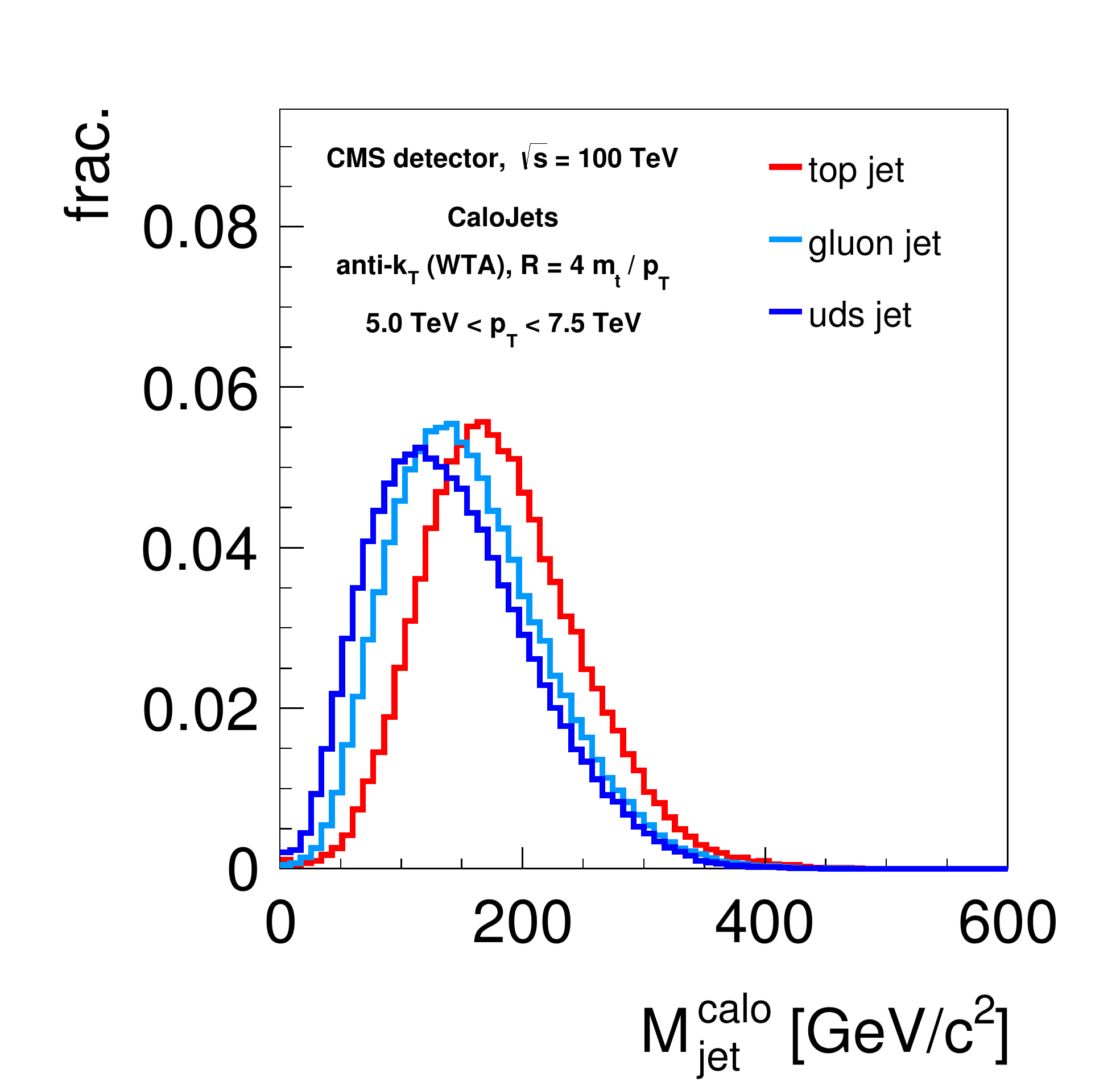}
}
\subfloat[]{
\includegraphics[width=6cm]{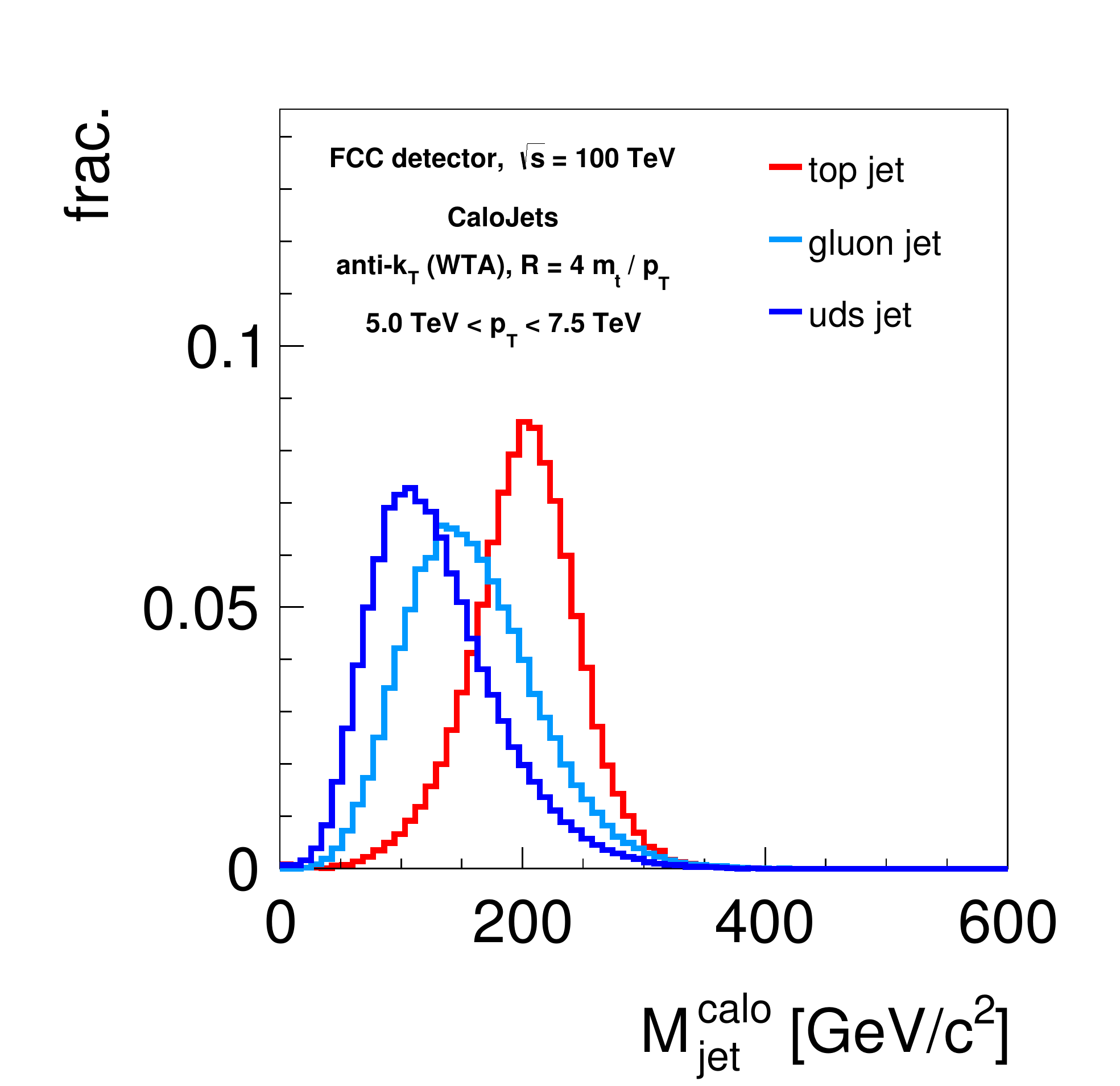}
}
\\
\subfloat[]{
\includegraphics[width=6cm]{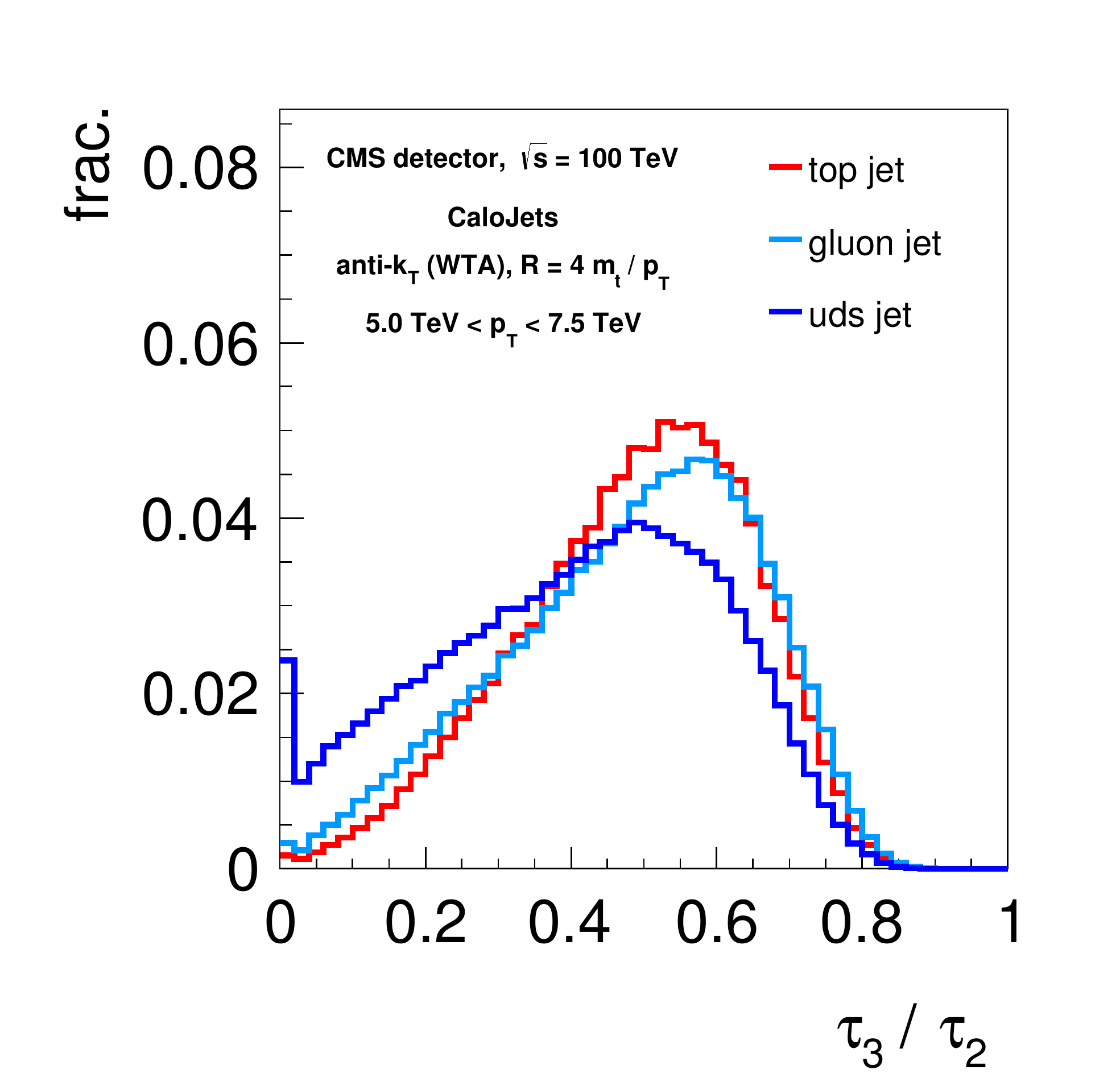}
}
\subfloat[]{
\includegraphics[width=6cm]{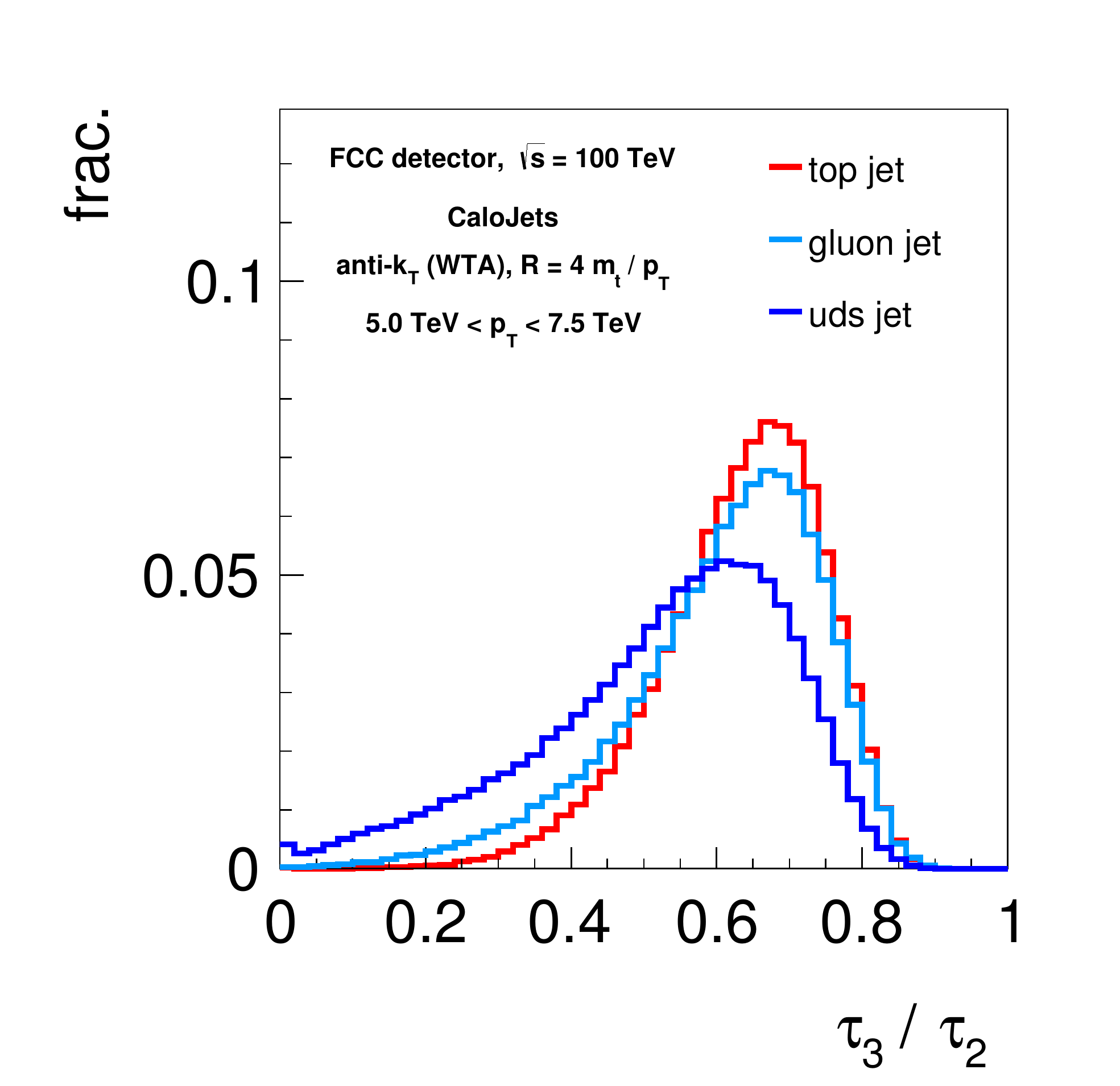}
}
\\
\subfloat[]{
\includegraphics[width=6cm]{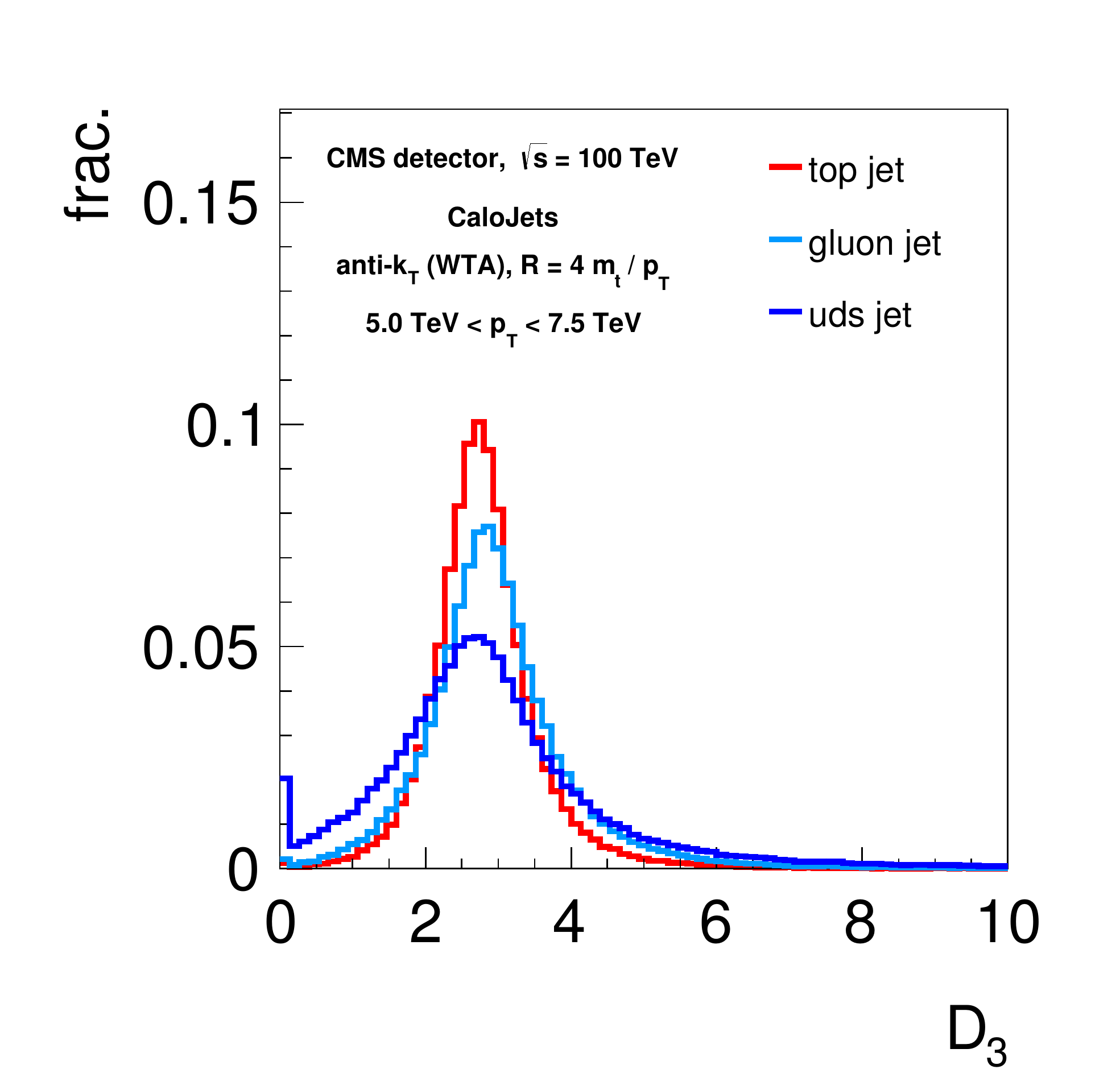}
}
\subfloat[]{
\includegraphics[width=6cm]{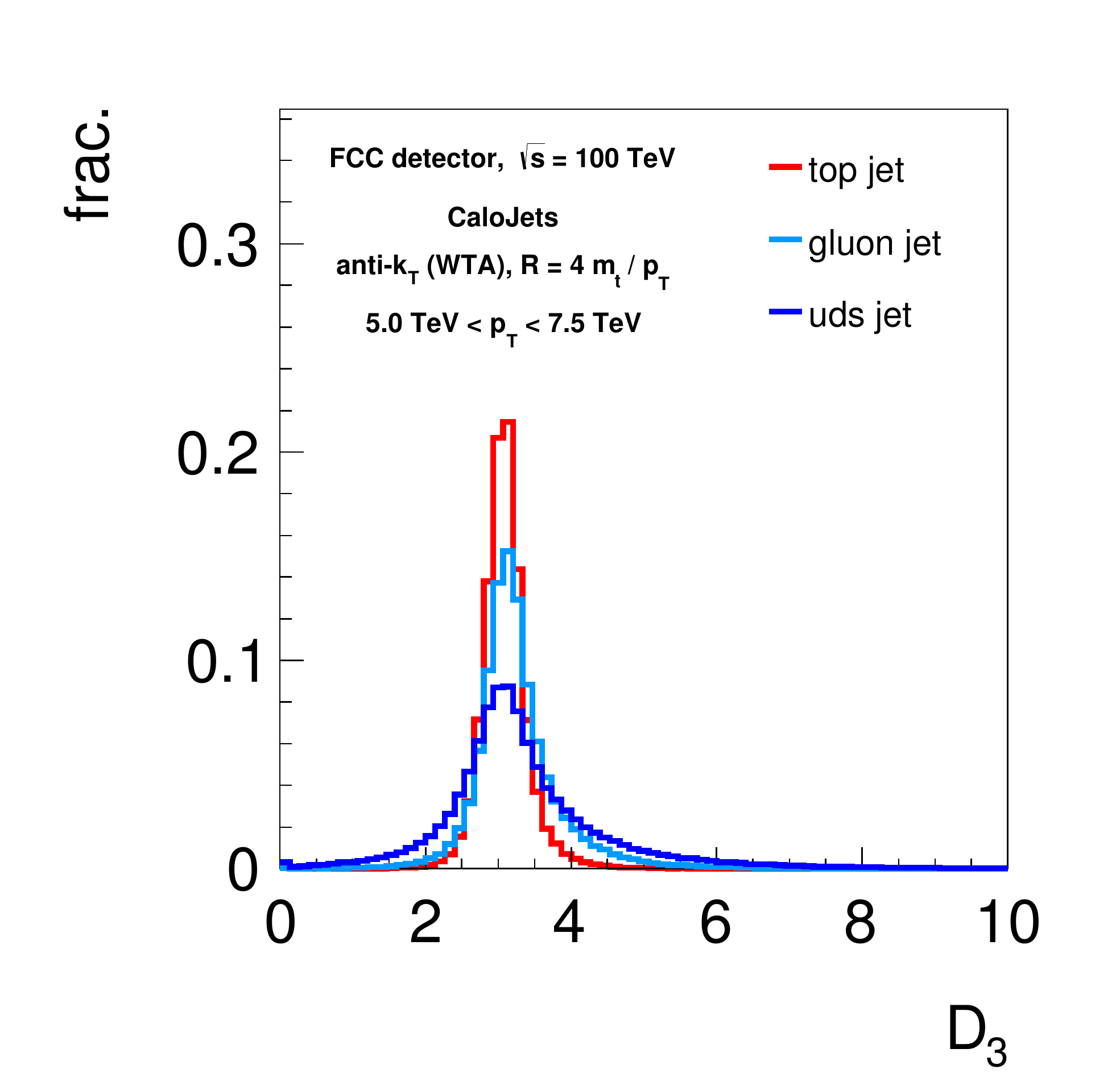}
}
\end{center}
\caption{
Distributions of the jet mass (top), $\tau_3 / \tau_2$ (middle), and $D_3$ (bottom) as measured on anti-$k_T$ jets with radius $R=4 m_\text{top}/p_T$ and $p_T \in [5.0,7.5]$ TeV on boosted top jets and QCD jets from light quarks and gluons.  (left) Distributions as measured from the CMS detector's calorimeter system.  (right) Distributions as measured from a future collider detector's calorimeter system.
}
\label{fig:cal_scale_r32_5}
\end{figure}

\begin{figure}[t]
\begin{center}
\subfloat[]{
\includegraphics[width=6cm]{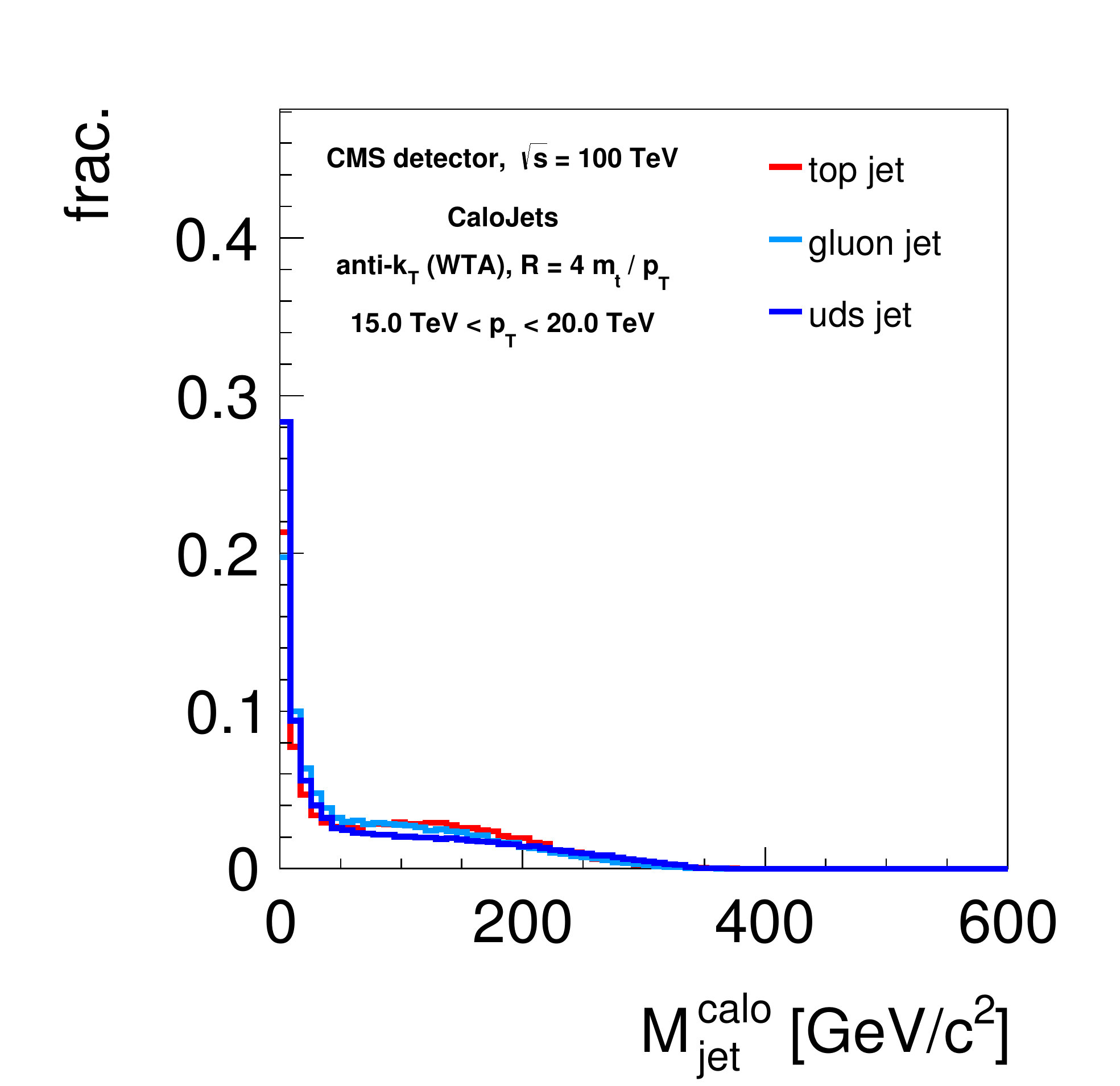}
}
\subfloat[]{
\includegraphics[width=6cm]{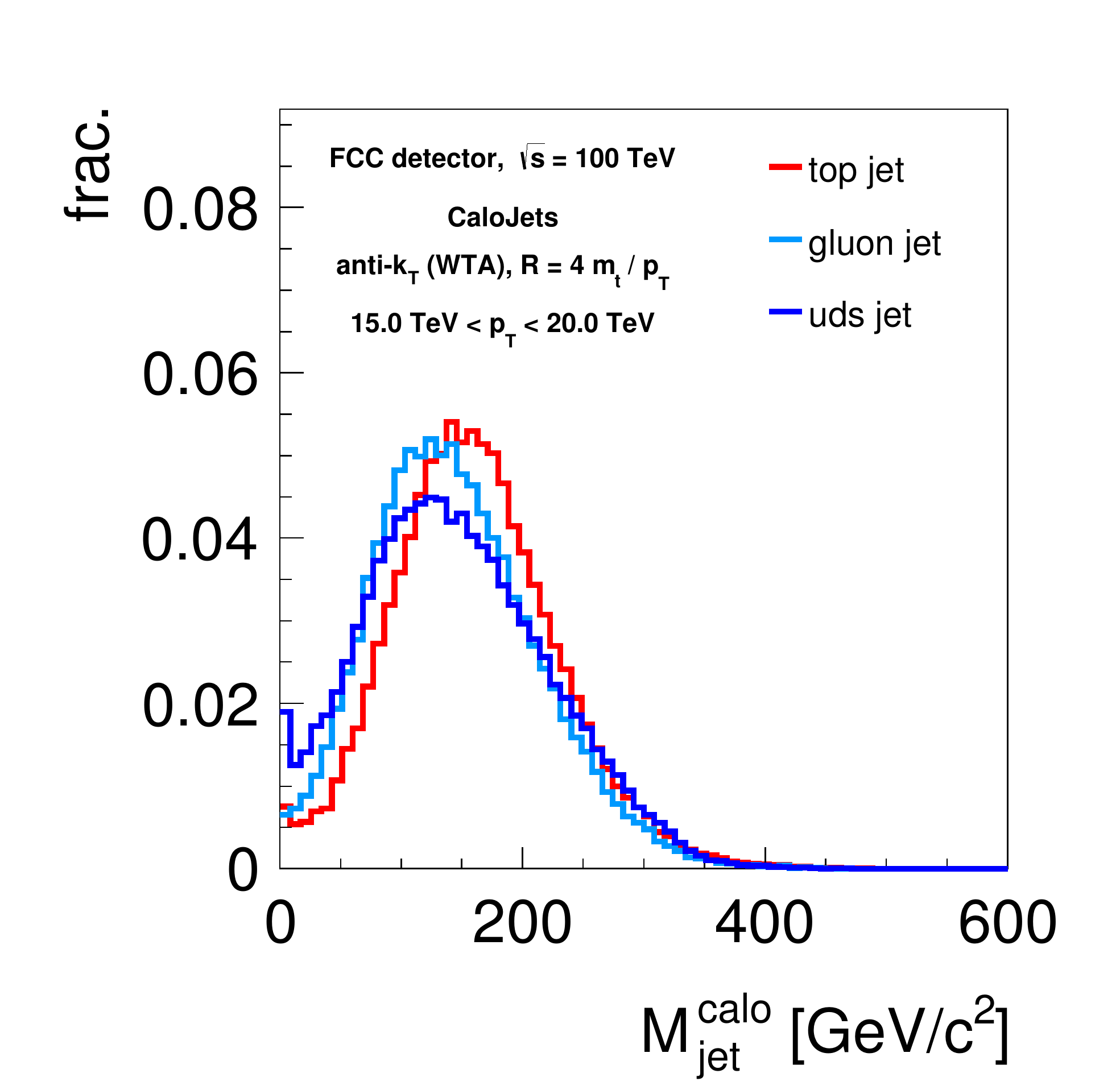}
}
\\
\subfloat[]{
\includegraphics[width=6cm]{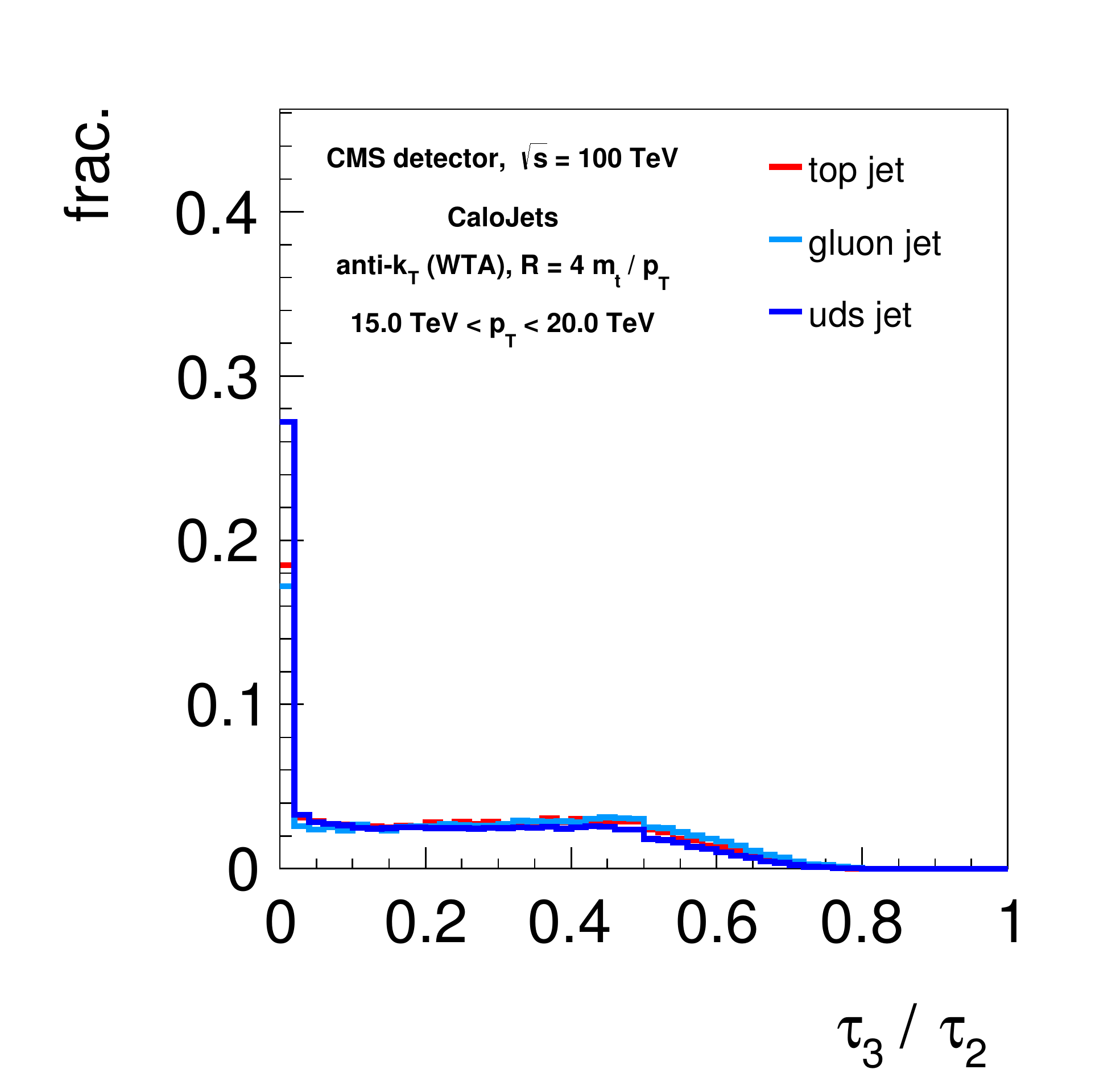}
}
\subfloat[]{
\includegraphics[width=6cm]{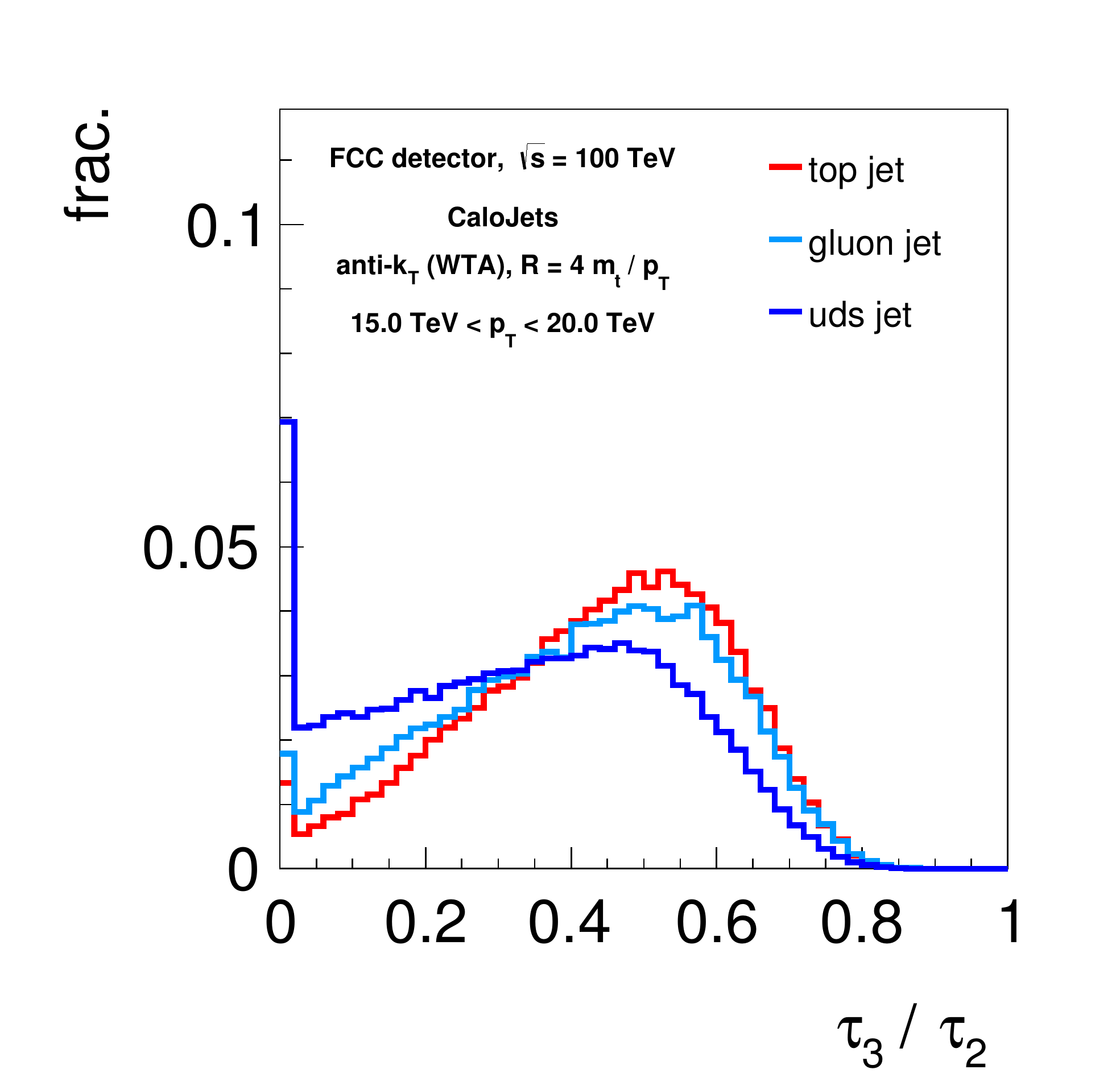}
}
\\
\subfloat[]{
\includegraphics[width=6cm]{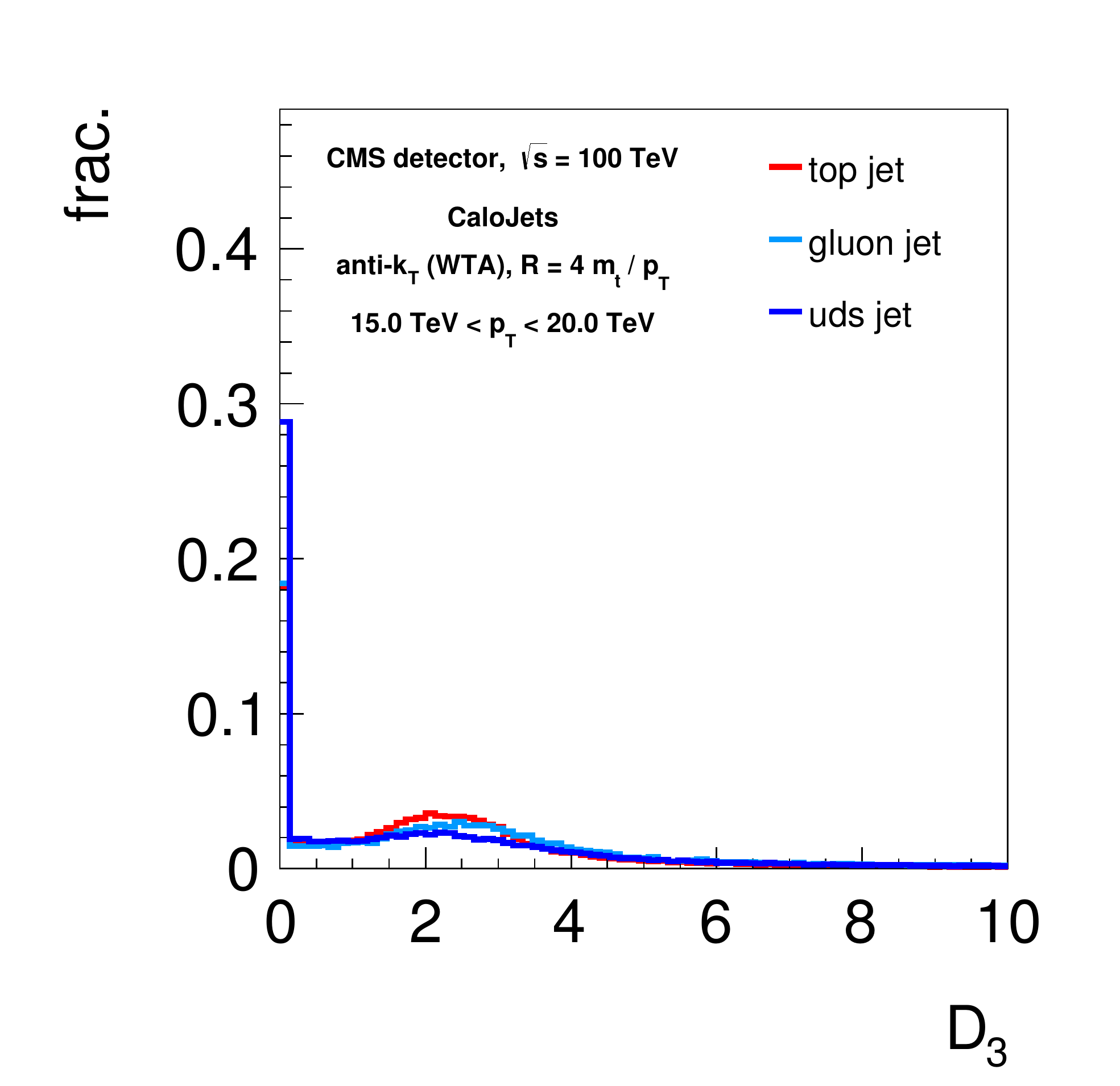}
}
\subfloat[]{
\includegraphics[width=6cm]{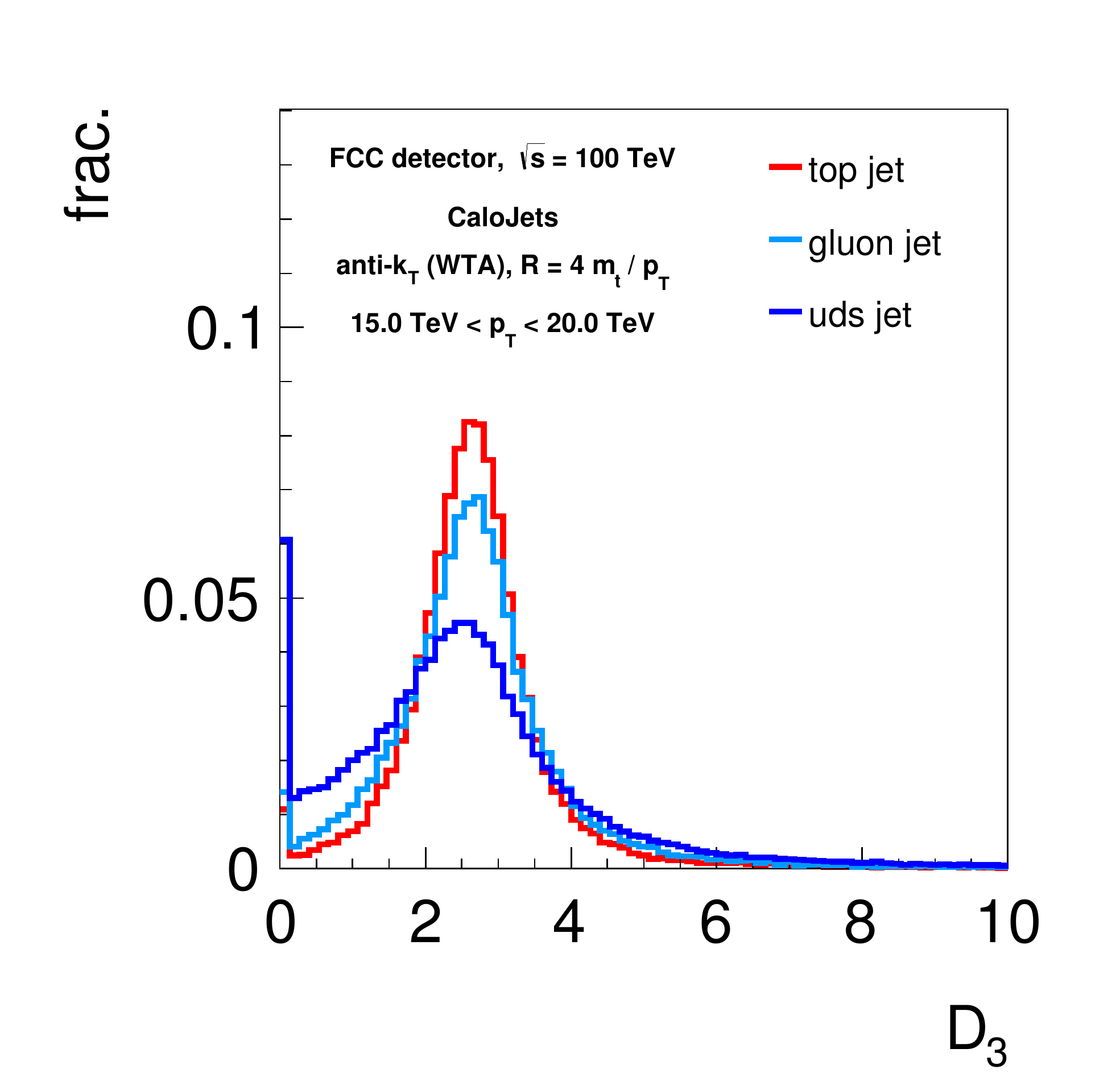}
}
\end{center}
\caption{
Distributions of the jet mass (top), $\tau_3 / \tau_2$ (middle), and $D_3$ (bottom) as measured on anti-$k_T$ jets with radius $R=4 m_\text{top}/p_T$ and $p_T \in [15,20]$ TeV on boosted top jets and QCD jets from light quarks and gluons.  (left) Distributions as measured from the CMS detector's calorimeter system.  (right) Distributions as measured from a future collider detector's calorimeter system.
}
\label{fig:cal_scale_r32_15}
\end{figure}

\begin{figure}[t]
\begin{center}
\subfloat[]{
\includegraphics[width=6cm]{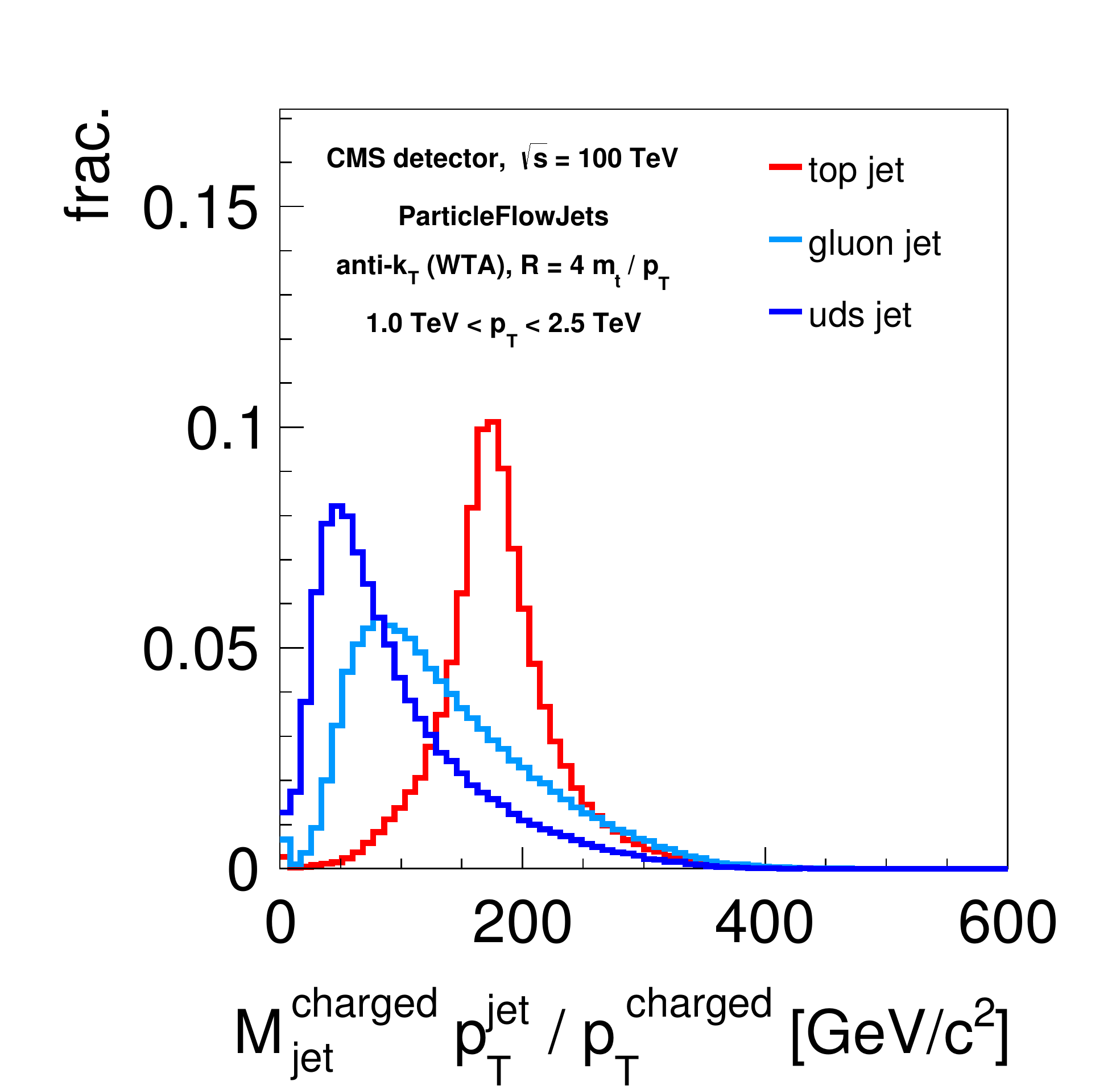}
}
\subfloat[]{
\includegraphics[width=6cm]{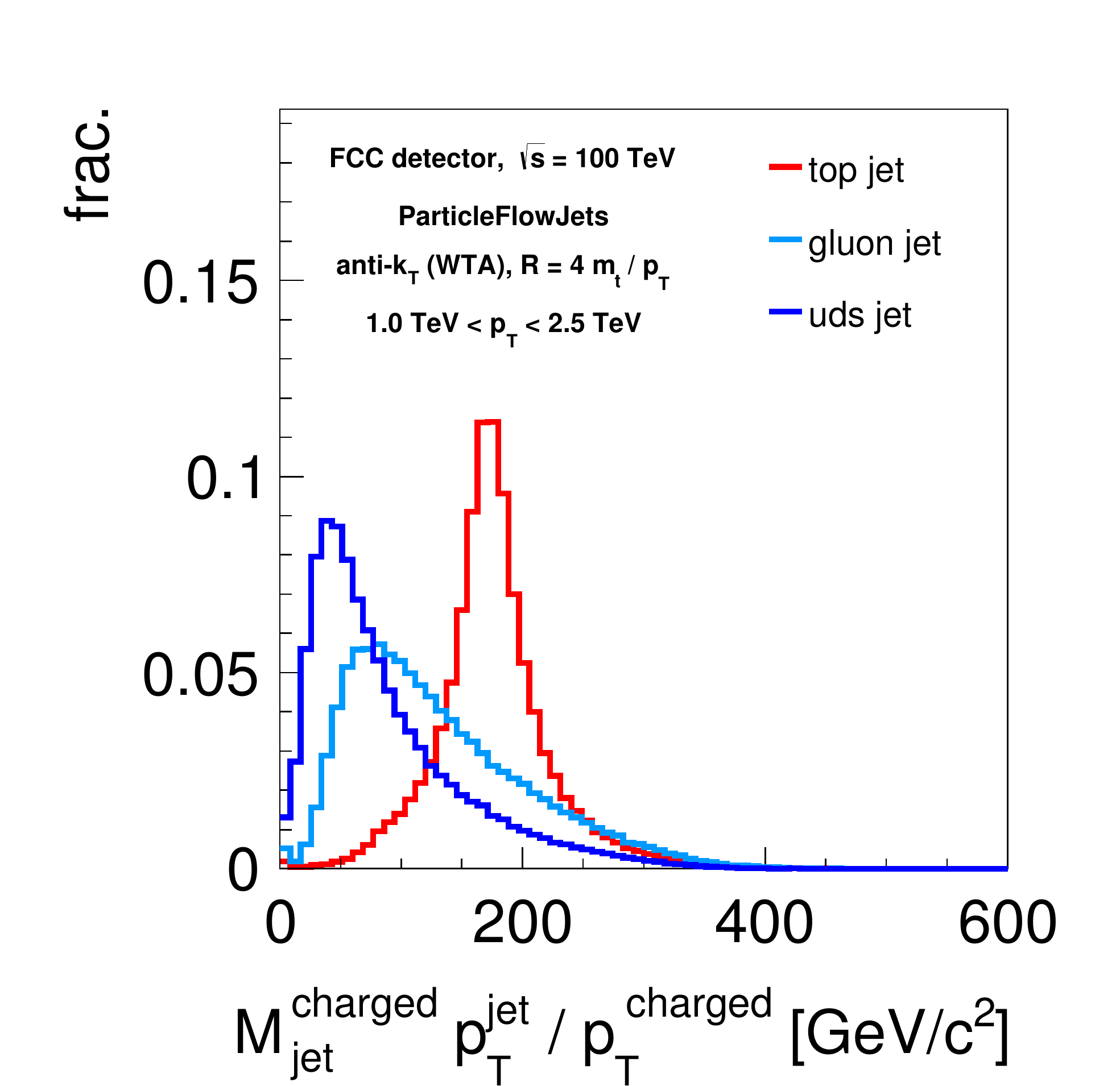}
}
\\
\subfloat[]{
\includegraphics[width=6cm]{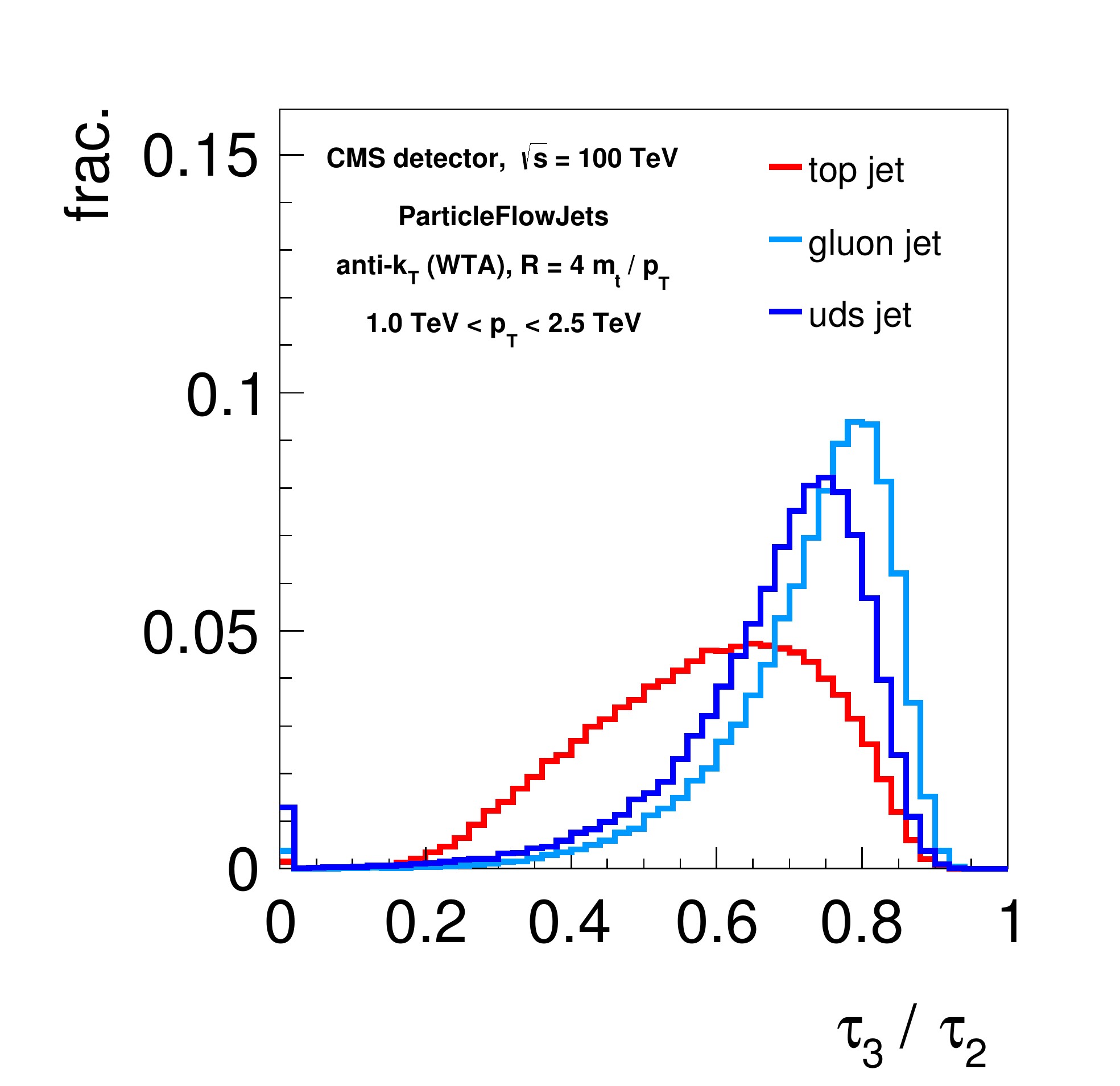}
}
\subfloat[]{
\includegraphics[width=6cm]{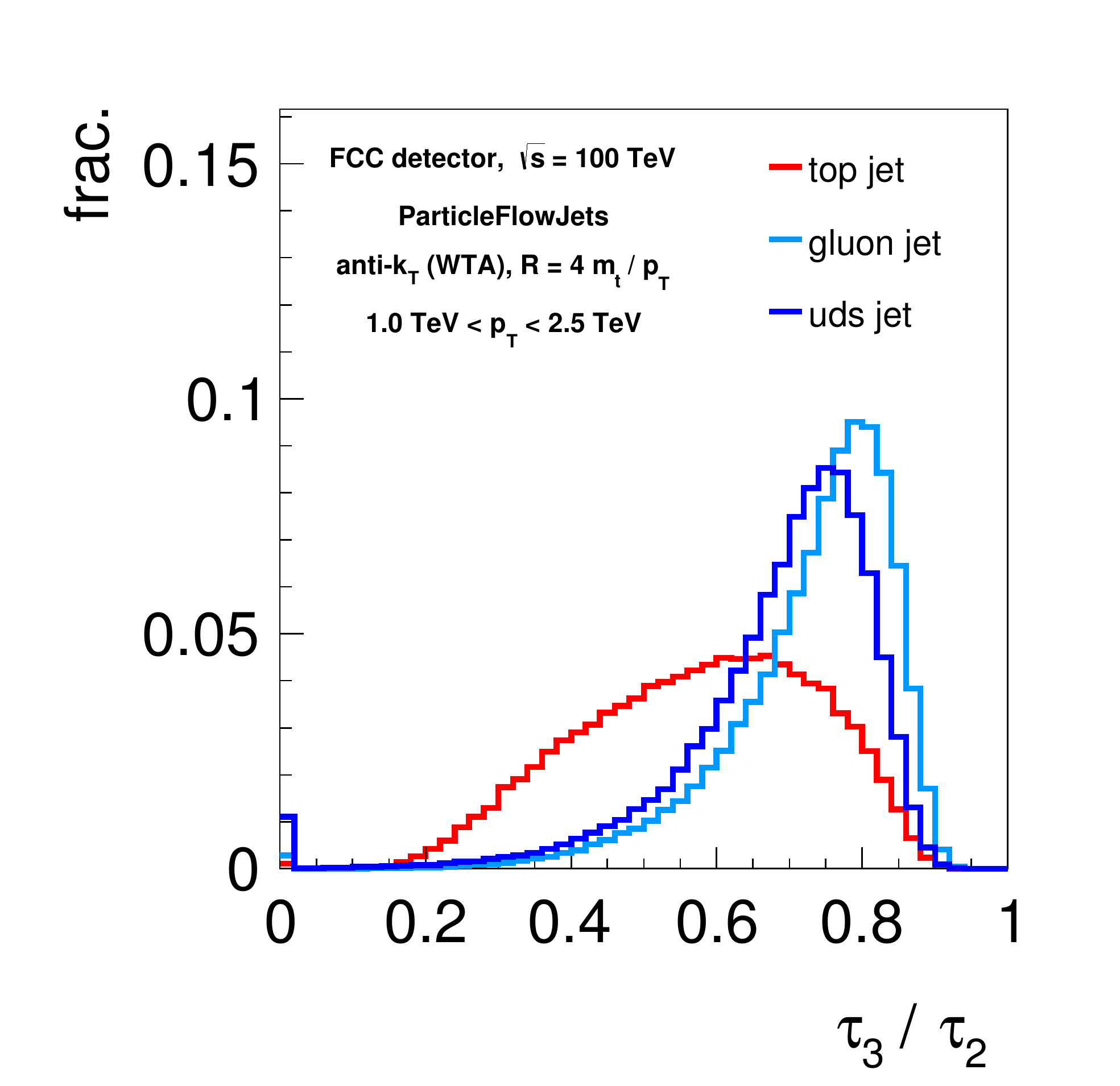}
}
\\
\subfloat[]{
\includegraphics[width=6cm]{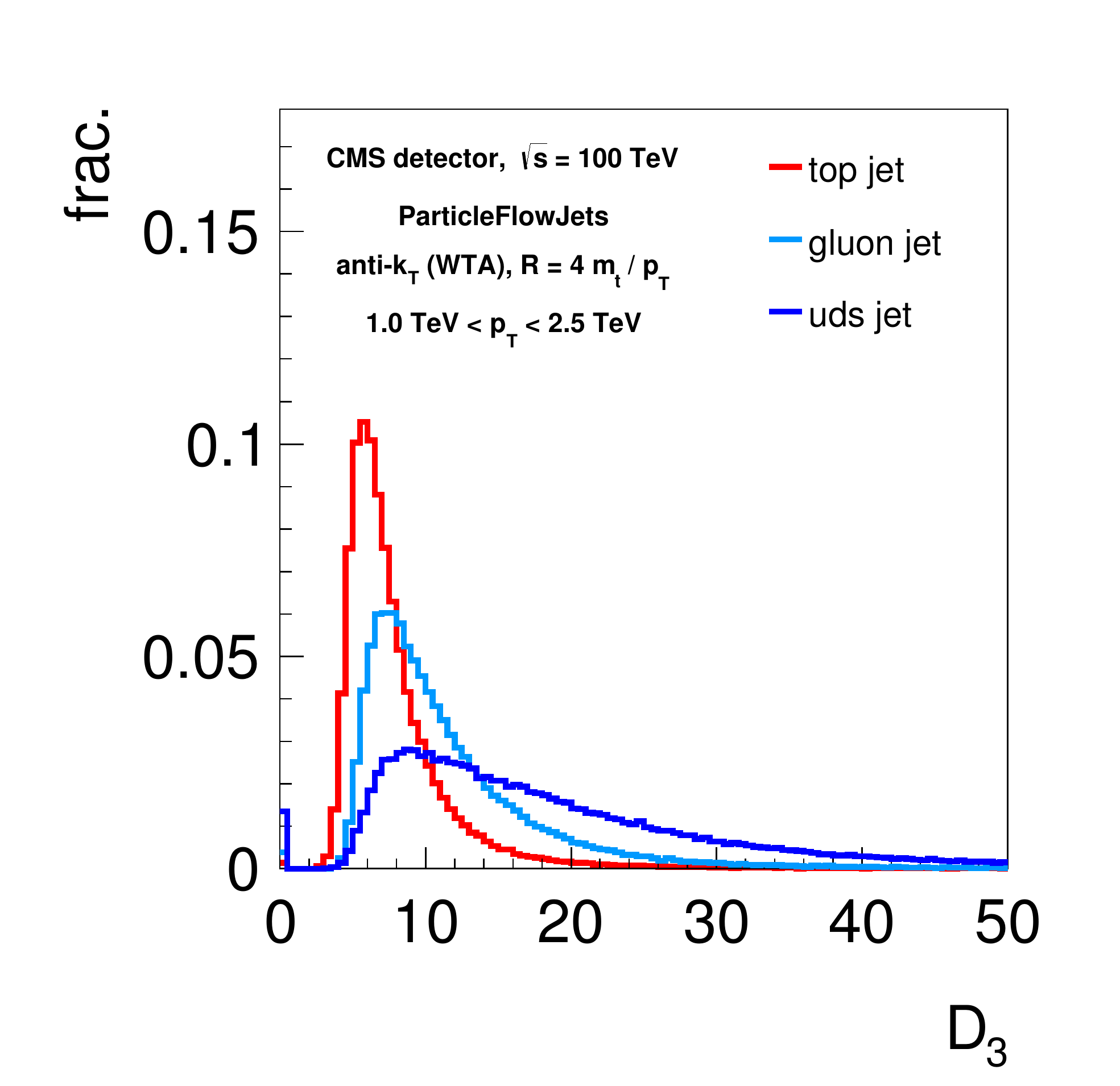}
}
\subfloat[]{
\includegraphics[width=6cm]{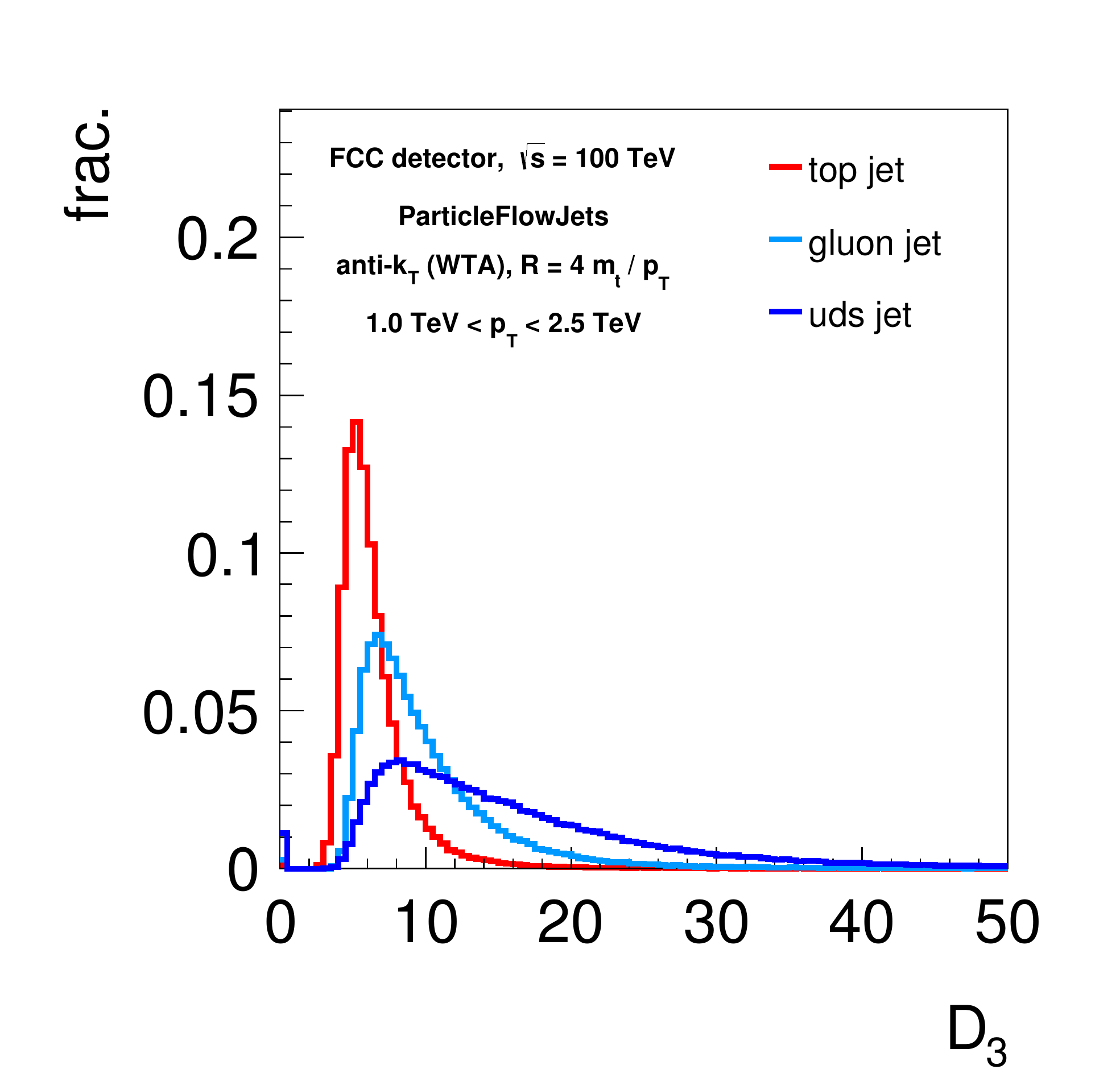}
}
\end{center}
\caption{
Distributions of the rescaled track-based jet mass (top), $\tau_3 / \tau_2$ (middle), and $D_3$ (bottom) as measured on anti-$k_T$ jets with radius $R=4 m_\text{top}/p_T$ and $p_T \in [1.0,2.5]$ TeV on boosted top jets and QCD jets from light quarks and gluons.  (left) Distributions as measured from the CMS detector's tracking system.  (right) Distributions as measured from a future collider detector's tracking system.
}
\label{fig:track_scale_r32_2}
\end{figure}

\begin{figure}[t]
\begin{center}
\subfloat[]{
\includegraphics[width=6cm]{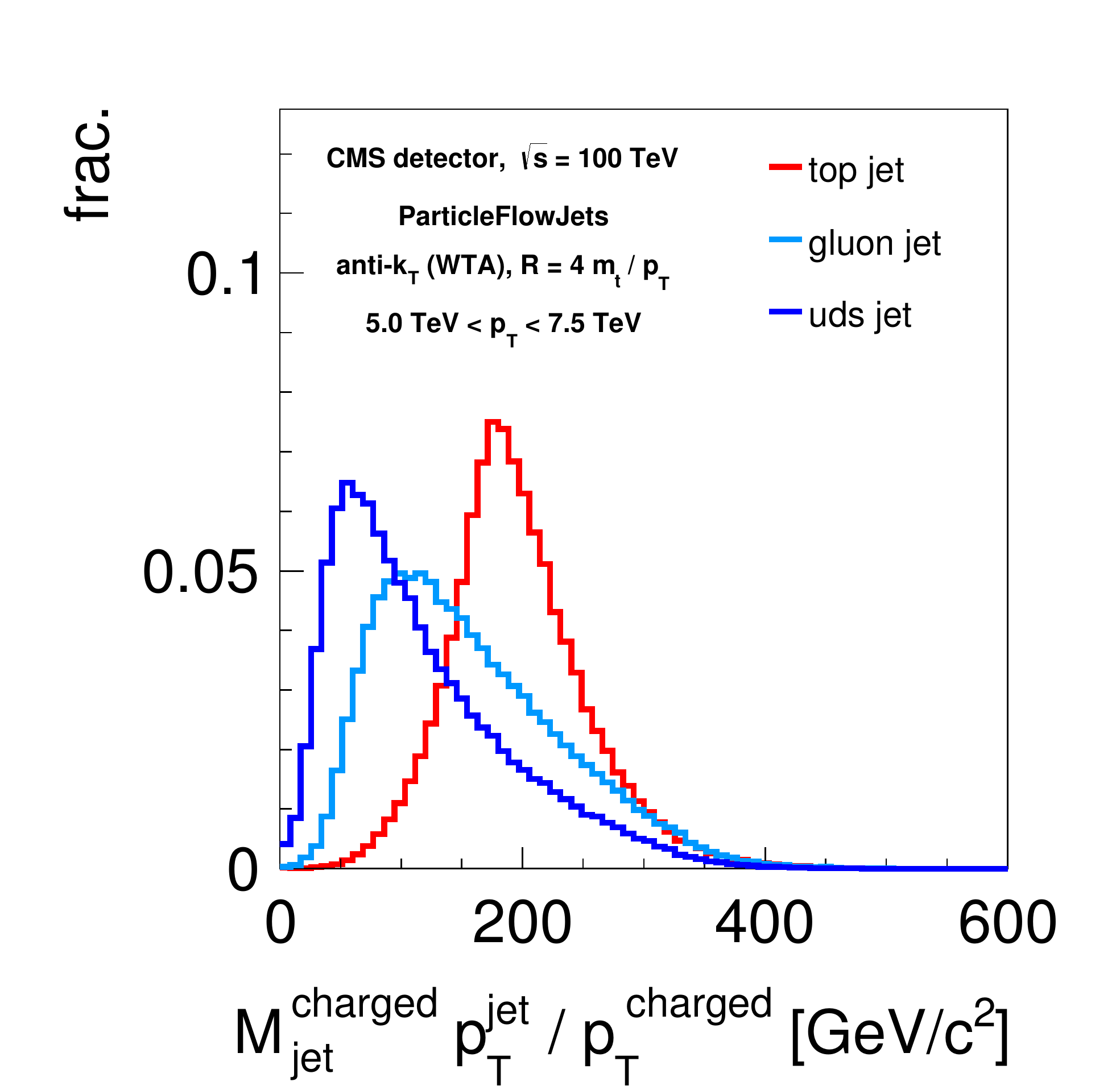}
}
\subfloat[]{
\includegraphics[width=6cm]{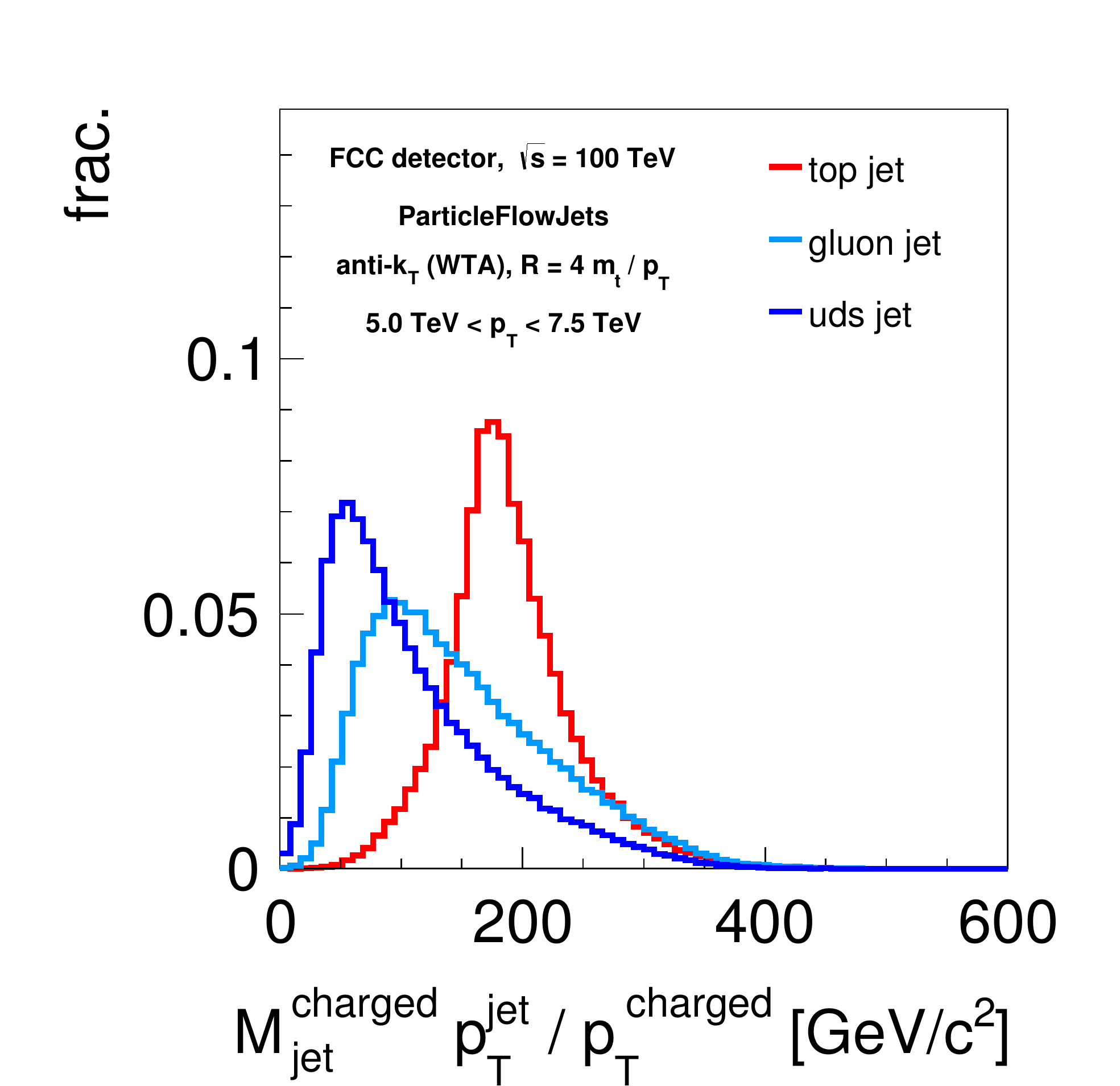}
}
\\
\subfloat[]{
\includegraphics[width=6cm]{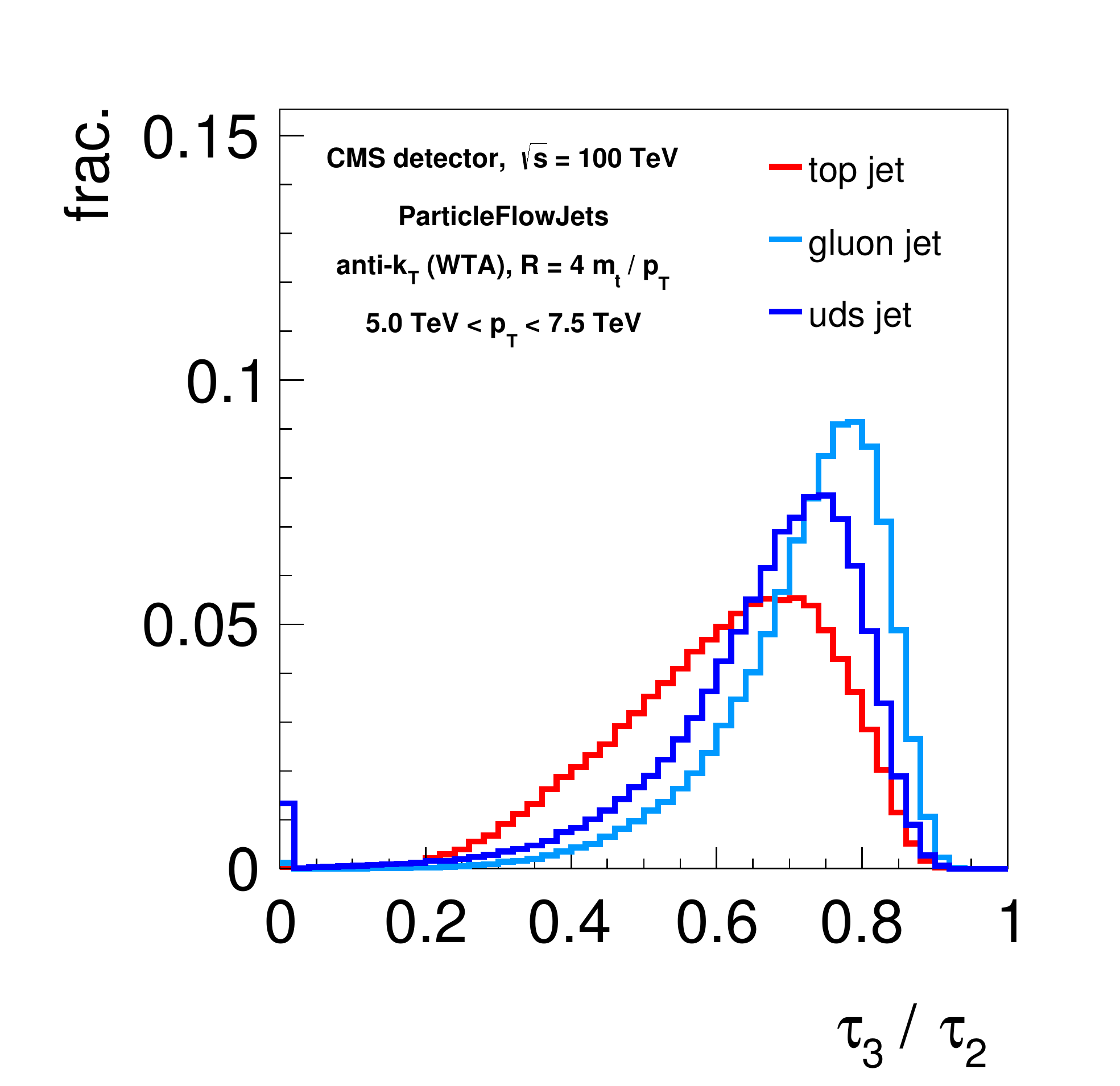}
}
\subfloat[]{
\includegraphics[width=6cm]{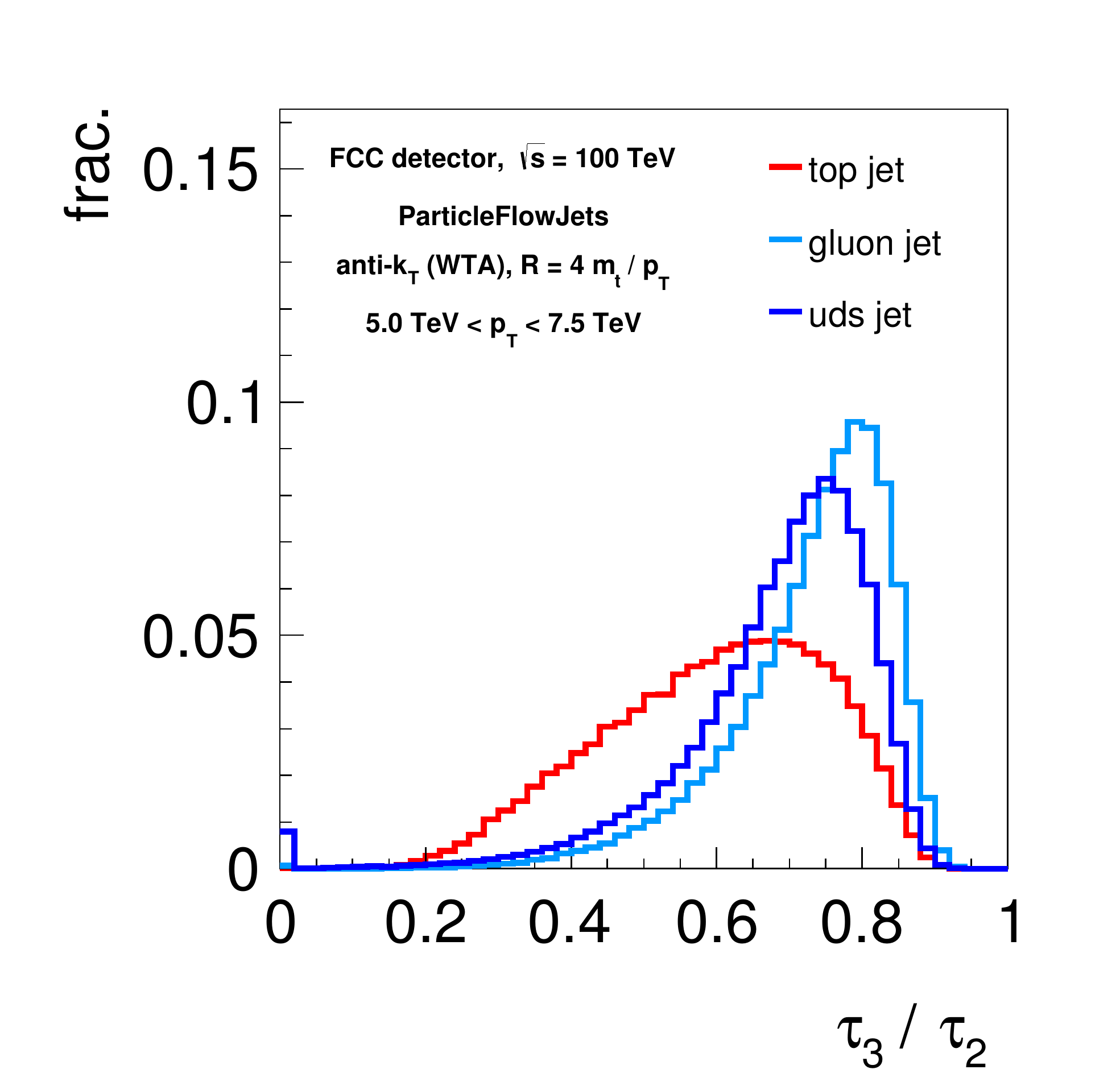}
}
\\
\subfloat[]{
\includegraphics[width=6cm]{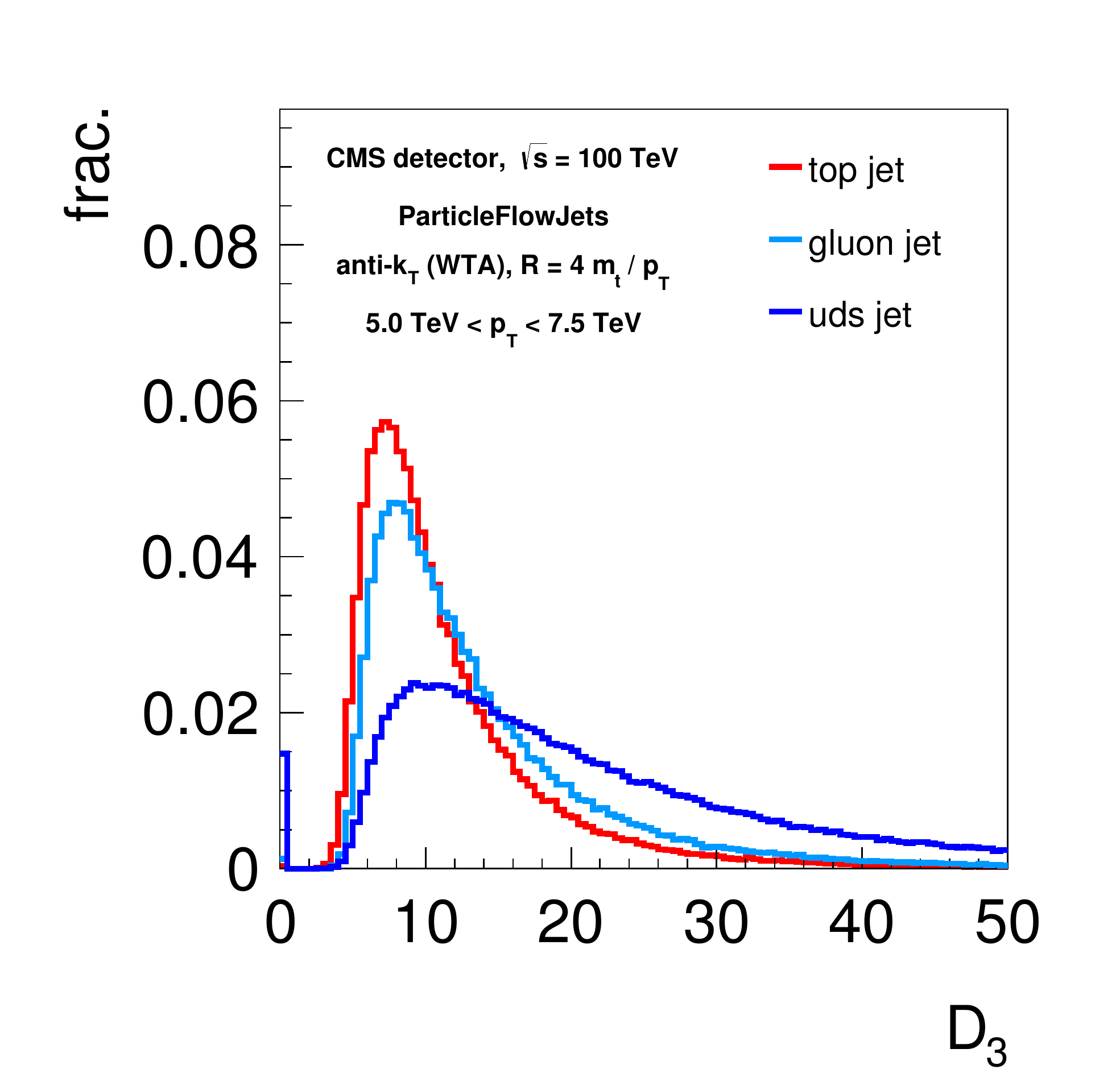}
}
\subfloat[]{
\includegraphics[width=6cm]{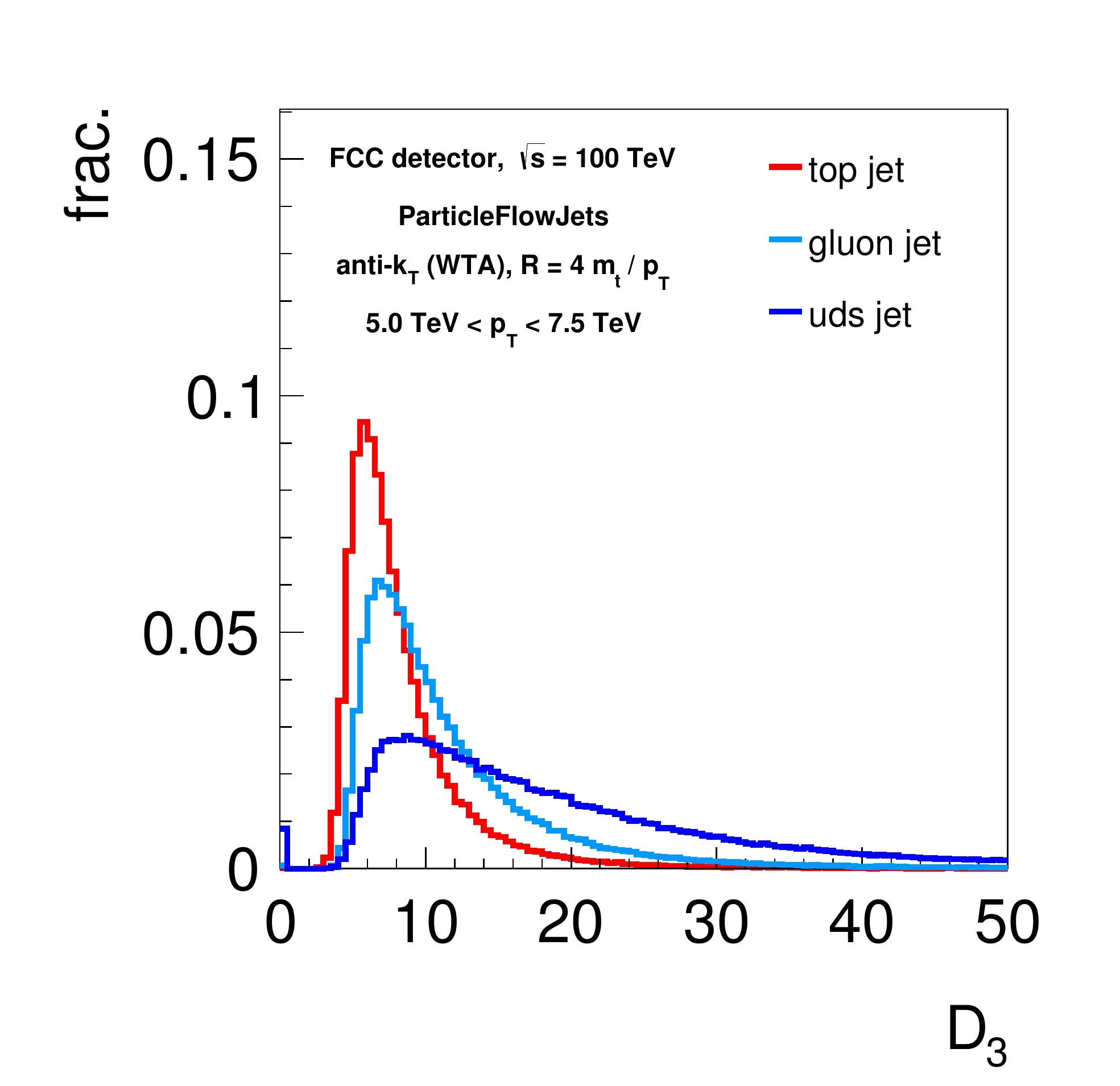}
}
\end{center}
\caption{
Distributions of the rescaled track-based jet mass (top), $\tau_3 / \tau_2$ (middle), and $D_3$ (bottom) as measured on anti-$k_T$ jets with radius $R=4 m_\text{top}/p_T$ and $p_T \in [5.0,7.5]$ TeV on boosted top jets and QCD jets from light quarks and gluons.  (left) Distributions as measured from the CMS detector's tracking system.  (right) Distributions as measured from a future collider detector's tracking system.
}
\label{fig:track_scale_r32_5}
\end{figure}

\begin{figure}[t]
\begin{center}
\subfloat[]{
\includegraphics[width=6cm]{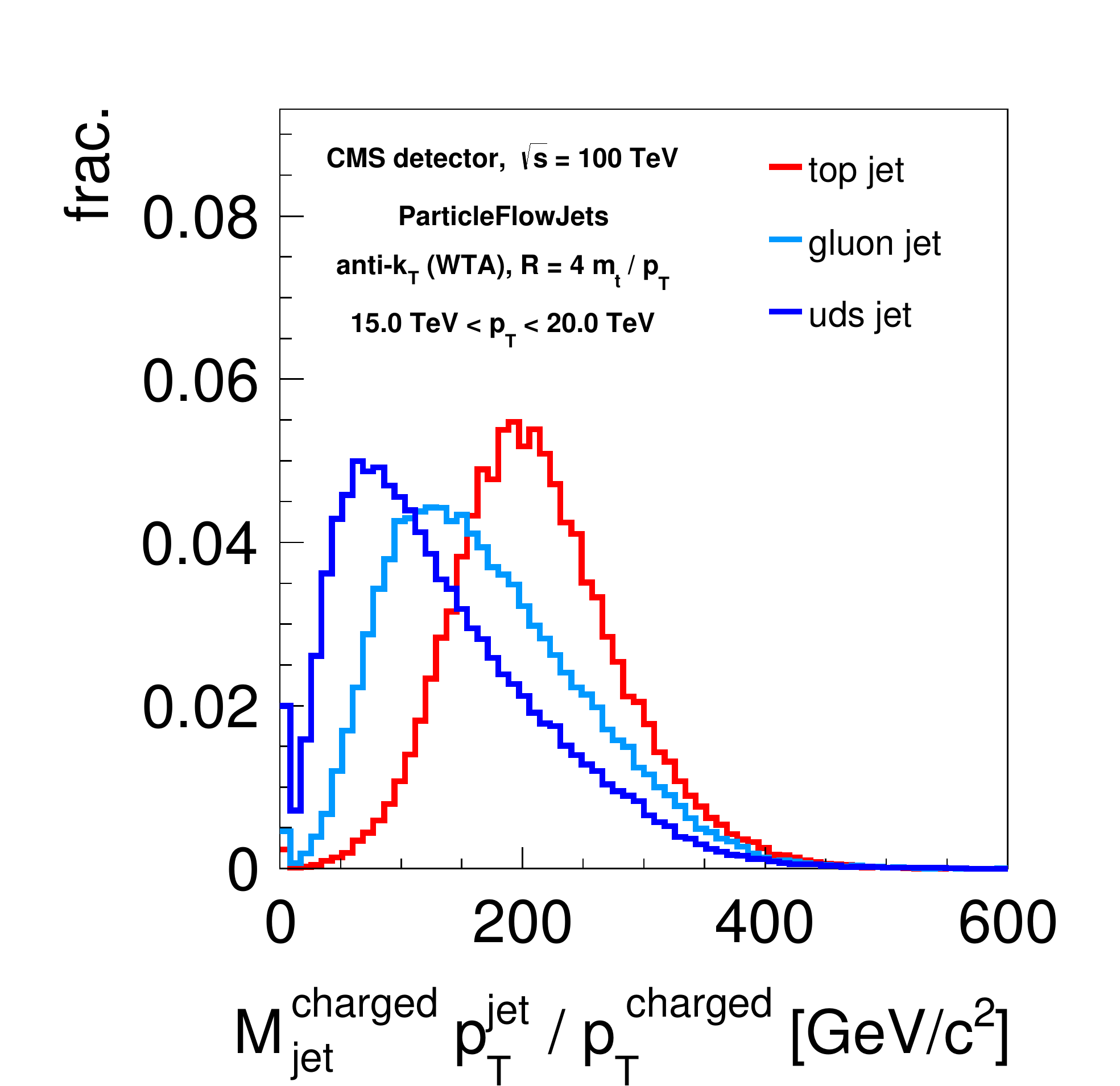}
}
\subfloat[]{
\includegraphics[width=6cm]{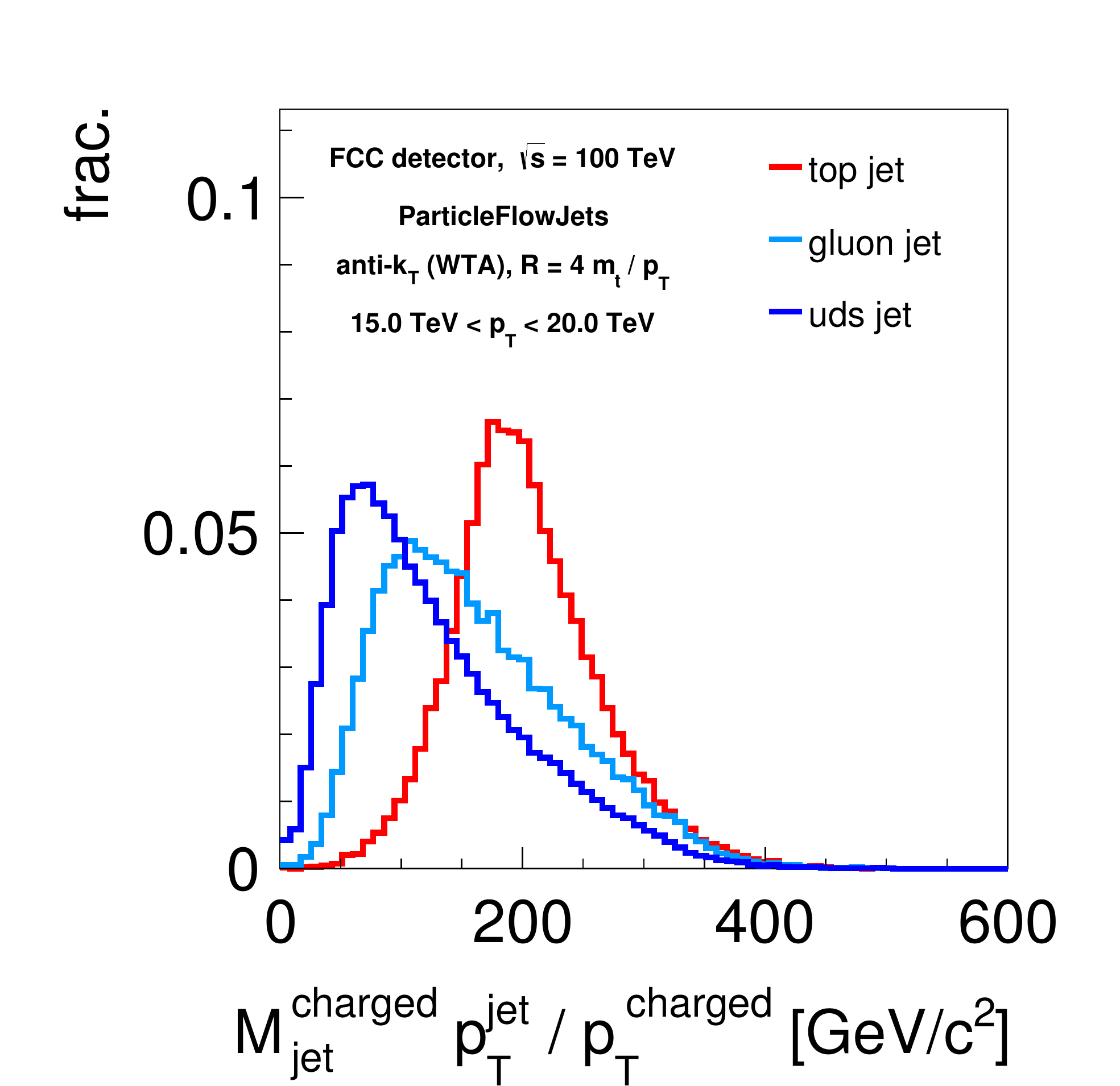}
}
\\
\subfloat[]{
\includegraphics[width=6cm]{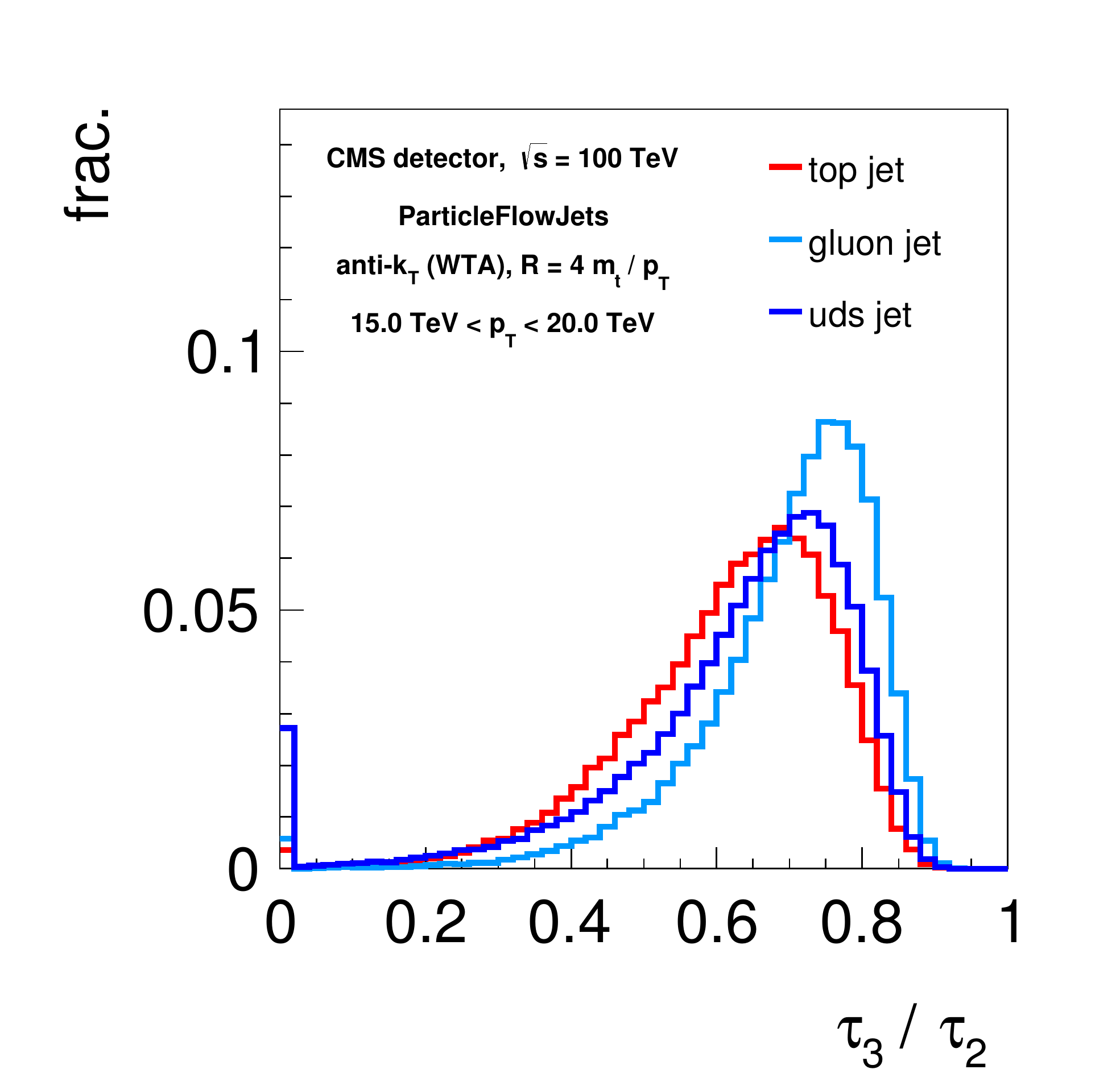}
}
\subfloat[]{
\includegraphics[width=6cm]{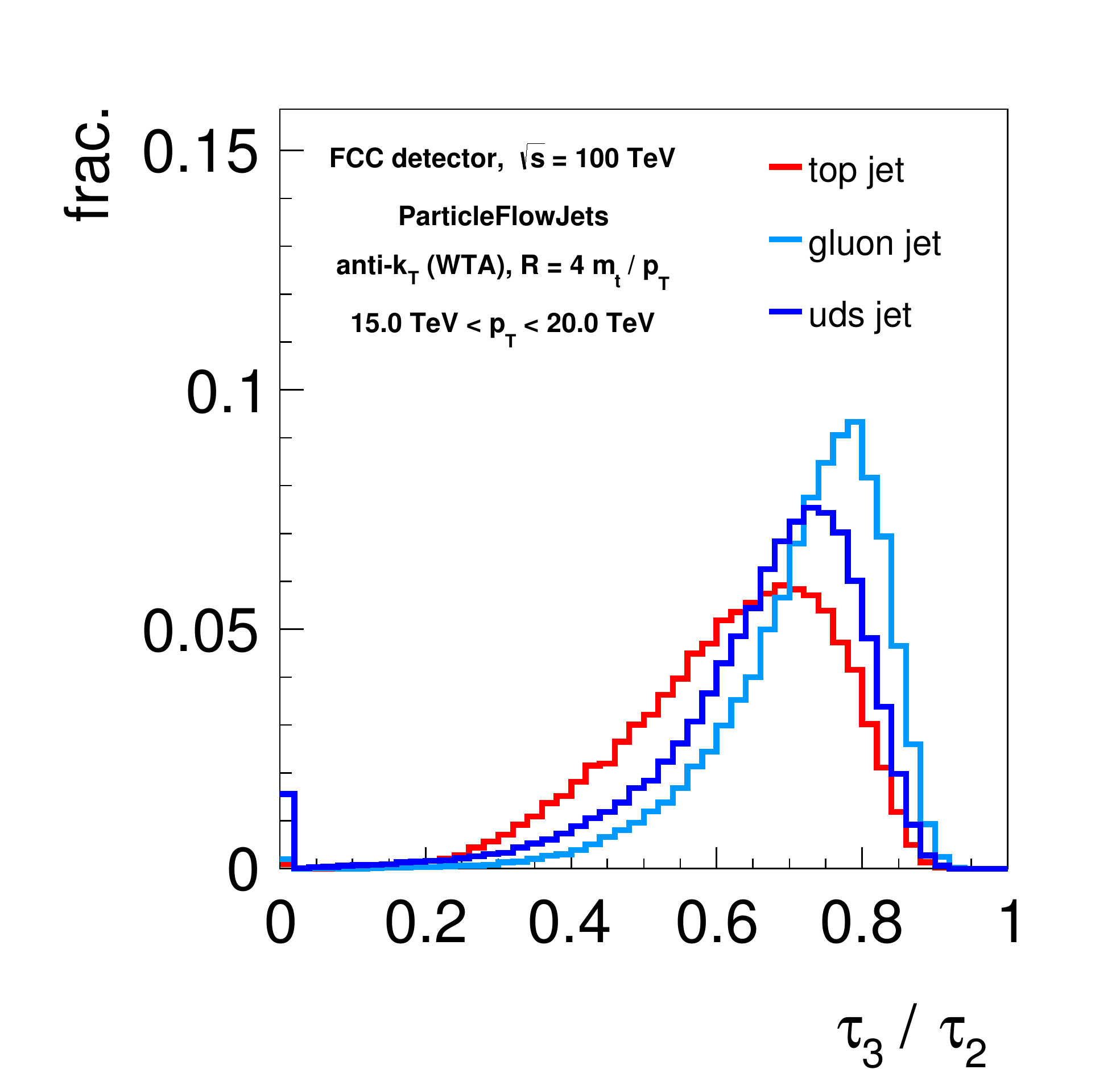}
}
\\
\subfloat[]{
\includegraphics[width=6cm]{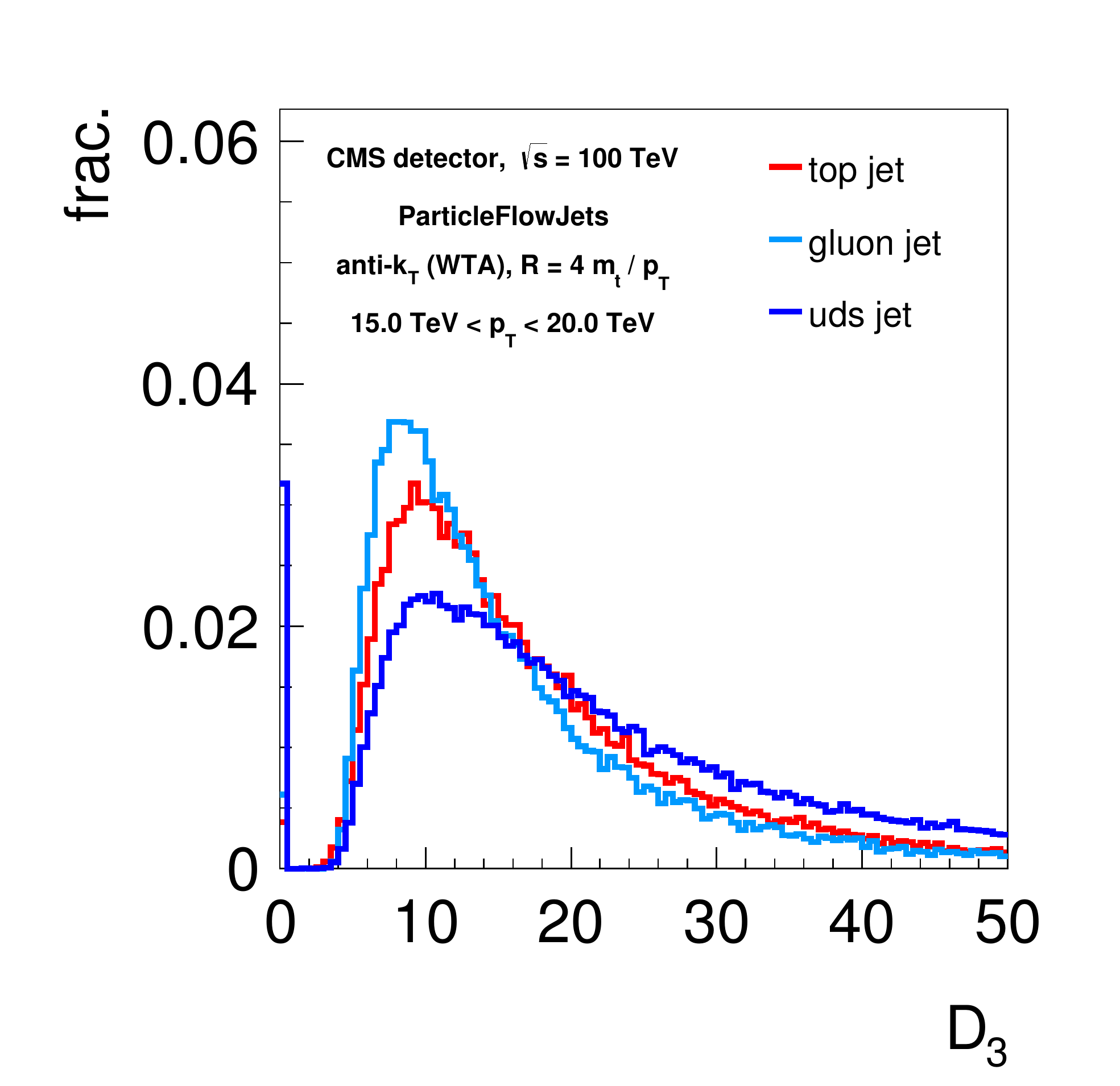}
}
\subfloat[]{
\includegraphics[width=6cm]{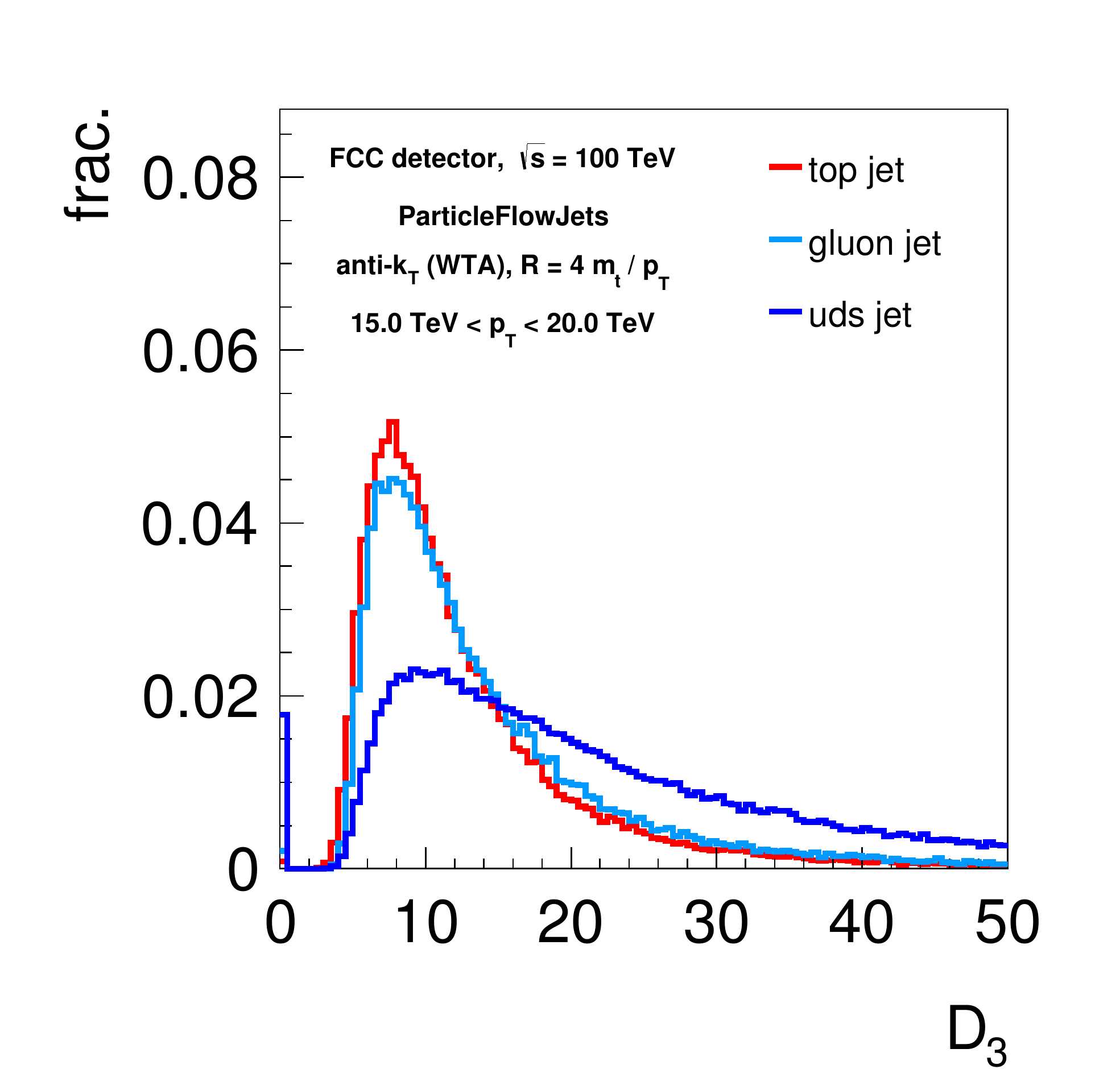}
}
\end{center}
\caption{
Distributions of the rescaled track-based jet mass (top), $\tau_3 / \tau_2$ (middle), and $D_3$ (bottom) as measured on anti-$k_T$ jets with radius $R=4 m_\text{top}/p_T$ and $p_T \in [15,20]$ TeV on boosted top jets and QCD jets from light quarks and gluons.  (left) Distributions as measured from the CMS detector's tracking system.  (right) Distributions as measured from a future collider detector's tracking system.
}
\label{fig:track_scale_r32_15}
\end{figure}

\begin{figure}
\begin{center}
\subfloat[]{
\includegraphics[width=7.5cm]{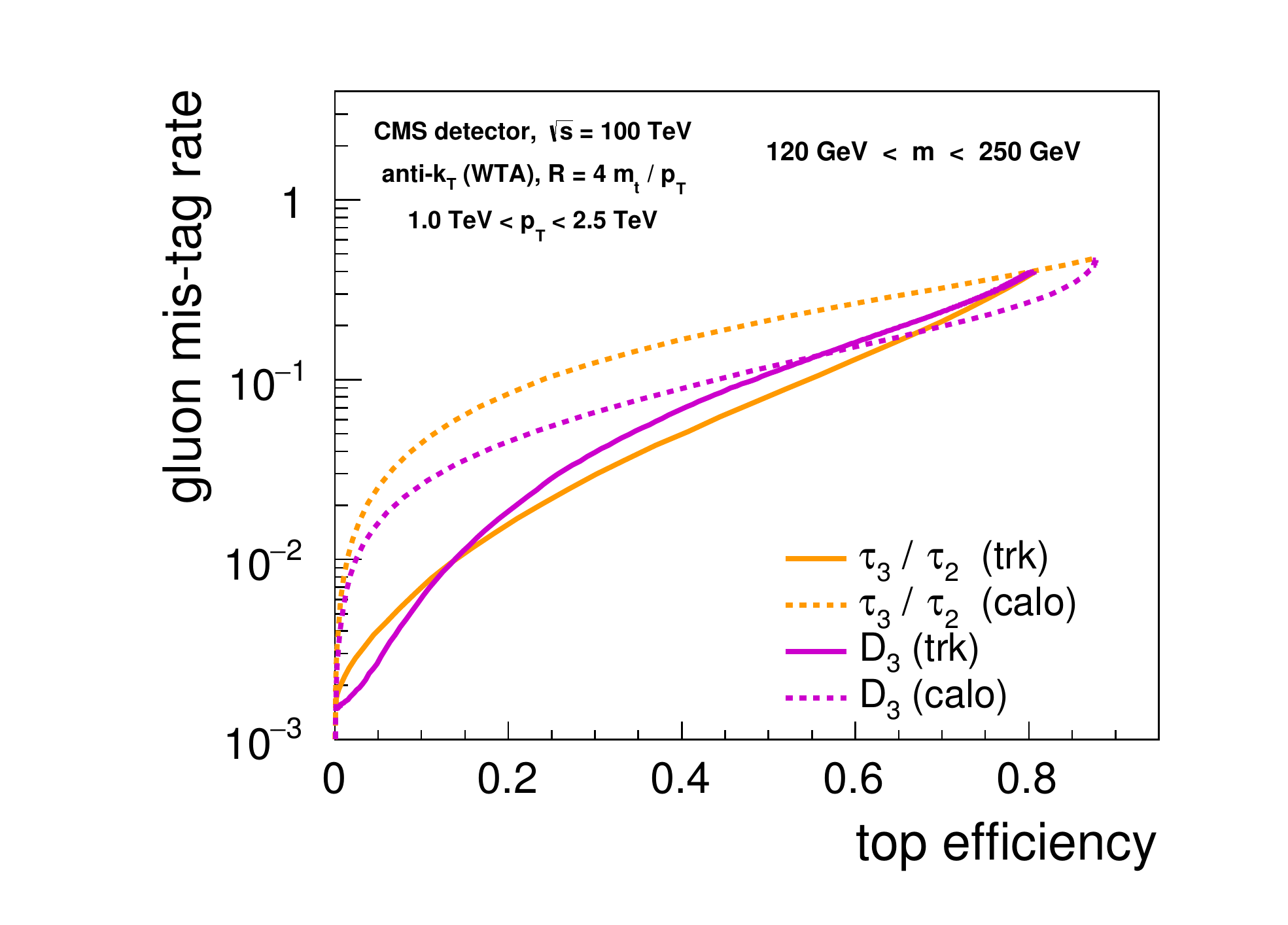}
}
\subfloat[]{
\includegraphics[width=7.5cm]{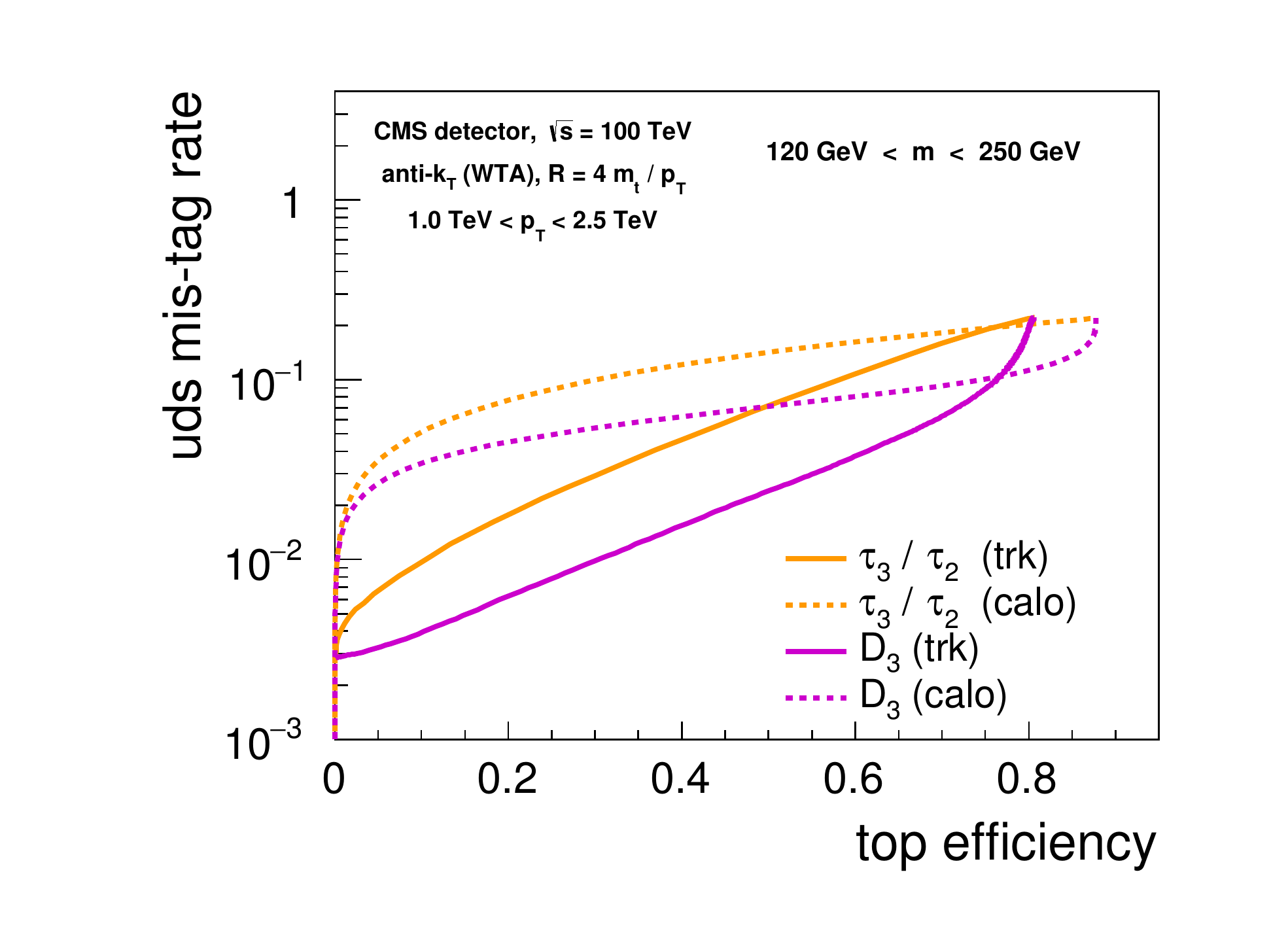}
}
\\
\subfloat[]{
\includegraphics[width=7.5cm]{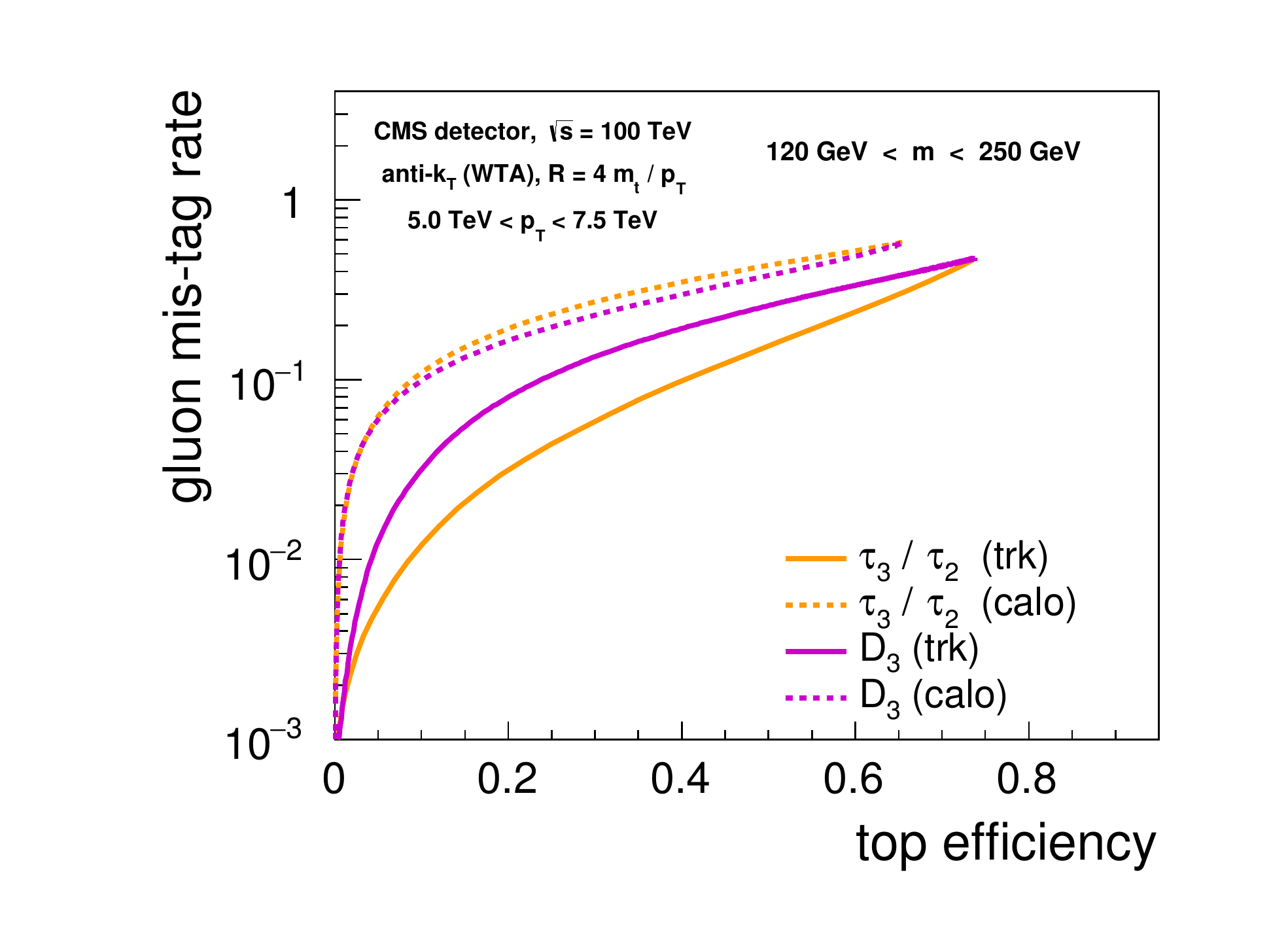}
}
\subfloat[]{
\includegraphics[width=7.5cm]{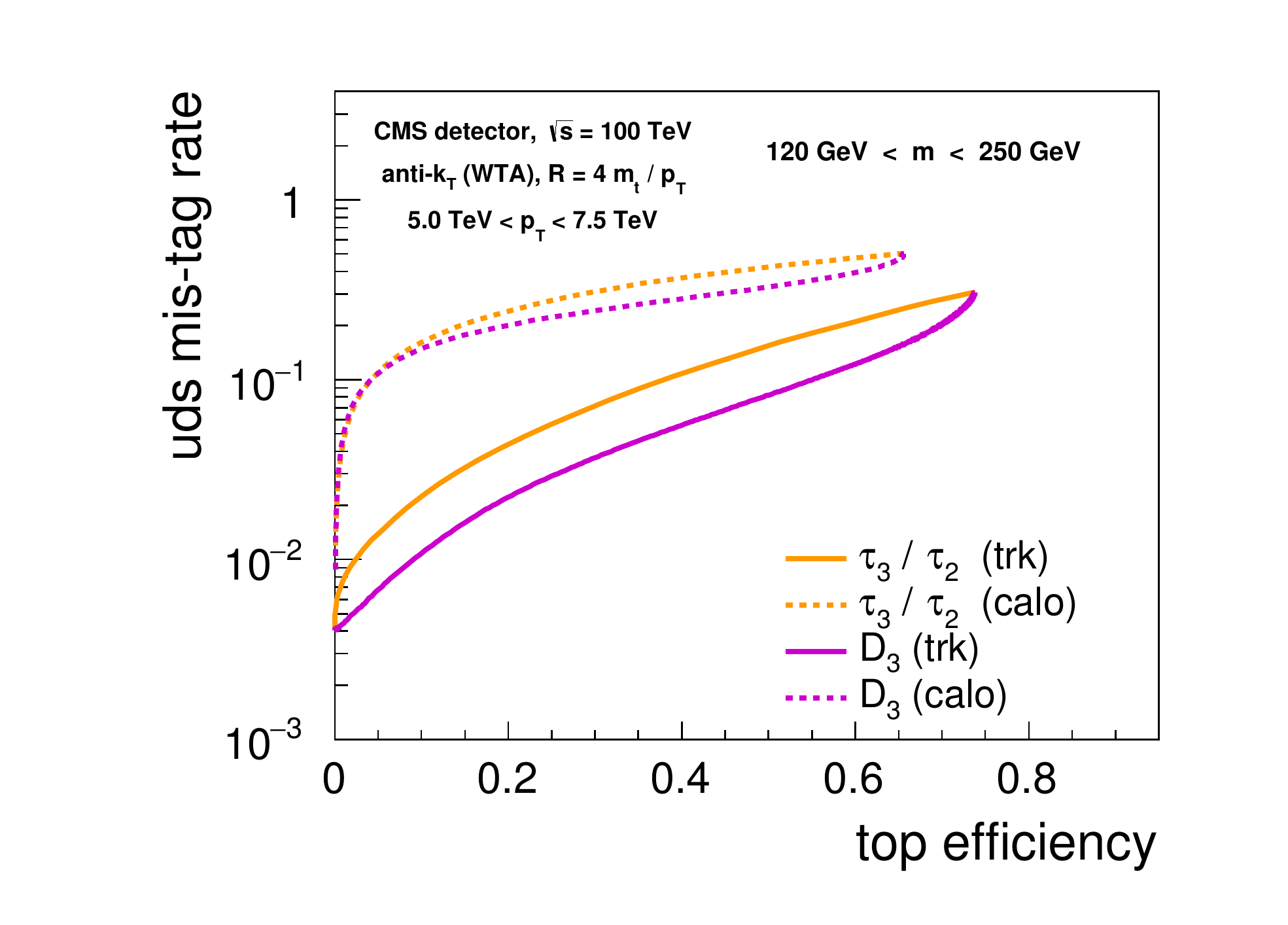}
}
\\
\subfloat[]{
\includegraphics[width=7.5cm]{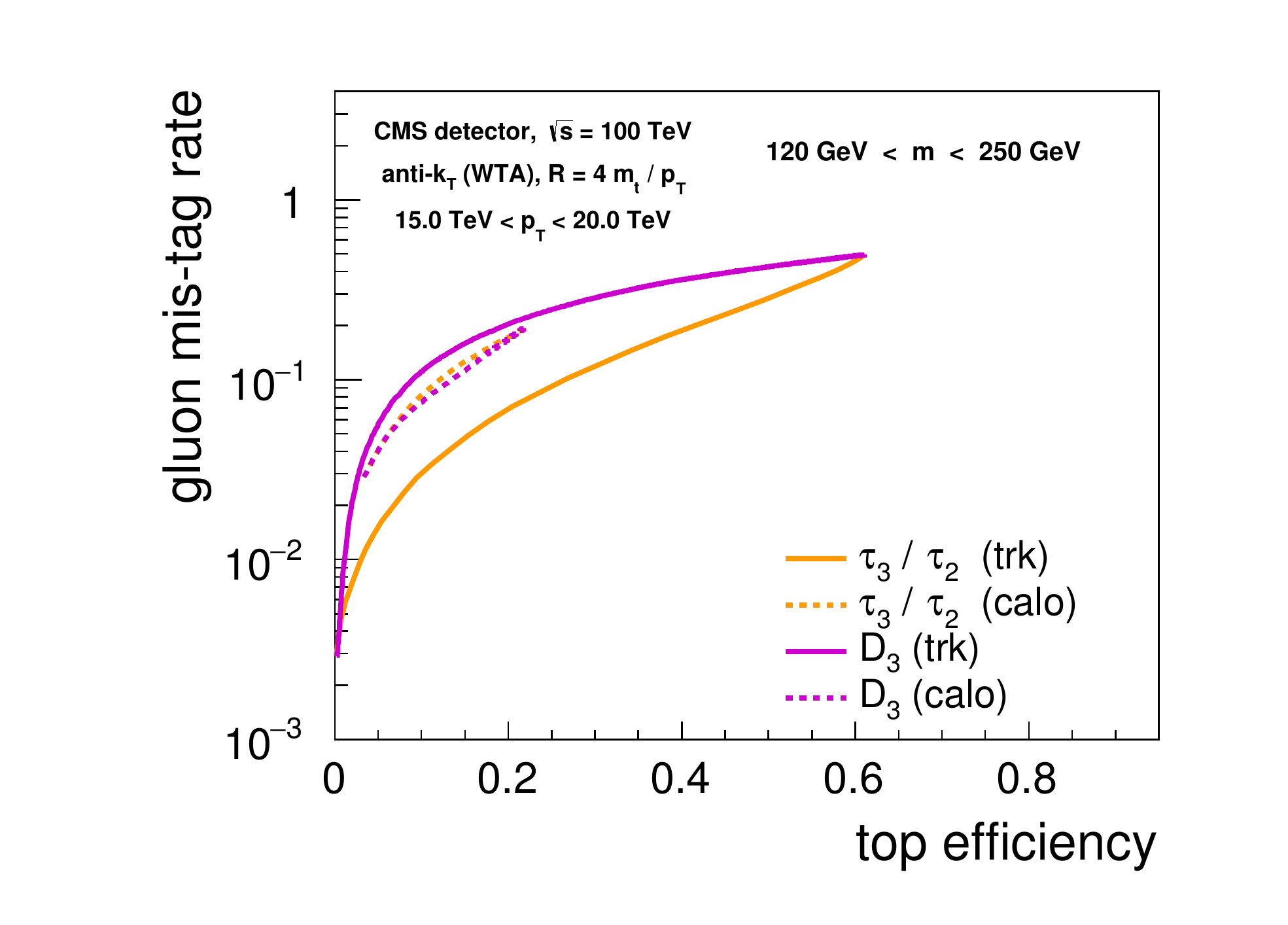}
}
\subfloat[]{
\includegraphics[width=7.5cm]{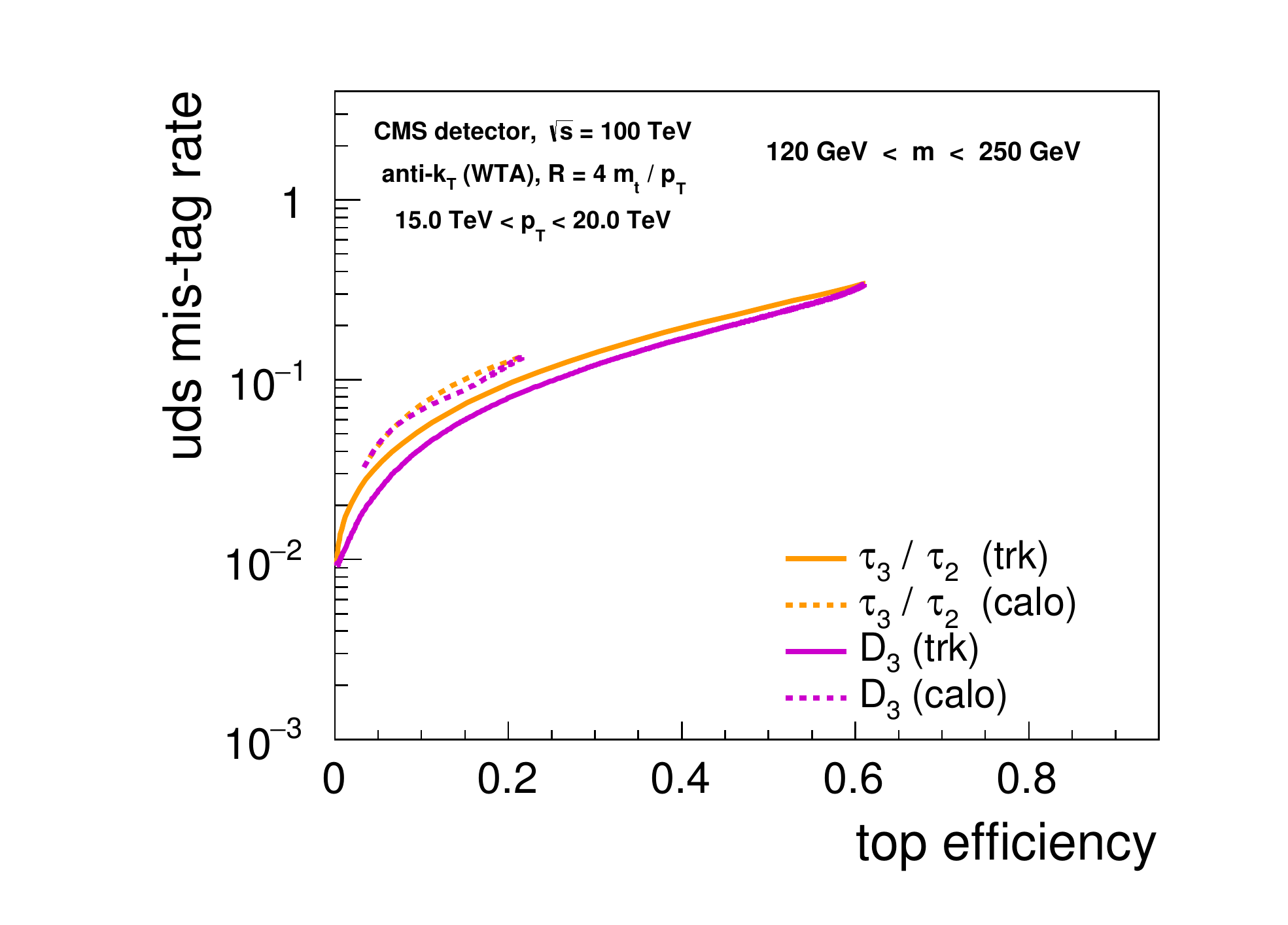}
}
\end{center}
\caption{
Signal vs.~background efficiency (ROC) curves for top quark identification from QCD background utilising $\tau_{3,2}$ and $D_3$ with the CMS detector for three $p_T$ bins: $[1.0,2.5]$ TeV (top), $[5.0,7.5]$ TeV (middle), $[15,20]$ TeV (bottom). (left) top quarks vs.~gluon jets, (right) top quarks vs.~light quark jets.  The cut on the jet mass of $m\in[120,250]$ GeV is included in the efficiencies.
}
\label{fig:roc_curves_CMS}
\end{figure}

In this appendix, we present the results of our top quark identification procedure over a wide range of jet $p_T$s at a 100 TeV future collider.  A discussion of the physics or analysis of the results will be limited, as these were discussed in detail in the text.  In Figs.~\ref{fig:cal_scale_r32_1}, \ref{fig:cal_scale_r32_5} and \ref{fig:cal_scale_r32_15}, we plot the jet mass, $\tau_{3,2}$ and $D_3$ distributions as measured on calorimeter jets with radius $R=4m_\text{top}/p_T$.  As the jet $p_T$ increases, finite resolution effects dominate, and especially in the CMS detector for jets in the highest $p_T$ bin, a large fraction of the jets consist of a single calorimeter cell.  In Figs.~\ref{fig:track_scale_r32_2}, \ref{fig:track_scale_r32_5} and \ref{fig:track_scale_r32_15}, we plot the rescaled track-based mass, $\tau_{3,2}$ and $D_3$ distributions on charged tracks for jets with a radius $R=4m_\text{top}/p_T$ for both signal and background jets.  \Figs{fig:roc_curves_CMS}{fig:roc_curves_FCC} show the signal vs.~background efficiency curves for a simulated CMS or FCC detector, respectively, at a future 100 TeV collider.  In these figures, we compare the discrimination power of the jet substructure observables $\tau_{3,2}$ and $D_3$ over a wide range of jet $p_T$s as measured on calorimeter or track-based jets.  As with the $p_T \in [7.5,10]$ TeV bin studied in \Sec{sec:onept}, discrimination power is significantly improved by working with track-based observables.  In addition, $\tau_{3,2}$ is a better discriminant over a wide range of signal efficiencies than $D_3$ for background gluon jets, but the improvement decreases as $p_T$ increases.  $D_3$, on the other hand, is a much more powerful observable for rejecting light quark background jets than $\tau_{3,2}$.  These features are further illustrated in \Tab{tab:rej_pt} where we list the rejection rates ($=100\%\,-\,$background efficiency) for benchmark signal efficiencies, for jets with $p_T$s ranging from 2.5 to 20 TeV.

\begin{figure}
\begin{center}
\subfloat[]{
\includegraphics[width=7.5cm]{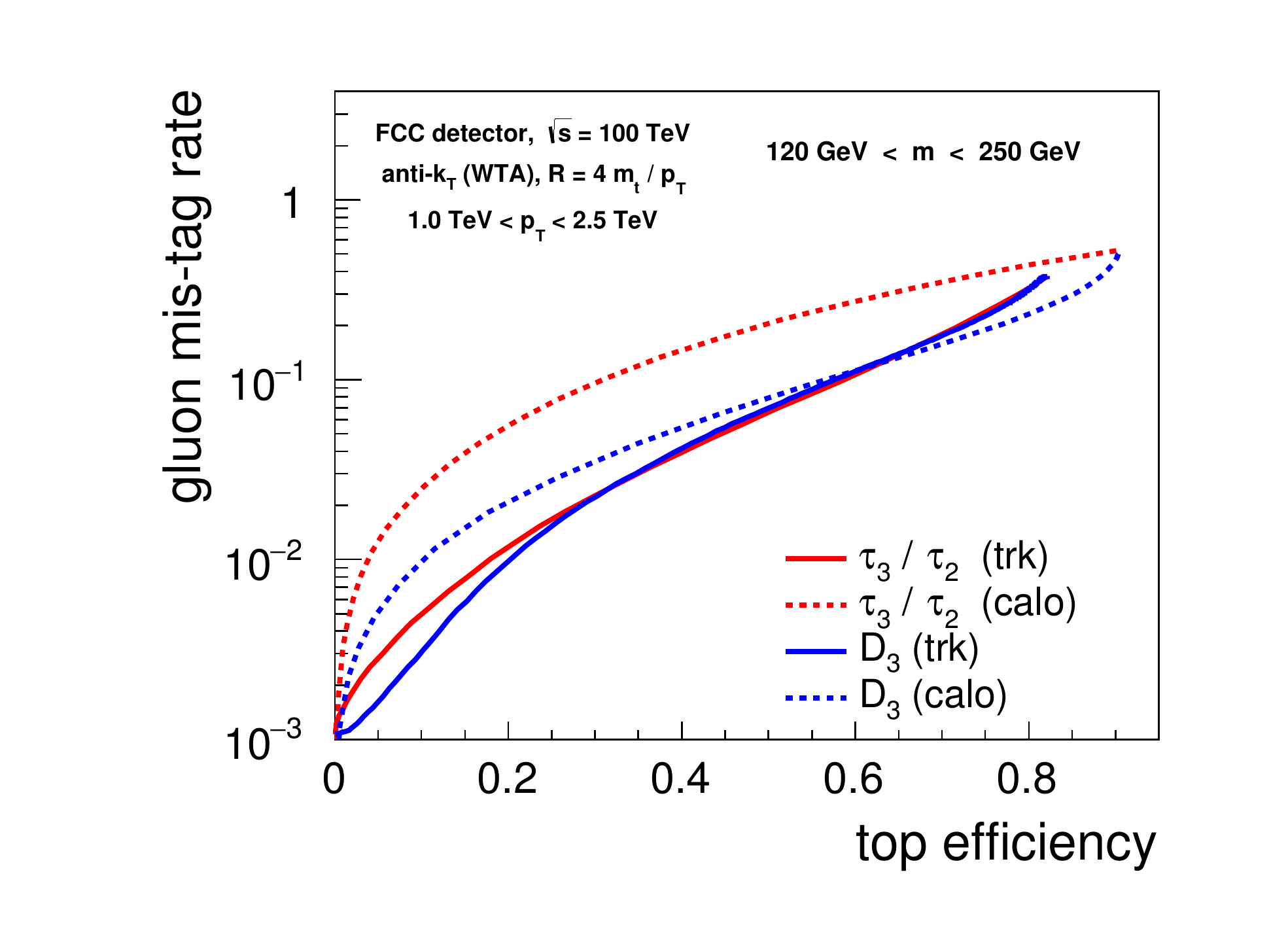}
}
\subfloat[]{
\includegraphics[width=7.5cm]{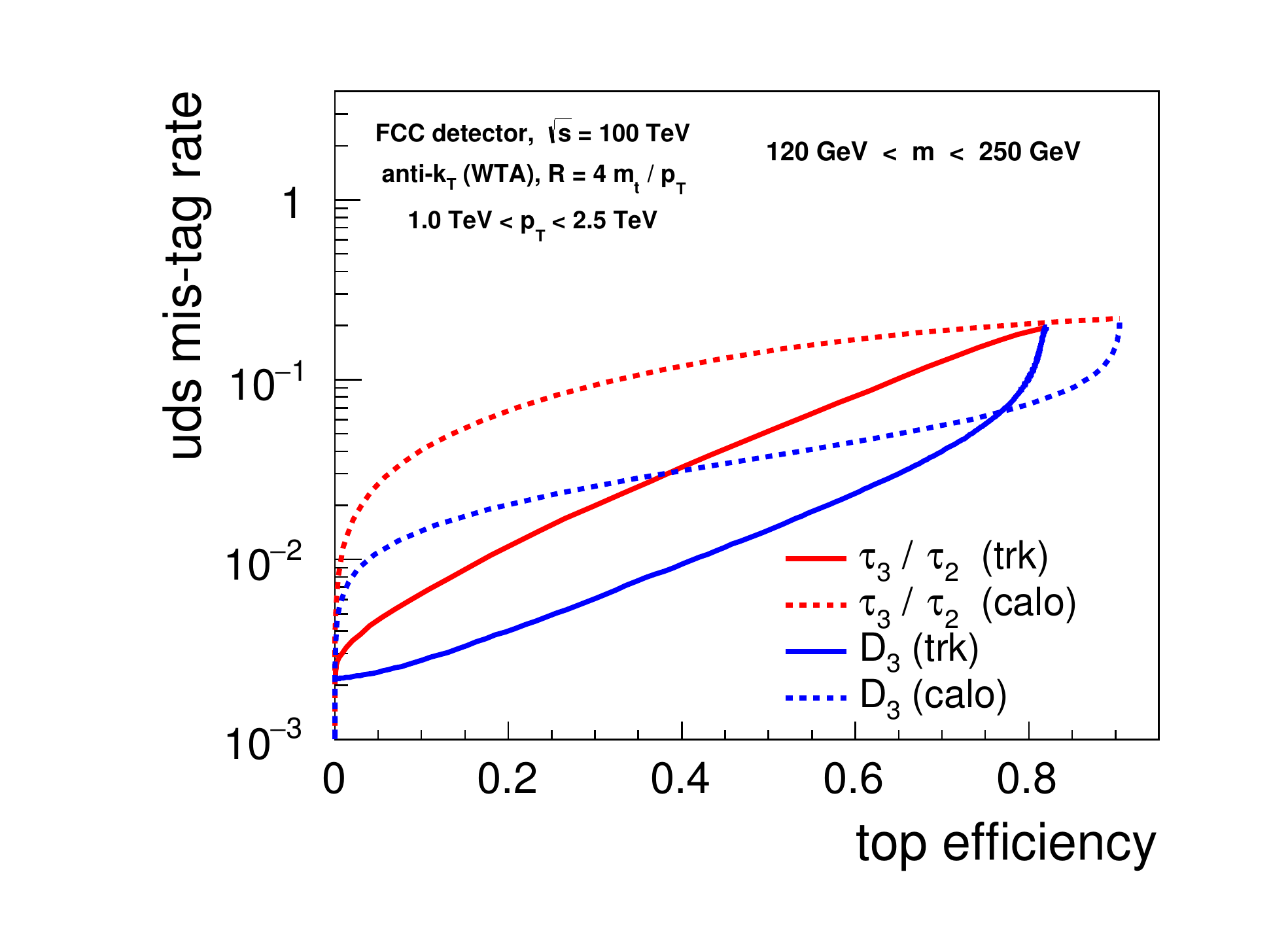}
}
\\
\subfloat[]{
\includegraphics[width=7.5cm]{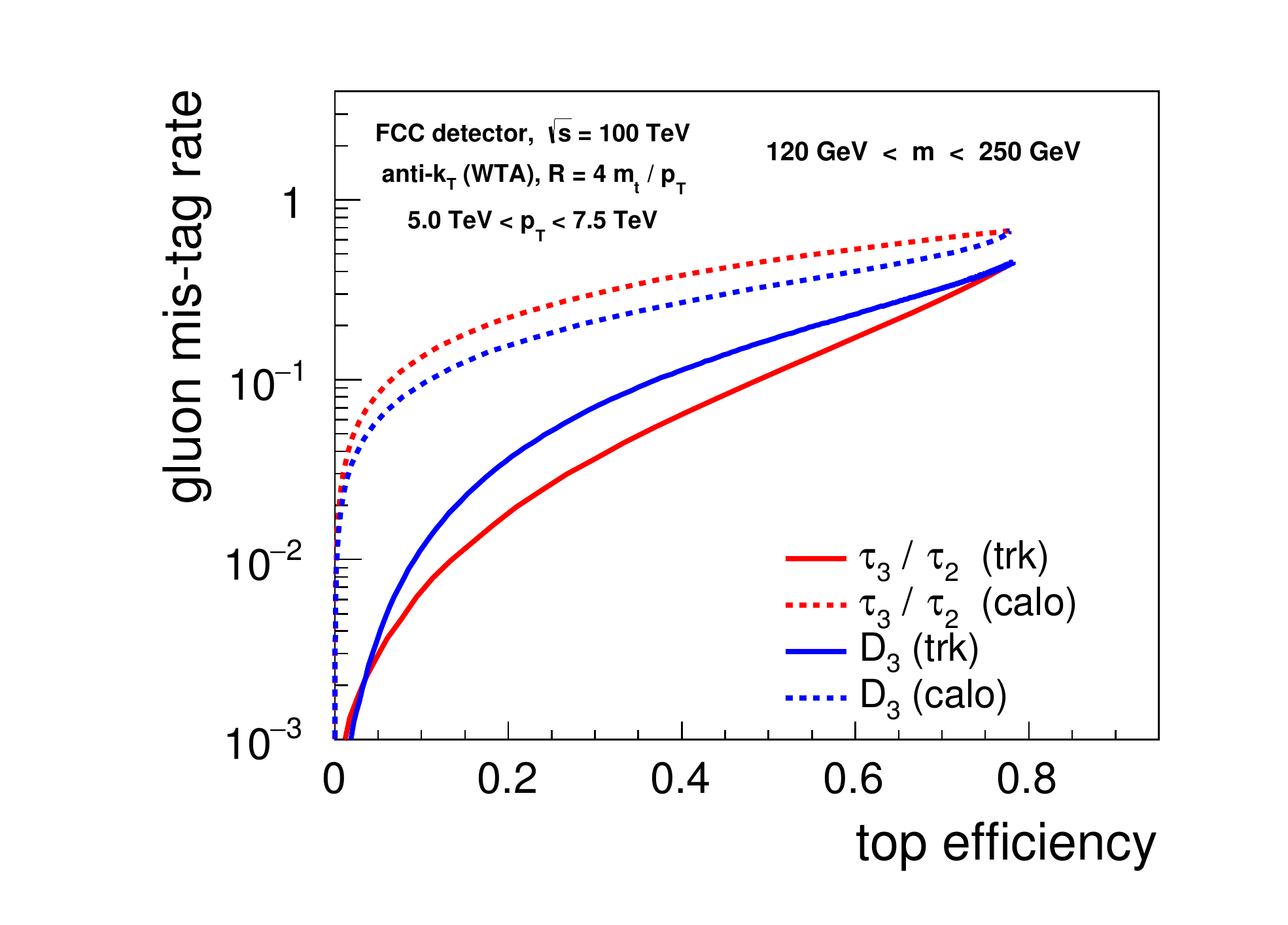}
}
\subfloat[]{
\includegraphics[width=7.5cm]{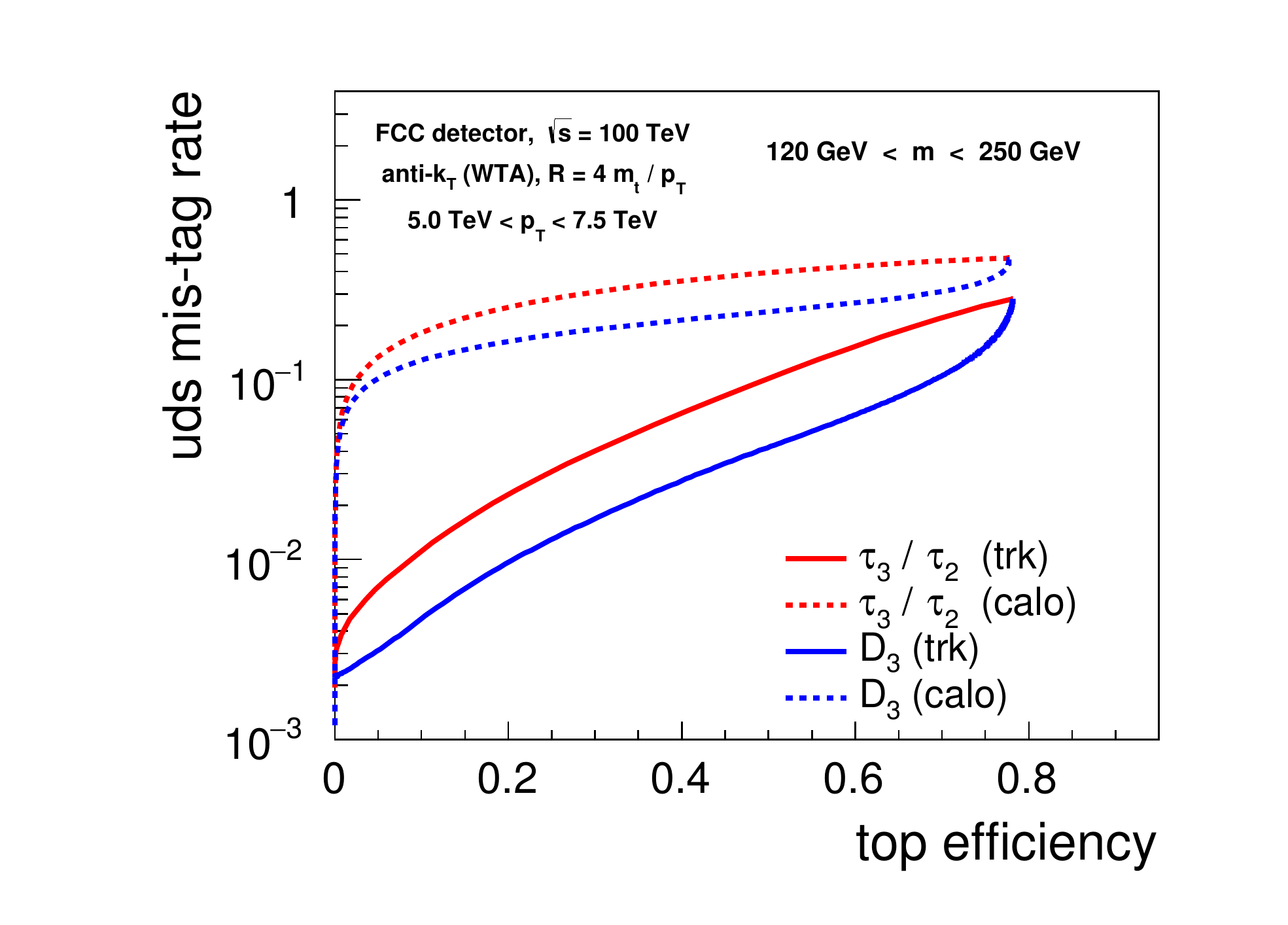}
}
\\
\subfloat[]{
\includegraphics[width=7.5cm]{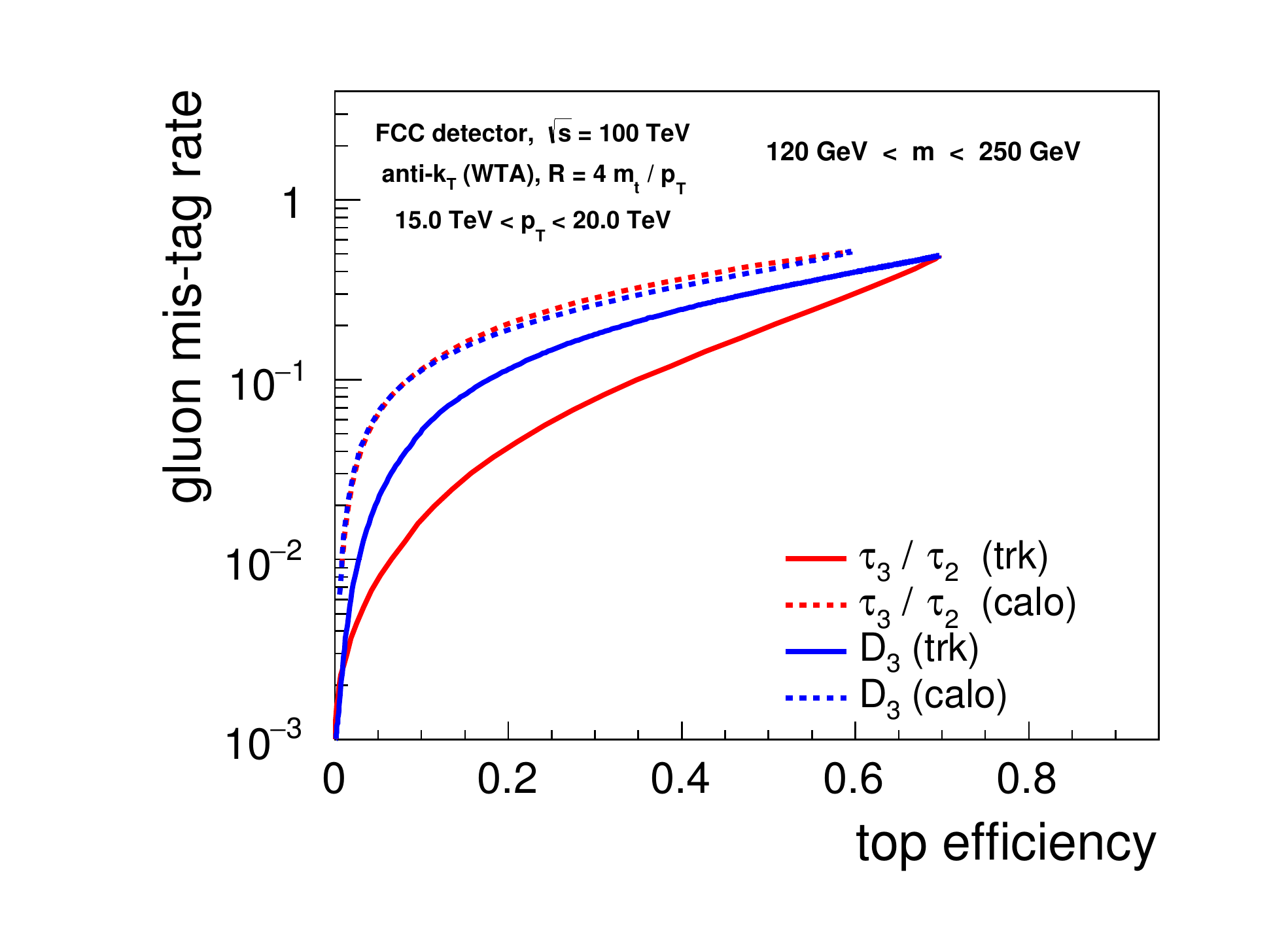}
}
\subfloat[]{
\includegraphics[width=7.5cm]{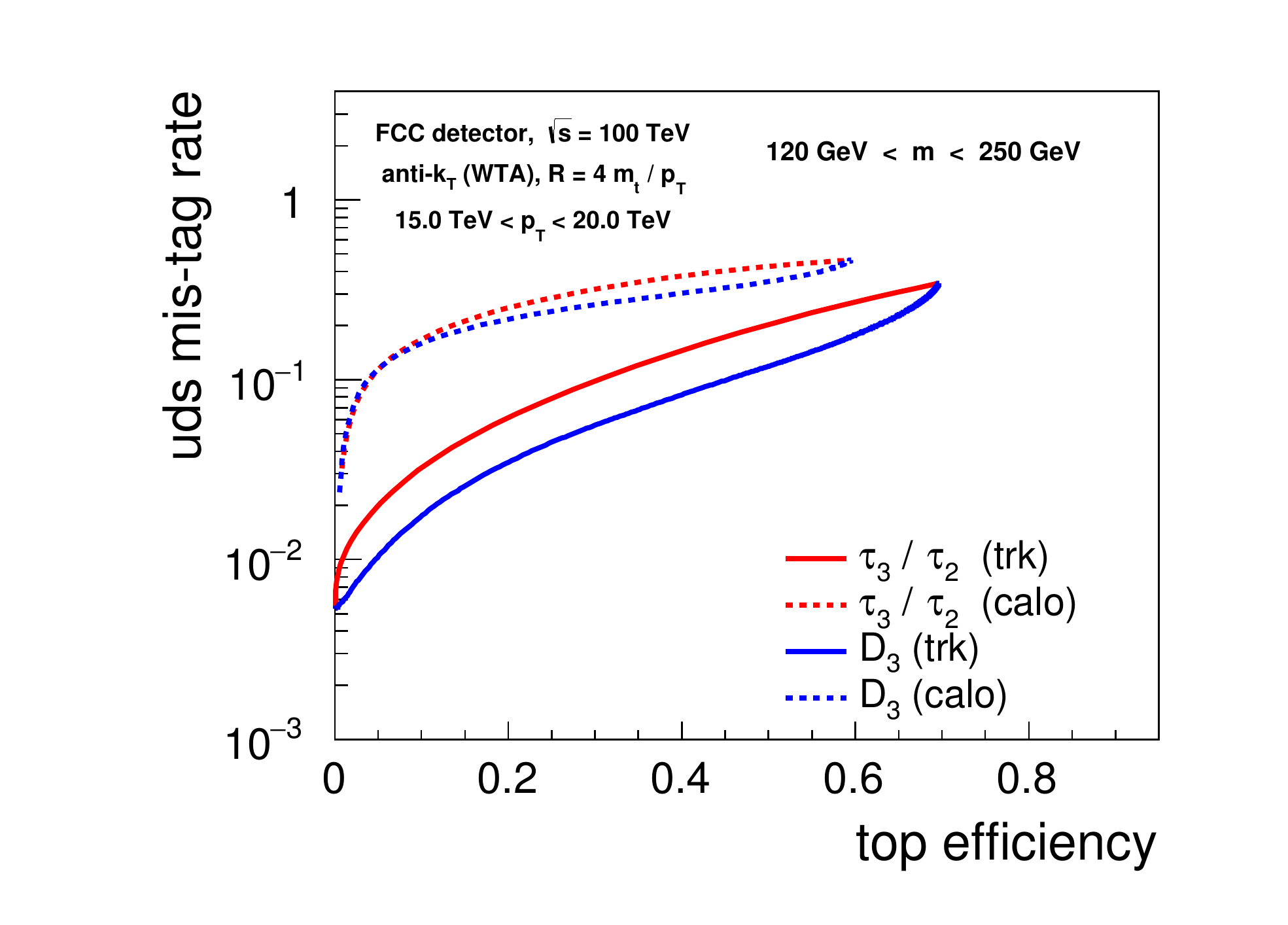}
}
\end{center}
\caption{
Signal vs.~background efficiency (ROC) curves for top quark identification from QCD background utilising $\tau_{3,2}$ and $D_3$ with the FCC detector for three $p_T$ bins: $[1.0,2.5]$ TeV (top), $[5.0,7.5]$ TeV (middle), $[15,20]$ TeV (bottom). (left) top quarks vs.~gluon jets, (right) top quarks vs.~light quark jets.  The cut on the jet mass of $m\in[120,250]$ GeV is included in the efficiencies.
}
\label{fig:roc_curves_FCC}
\end{figure}

\bibliography{BoostedTops}

\end{document}